\documentclass[a4paper,pra,twocolumn,nofootinbib]{revtex4}
\usepackage[T1]{fontenc}
\usepackage{times}

\usepackage{amsmath}
\usepackage{amsthm}
\usepackage{amssymb}
\usepackage[utf8]{inputenc}
\usepackage{color}
\usepackage{hyperref}

\usepackage{tikz}
\usetikzlibrary{arrows.meta}

\usetikzlibrary{decorations.markings}
\tikzset{->-/.style={decoration={
  markings,
  mark=at position #1 with {\arrow{>}}},postaction={decorate}}}
\tikzset{-<-/.style={decoration={
  markings,
  mark=at position #1 with {\arrow{<}}},postaction={decorate}}}
\usetikzlibrary{external}
\def\Real{{\mathbb R}}

\def\innerprod(#1,#2){{\left<#1\,,\,#2\right>}}
\def\Set#1{{\left\{#1\right\}}}
\def\qquadtext#1{\qquad\textup{#1}\qquad}
\def\qquadand{\qquadtext{and}}
\def\quadtext#1{\quad\textup{#1}\quad}
\def\quadand{\quadtext{and}}
\def\pfrac#1#2{\frac{\partial #1}{\partial #2}}

\def\pqfrac#1#2#3{\frac{\partial^2 #1}{\partial {#2}\partial{#3}}}
\def\dfrac#1#2{\frac{d #1}{d #2}}

\def\VAct#1{\!\left\langle #1 \right\rangle}
\def\Vp{{\boldsymbol{p}}}

\def\xhat{{\hat x}}

\newcommand{\XDOI}[1]{\href{http://dx.doi.org/#1}{doi:#1}}
\newcommand{\XARXIV}[1]{\href{http://arxiv.org/abs/#1}{arXiv:#1}}
\newcommand{\XWEB}[1]{\href{#1}{#1}}

\def\grad{\nabla}
\def\Vzero{{\boldsymbol 0}}

\def\emKappa{\kappa}   % the EM constitutive tensor

\newcommand{\pVec}[1]{\boldsymbol{#1}}

\def\pE{E} % a scalar electric field
\def\pB{B} % a scalar magnetic field
\def\pJ{J} % a scalar current, probably a free current
\def\pS{C} % a scalar electric charge density
\def\pA{A} % a scalar electric charge polarization density

\def\VE{\pVec{\pE}} % the EM electric field
\def\VB{\pVec{\pB}} % the EM magnetic field
\def\VD{\pVec{D}} % the EM electric excitation field (displacement field)
\def\VH{\pVec{H}} % the EM magnetic excitation field
\def\VA{\pVec{\pA}} % electromagnetic, vector potential
\def\VJ{\pVec{\pJ}} % electromagnetic, current vector
\def\Vk{\pVec{k}}
\def\Vzeta{\pVec{\zeta}}
\def\gauge{{\textup{g}}}
\def\axion{{\textup{ax\,}}}

\def\EMepsilon{\varepsilon}  % the EM electric permittivity
\def\EMmu{\mu}               % the EM magnetic premeability
\def\epsilonTen{{{\EMepsilon}\,}}
\def\muTen{{{\EMmu}\,}}

\def\emScalar{\phi} % the EM vector potential
\def\emVector{\VA}  % the EM scalar potential

\def\gScalar{\phi_{\gauge}} % scalar gauge for D,H
\def\gVector{\VA_{\gauge}}  % vector gauge for D,H

\def\eCharge{\rho} % the EM charge density
\def\eCurrent{\VJ} % the EM current density

\newcommand\xpair[2]{($#1$, $#2$)}
\newcommand\xtrio[3]{($#1$, $#2$, $#3$)}
\newcommand\xquad[4]{($#1$, $#2$, $#3$, $#4$)}

\newcommand\dTime{\partial_t}        % old partials
\newcommand\dtime[1]{\partial_t{#1}} % old over-dots

\def\Stress{{S}}
\def\TenStress{\bar{\bar{\Stress}}}

\def\PsiAny{{\Psi}}
\newcommand\xPsi[2]{{\PsiAny}^{#1}_{#2}}

\def\PsiErho{\xPsi{E}{\eCharge}}
\def\PsiBrho{\xPsi{B}{\eCharge}}
\def\PsiEJ{\xPsi{E}{\pJ}}
\def\PsiBJ{\xPsi{B}{\pJ}}

\newcommand{\NullPsi}{%
  \mathrel{\vbox{\offinterlineskip\ialign{%
    \hfil##\hfil\cr
    $\scriptscriptstyle\circ$\cr
    \noalign{\kern0.1ex}
    $\PsiAny$\cr
}}}}

\newcommand{\NullPsiIndices}[1]{{%
  \mathrel{\vbox{\offinterlineskip\ialign{%
    \hfil##\hfil\cr
    $\scriptscriptstyle\circ$\cr
    \noalign{\kern0.1ex}
    $\PsiAny$\cr
}}}}\vphantom{\PsiAny}^{#1}}

\newcommand\nPsi[2]{{\NullPsi}{}^{#1}_{#2}}
\def\NullPsiErho{\nPsi{E}{\rho}}
\def\NullPsiBrho{\nPsi{B}{\rho}}
\def\NullPsiEJ{\nPsi{E}{J}}
\def\NullPsiBJ{\nPsi{B}{J}}

\def\TenKappa{\bar{\bar{\kappa}}}
\def\kappaDE{\TenKappa^{\textup{DE}}}
\def\kappaDB{\TenKappa^{\textup{DB}}}
\def\kappaHE{\TenKappa^{\textup{HE}}}
\def\kappaHB{\TenKappa^{\textup{HB}}}

\def\Jhat{{\hat{J}}}
\def\Psihat{{\hat{\Psi}}}
\def\Fhat{{\hat{F}}}

\def\ahat{{\hat{a}}}
\def\bhat{{\hat{b}}}
\def\chat{{\hat{c}}}
\def\dhat{{\hat{d}}}

\def\AxisR{\hat{\boldsymbol{r}}}          % r unit vector, was \Vrhat
\def\AxisTheta{\hat{\boldsymbol{\theta}}} % theta unit vector, was \Vthetahat
\def\AxisX{\hat{\boldsymbol{x}}}     % was \Vi
\def\AxisY{\hat{\boldsymbol{y}}}     % was \Vj
\def\AxisZ{\hat{\boldsymbol{z}}}     % was \Vk

\definecolor{XRED}{rgb}{0.70, 0.01, 0.01}
\definecolor{XBLUE}{rgb}{0.01, 0.01, 0.70}

\def\axiva{\zeta}
\def\axivav{\Vzeta}
\def\axivar{\axiva_r}
\def\axivaa{\axiva_\theta}
\def\axivaz{\axiva_z}
\def\axivat{\axiva_t}
\def\axivax{\axiva_x}
\def\axivay{\axiva_y}

\def\axipot{{\kappa}_{\textup{ax}}}  % an axion field
\def\axiphi{{\Xi}}  % an axion field

\def\pAxionResponse{axionic response}
\def\pAxionREA{{\pAxionResponse} material}

\newcommand\ResubOne[1]{#1} %{\color{red}#1}}

\begin{document}

\title{Electromagnetism, Axions, and Topology: \\
 a first-order operator approach
          to constitutive responses provides greater freedom}

\author{Jonathan Gratus$^{1,2}$}
\homepage[]{https://orcid.org/0000-0003-1597-6084}
\author{Martin W. McCall$^3$}
\homepage[]{https://orcid.org/0000-0003-0643-7169}
\author{Paul Kinsler$^{1,2}$}
\homepage[]{https://orcid.org/0000-0001-5744-8146}

\affiliation{$^1$ %\address{%
  Department of Physics,
  Lancaster University,
  Lancaster LA1 4YB,
  United Kingdom,
}
\affiliation{$^2$ %\address{%
The Cockcroft Institute,
Sci-Tech Daresbury,
Daresbury WA4 4AD,
% Warrington WA4 4AD
United Kingdom.
}
\affiliation{$^3$ %\address{%
  %Blackett Laboratory, Imperial College,
  Department of Physics,
  Imperial College London,
  Prince Consort Road,
  London SW7 2AZ,
  United Kingdom.
}

\begin{abstract}

We show how the standard constitutive assumptions
 for the macroscopic Maxwell equations can be relaxed.
This is done by arguing that the
 Maxwellian excitation fields {\xpair{\VD}{\VH}}
 should be dispensed with,
 on the grounds that they
 (a) cannot be measured,
 and
 (b) act solely as gauge potentials for the charge and current.
In the resulting theory,
 it is only the links between the fields {\xpair{\VE}{\VB}}
 and the charge and current {\xpair{\eCharge}{\eCurrent}} that matter;
 and so we introduce appropriate linear operator equations
 that combine the Gauss and Maxwell-Amp{\`e}re equations with 
 the constitutive relations, eliminating {\xpair{\VD}{\VH}}.
The result is that we can admit
 more types of electromagnetic media --
 notably,
 the new relations can allow coupling
 in the bulk to a homogeneous axionic material;
 in contrast to standard electromagnetism (EM)
 where any homogeneous axion-like field
 is completely decoupled in the bulk,
 and only accessible at boundaries.
We also consider a wider context,
 including the role of topology,
 extended non-axionic constitutive parameters,
 and treatment of Ohmic currents.
A range of examples including an {\pAxionREA} is presented,
 including static electromagnetic scenarios,
 a possible metamaterial implementation,
 and how the transformation optics paradigm would be modified.
Notably, 
 these examples include one where 
 topological considerations make it
 impossible to model using {\xpair{\VD}{\VH}}.

\end{abstract}

\maketitle

\vspace{-9mm}

{\small There is a popular summary in appendix \ref{S-popular}.}
\footnote[0]{Published as 
  \href{https://doi.org/10.1103/PhysRevA.101.043804}{Phys. Rev. A\textbf{101}, 043804 (2020)}
\\
Preprint version at
 \href{https://arxiv.org/abs/1911.12631}{ArXiv:1911.12631}.}

%
% ======================================================================
\section{Introduction}
\label{ch_IN}

Maxwell's equations rely
 \cite{Jackson-ClassicalED,RMC}
 on the electromagnetic
 constitutive relations (EMCR),
 which provide the crucial information relating
 the excitation fields {\xpair{\VD}{\VH}}
 to the electromagnetic fields {\xpair{\VE}{\VB}}.
In the simplest cases,
 these constitutive relations are expressed
 as a simple --
 homogeneous and isotropic --
 permittivity and permeability,
 but the full EMCR allow a much greater freedom.
Arguably,
 the EMCR are the unsung heroes of electromagnetics:
 without them Maxwell's equations would be underdetermined and
 lose predictive power.
Because of this central role for the EMCR,
 a re-examination of the fundamental assumptions behind them
 has a significant potential
 to open up new opportunities for electromagnetic metamaterials.

Here,
 we wish to avoid using {\xpair{\VD}{\VH}}
 because not only is
 their measurability controversial
 \cite{Gratus-KM-2019ejp-dhfield}, 
 but
 they may also have a multi-valued nature
 \cite{Gratus-KM-2019foop-nocharge}.
In this article, 
 we present a minimal
 extension to standard Maxwell theory
 which
combines the usual constitutive relations
 and both the Gauss's and the Maxwell-Amp{\`e}re laws.
The resulting theory
 uses new ``first order'' operators
 {\xquad{\PsiErho}{\PsiBrho}{\PsiEJ}{\PsiBJ}}
 that only act on the
 measurable
 \cite{Gratus-KM-2019ejp-dhfield,Gratus-KM-2019foop-nocharge}
 Maxwell fields {\xpair{\VE}{\VB}}, 
 connecting them to the sources {\xpair{\eCharge}{\VJ}}.
As a result
 the Maxwellian excitation fields {\xpair{\VD}{\VH}}
 are eliminated.

A significant feature of our approach
 is that
 our program admits constitutive relations
 that allow coupling to axion-like terms
 in a less restrictive way than is usual;
 notably in the case of homogeneous systems
 or those with a non-trivial topology.
In the latter case,
 we give an example where it is \emph{not} possible to model the material
 using standard constitutive relations. 
We also suggest an experimental scenario
 where such an axionic field can be emulated.
Although a
 discussion of axions in the context of EM is
 an established area of research
 \cite{Tobar-MG-2019pdu,Visinelli-2013mpla},
 this usually occurs in the context
 of an added coupling between the Maxwell fields
 and the field of an axion particle.
This contrasts with our
 {\pAxionResponse} terms,
 which result from a constitutive property
 of the background medium.
Notably,
 the Lagrangian for coupling EM to a
 particle physics type
 axion field ($\axiphi$)  has the form
 \cite{Tobar-MG-2019pdu,Visinelli-2013mpla}
~
\begin{align}
  L
&=
    \tfrac12
    \left(
      \VE \cdot \VE - \VB \cdot \VB
    \right)
   +
    \VA \cdot \VJ
\nonumber
\\
& \qquad %?? ??
 +
    g_{\axiphi} \,\axiphi\,
   \VE \cdot \VB
   \pm
    \tfrac12 g_{\axiphi}^2
    \left[
      \left(\dTime{} \axiphi\right)^2
     -
      \left({\nabla \axiphi}\right)^2
    \right]
.
\label{IN_Axio_Lagrangian}
\end{align}
 where $g_{\axiphi}$ is the massless axion coupling.
This leads to a coupled Maxwell and axion dynamics,
 where a background axion field can now influence
 the EM behaviour.

In the traditional description of a material,
 the axionic influences appear
 via a (pseudo-)scalar quantity \cite{HehlObukhov} 
 representing a non-zero trace part
 of the 4-dimensional constitutive tensor.
In 3-vector notation,
 axionic contributions to the excitation fields appear as
~
\begin{align}
  \VD_{\axion}
=
  \axipot \VB
\qquadand
  \VH_{\axion}
=
 -
  \axipot \VE
.
\label{M_DH_axion}
\end{align}
 where $\axipot$ is a constant representing the {\pAxionResponse}.
A simple application of \eqref{M_DH_axion}
 into Maxwell's equations gives
 the contribution $\rho_{\axion}$ and $\VJ_{\axion}$ to the charge and current,
 due to the axionic field,
 as
%[
\begin{align}
  \left(\nabla \axipot\right) 
 \cdot
  \VB
 =
  \rho_{\axion}
\quadand
  \left( \nabla \axipot\right)
  \times\VE
 +
  \left( \partial_t \axipot \right)
  \VB
 =
 -
  \VJ_{\axion} 
\label{Intro_Axion_current}
\end{align}
%]
Thus for blocks of homogeneous static materials,
 the effects of $\axipot$ only appears as a surface term
 at the boundary of the material.

This response can be considered a special case
 of a duality rotation \cite{Visinelli-2013mpla} of {\xpair{\VE}{\VB}},
 in which
 the rotation is just $\pi/2$.
On the basis of theoretical arguments,
 Post \cite{Post-FSEM}
 suggested %on the basis of theoretical arguments
 that
 the completely antisymmetric part of the constitutive tensor
 vanishes (i.e. $\axipot=0$)
 for all naturally occurring media.
Subsequently,
 Lakhtakia and Weiglhofer proposed
 that this so-called Post constraint was fundamental
 and applied to \emph{all} electromagnetic responses
 \cite{Lakhtakia-W-1996pla,Lakhtakia-M-2015spie,Lakhtakia-M-2016jn}.
Indeed,
 provided $\axipot$ does not depend on either position or time
 then Maxwell's equations are unaffected
 by the presence of $\axipot$.
However,
 a \emph{piecewise} constant {\pAxionResponse}
 \emph{is} detectable at boundaries \cite{Obukhov-H-2005pla}
 where the response can be equivalently cast as either
 a perfect electrical,
 or perfect magnetic conductor \cite{Lindell-S-2013ieeetap}.
Axionic responses were apparently observed experimentally
 \cite{Hehl-ORS-2008pla-relativisitic}
 via the magneto-electric effect
 in Cr$_2$O$_3$.
More recently {\pAxionResponse}s
 have also been proposed \cite{Li-WQZ-2010np-dynamical}
 and observed \cite{Wu-SKMO-2016s-quantised} in topological insulators.
Observations are,
 however,
 still controversial,
 with claims that evidence of Post violation 
 can be explained in other ways \cite{Lakhtakia-M-2015spie}.
In the domain of particle physics,
 axions have been proposed as candidates for dark matter
 \cite{Feng-2010araa-dark}.

In this paper,
 since we generalise the definition of EMCR,
 we can treat axion-like effects
 in the manner of bound charge and current sources,
 which allows them to be seen in the bulk.
We will call these constitutive properties
 an {``{\pAxionResponse}''}.
Moreover,
 our approach allows the number of potential
 material parameters that represent an {\pAxionResponse}
 to increase from one to four.

Our paper is organised as follows:
In section \ref{ch_GH}
 we briefly summarise how constitutive properties usually appear
 in the context of Maxwell's equations,
 and in section \ref{ch_EMCR} we introduce our new approach;
 in
 section \ref{S-axion-topological}
 we then discuss two important topological consequences.
Next,
 in section \ref{ch_Examples}
 we give some examples
 in media incorporating the new {\pAxionResponse} constitutive effects.
We then briefly propose
 a possible metamechanical {\pAxionResponse} element,
 introduce how the transformation optics paradigm needs to be modified,
 and discuss an extension handling Ohmic current and resistance,
 before concluding in section \ref{ch_Conclusion}.
There are also a number of appendices covering further
 mathematical details,
 including a presentation of a spacetime formalism (\ref{ch_GR}),
 proofs (\ref{ch_Proofs}),
 and coordinate-free notation (\ref{ch_Coordfree}).

%
% ======================================================================
\section{Maxwell's Equations and the constitutive tensor}
\label{ch_GH}

The macroscopic Maxwell's equations
 are perhaps most elegantly expressed in a fully four dimensional
 spacetime form using exterior calculus \cite{HehlObukhov},
 although spacetime formulations using tensors are also popular
 (see e.g. \cite{Heras-B-2009ejp}).
Nevertheless,
 it is the very familiar
 Gibbs-Heaviside vector calculus form which is most widely used
 in practical calculations,
 where they are written as
%[
\begin{align}
  \nabla
  \cdot
  \VB
&=
  0
,
\label{GH_Max_NoMono}
\\
  \nabla
  \times
  \VE
 +
  \dtime{\VB}
&=
  \Vzero
,
\label{GH_Max_Faraday}
\\
  \nabla
  \cdot
  \VD
&=
  {\eCharge}
,\qquad\qquad
\label{GH_Max_Gibbs}
\\
\text{and}\qquad
  \nabla
  \times
  \VH
 -
  \dtime{\VD}
&=
  {\eCurrent}
.
\label{GH_Max_Amp}
\end{align}
%]
These are augmented with electromagnetic constitutive relations (EMCR)
 which relate the excitation fields {\xpair{\VD}{\VH}}
 to the electromagnetic fields {\xpair{\VE}{\VB}}.
With the possible \footnote{Nonlinear and higher order models
 of the EMCR vacuum \cite{Gratus-PT-2015jpa}
 exist for which the standard
 $\EMepsilon_0$, $\EMmu_0$ are simply an approximation.} 
 exception of the vacuum,
 EMCR are always an approximation as the underlying structure
 is either unknown or too complicated to analyse fully.

Usually,
 the choice of models for the EMCR
 is limited only by the imagination of the researcher,
 and the skill of the experimentalist to fabricate and measure.
Traditionally these might include fixed values
 for permittivity and permeability,
 dynamical models which generate
 a frequency dependence \cite{Jackson-ClassicalED,RMC},
 an accommodation of anisotropy and birefringence \cite{Nye-PROPX},
 magnetoelectricity \cite{Fiebig-2005jpd,Eerenstein-MS-2006n},
 chirality \cite{Hillion-1993pre,Wang-ZKKS-2009joa,McCall-2009joa},
 nonlinearity \cite{New-INLO,Boyd-NLO},
 a dependence on temperature or stress \cite{Nye-PROPX},
 or even {spatial dispersion}
 \cite{AgranoGinsberg,Belov-TV-2002jewa,Ciraci-PS-2013cphc,Kinsler-2019arxiv-spatype} --
 all these can provide good matches to materials found in nature.
More complicated empirical models can also be used,
 with parameters being estimated from or fitted to experimental data.
Such descriptions are often remarkably accurate and useful
 within their own domain,
 providing us with vital information
 about the underlying electromagnetic medium,
 such as the resonances of the individual atoms.

Although a common simple case
is where the permittivity and permeability are constant,
 for anisotropic media these are replaced by tensors
 $\epsilonTen$ and $\muTen$,
 and can even be generalised to include magnetoelectric terms.
This general tensor form can be written
%[
\begin{align}
  \VD
&=
  \kappaDE(\VE) + \kappaDB(\VB)
,
\label{DEDB_kappa}
\\
  \VH
&=
  \kappaHE(\VE) + \kappaHB(\VB)
.
\label{GH_kappa}
\end{align}
%]
Here $\kappaDE = \EMepsilon$ is the permittivity tensor,
 $\kappaHB = \EMmu^{-1}$ is the (inverse) permeability tensor
 and $\kappaDB$ and $\kappaHE$ are magnetoelectric tensors.
The number of parameters appearing in these four tensors is 36.
In general the tensors {\xquad{\kappaDE}{\kappaDB}{\kappaHE}{\kappaHB}}
 may depend on position,
 but in a homogeneous medium they are constant.
They may also have temporal and spatial dispersion,
 i.e. if they depend on {\xpair{\omega}{\Vk}}.
But for a non-dispersive
 medium,
 as we consider here,
 the tensors {\xquad{\kappaDE}{\kappaDB}{\kappaHE}{\kappaHB}}
 have no additional dependence.

\ResubOne{Note that any non-dispersive medium
 is automatically causal in the Kramers-Kronig sense
 \cite{Bohren-2010ejp,Kinsler-2011ejp}. 
Notably, Jackson \cite{Jackson-ClassicalED}
 shows that `generally correct' constitutive relations
 can be based on convolution integrals
 with respect to their historical or spatial environment.
However,
 whether in the standard EMCR,
 or in our CMCR defined in the next section,
 any non-vacuum response in a medium is a result of both 
 (a) its internal dynamics,
 and
 (b) its coupling to the EM fields,
 these must satisfy causality;
 and thus necessarily be dispersive.
However,
 in appropriately chosen situations,
 the dispersion may be small enough
 so that it can reasonably be approximated as negligible.}

The nature of the excitation fields {\xpair{\VD}{\VH}}
 is very different to the electromagnetic fields {\xpair{\VE}{\VB}},
 and indeed
 there is a debate in the literature as to whether or not they
 are actually physical quantities
 \cite{Heras-2011ajp,Gratus-KM-2019foop-nocharge,Gratus-KM-2019ejp-dhfield}.
Notably,
 it is easy to see that Maxwell's equations are invariant
 by adding a gauge {\xpair{\gScalar}{\gVector}}
 to the excitation fields,
 with the replacements
%[
\begin{align}
\VD\to\VD+\nabla\times \gVector
\quadand
\VH  \to  \VH - \nabla \gScalar + \dtime{\gVector}
.
\label{M_DH_Gauge}
\end{align}
%]
This gauge freedom is distinct
 from the usual gauge freedom which is associated only
 with the potential {\xpair{\emScalar}{\emVector}}
 for the {\xpair{\VE}{\VB}} fields.
These gauge freedoms
 mean that since Maxwell's equations
 \emph{only} couple with derivatives of
 {\xpair{\VD}{\VH}} or {\xpair{\VE}{\VB}}
 one cannot \emph{a priori} claim that either are measurable.
However,
 the electromagnetic fields {\xpair{\VE}{\VB}} can be directly measured,
 either directly using the Lorentz force equation
 or non-locally using the Aharonov-Bohm effect
 \cite{Ehrenberg-S-1949prsb,Aharonov-B-1959pr,Matteucci-IB-2003fp}.
This second case is particularly useful
 as it enables one to measure the electromagnetic fields
 inside a medium where the Lorentz force may not be useful due to
 collisions with atoms.
In
 Aharonov-Bohm tests,
 the electrons only need travel in vacuum {outside} the medium.
In contrast the excitation fields {\xpair{\VD}{\VH}}
 remain not directly measurable,
 as there is
 no accepted native Lorentz force-like equation\footnote{However,
   note that  --
   on the presumption that magnetic monopole might actually exist --
   proposals for such a force have been advanced \cite{Rindler-1989ajp}.}
 dependent on {\xpair{\VD}{\VH}},
 nor is there any analogous Aharonov-Bohm-like effect for them.
Consequently,
 whenever making claims about the
 measurability of {\xpair{\VD}{\VH}},
 one has to make assumptions about their nature,
 for example that they are linearly and locally related {\xpair{\VE}{\VB}}.

One consequence of the {gauge freedom} for {\xpair{\VD}{\VH}} is that
 for a \emph{homogeneous} medium,
 one of the parameters $\axipot$
 in the constitutive tensors $\kappaDB$ and $\kappaHE$ can be ignored.
This is the purely axionic field given by \eqref{M_DH_axion},
 and
 can be removed by setting
~
\begin{align}
  \gScalar
&=
  \axipot \emScalar,
\quad
   \gVector = -\axipot \emVector.
\label{g_em_axion}
\end{align}
Thus in this case there are only 35 \emph{free} parameters
 in the constitutive tensor.

%
% ======================================================================
\section{Proposed Operator Constitutive Relations}
\label{ch_EMCR}

Our minimal extension to standard Maxwell theory
 is motivated by a single crucial step:
 we decide
 that the charge and current {\xpair{\eCharge}{\eCurrent}}
 are \emph{the} most important components of
 Gauss' Law \eqref{GH_Max_Gibbs}
 and the Maxwell-Amp{\`e}re equation \eqref{GH_Max_Amp}.
This enables a simple generalisation of
 the constitutive properties which enables us to
 completely remove the non-measurable excitation fields {\xpair{\VD}{\VH}}
 from the description.
Therefore we start by rewriting \eqref{GH_Max_Gibbs} and \eqref{GH_Max_Amp}
 as
~
\begin{align}
  {\eCharge}
&=
  \nabla\cdot\VD
\label{GH_Max_Gibbs_rewrite}
,
\\
  {\eCurrent}
&=
  \nabla\times\VH-\dtime{\VD}
.
\label{GH_Max_Amp_rewrite}
\end{align}
Now
 we replace the right hand side (RHS)
 of these with some new operators
 {\xquad{\PsiErho}{\PsiBrho}{\PsiEJ}{\PsiBJ}}
 acting directly on {\xpair{\VE}{\VB}}
 rather than --
 as is usual --
 a differential operator acting on  {\xpair{\VD}{\VH}}.
This replacement is consistent with our
 discussion above where {\xpair{\VD}{\VH}} were seen as a gauge field
 for the charge and current.
A logical consequence of this is to
 dispense with {\xpair{\VD}{\VH}},
 and directly connect the electromagnetic field
 to the current.
We now implement this idea.

The resulting
 constitutive relations
 for media generalised in this way
 combine Gauss's Law \eqref{GH_Max_Gibbs},
 the Maxwell-Amp{\`e}re equation \eqref{GH_Max_Amp},
 and the constitutive tensors \eqref{GH_kappa}.
They are
%[
\begin{align}
  {\eCharge}
&=
\PsiErho\VAct{\VE} + \PsiBrho\VAct{\VB}
,
\label{M_New_Max_CR_rho}
\\
\quadand
{\eCurrent}
&=
\PsiEJ\VAct{\VE} + \PsiBJ\VAct{\VB}
,
\label{M_New_Max_CR_J}
\end{align}
%]
which we call the
 \emph{combined Maxwell and constitutive relation} (CMCR) equations,
 and where the angle brackets are used to emphasise that
 the CMCR operators are not tensors,
 but may also involve the first derivatives of their arguments.
These four %``$\PsiAny$'' CMCR
 operators {\xquad{\PsiErho}{\PsiBrho}{\PsiEJ}{\PsiBJ}}
 take vector fields and output
 either a scalar or a vector field;
 their properties are given below
 in section \ref{sch_New_CR}.

\ResubOne{Here $\eCharge$ and $\eCurrent$ 
 are the \emph{free} charges and currents,
 and \eqref{M_New_Max_CR_rho}, \eqref{M_New_Max_CR_J}
 are our new \emph{microscopic} replacements
 for \eqref{GH_Max_Gibbs_rewrite}, \eqref{GH_Max_Amp_rewrite}.
As is usual,
 it may in some particular circumstance be useful
 to move some of the bound charge or (current) effects
 from the RHS of \eqref{M_New_Max_CR_rho}, \eqref{M_New_Max_CR_J}
 over to the LHS,
 re-imagining them as free charges (currents). 
Indeed,
 we have complete freedom to split the total charges and currents
 into free and bound contributions
 according to our preferred material models; 
 the splitting is not unique 
 (see e.g. 
 \cite{Gratus-KM-2019ejp-dhfield,Gratus-KM-2019foop-nocharge}).}

Clearly in \eqref{M_New_Max_CR_rho},
 $\PsiErho\VAct{\VE}$ is the generalisation of the divergence
 $\nabla\cdot\left(\kappaDE(\VE)\right)$,
 and $\PsiBrho\VAct{\VB}$ is a magnetoelectric term.
In \eqref{M_New_Max_CR_J},
 $\PsiEJ\VAct{\VE}$ is the generalisation of $\dTime{}(\kappaDE(\VE))$
 but now can also contain magnetoelectric terms.
A careful analysis of the symmetries of the CMCR operators
 (see appendix \ref{ch_Proofs})
 show that in the homogeneous non-dispersive case
 they possess 55 free parameters,
 20 more than the constitutive tensor $\kappa$ {can ever allow.}
Note that \eqref{M_New_Max_CR_rho} and \eqref{M_New_Max_CR_J}
 reduce to Gauss's Law and the Maxwell-Amp{\`e}re equation
 in a vacuum 
 if we set
 $\PsiErho\VAct{\VE} = \EMepsilon_0 \VE$, 
 $\PsiBJ\VAct{\VB} = \grad \times \EMmu_0^{-1} \VB$, 
 $\PsiEJ\VAct{\VE}= \partial_t \EMepsilon_0 \VE$,
 and 
 $\PsiBrho\VAct{\VB}=0$.

\ResubOne{We do not consider nonlinear responses in this paper. 
Nevertheless,
 one can certainly imagine nonlinear generalizations,
 although the number of potential CMCR-like constitutive terms
 would be very large,
 even for low-order nonlinearities.  
Our CMCR generalization is a strict superset of the EMCR,
 and all possible EMCR responses are expressible under our CMCR scheme.}

%
% ----------------------------------------------------------------------
\subsection{The CMCR operators}
\label{sch_New_CR}

These new CMCR operators {\xquad{\PsiErho}{\PsiBrho}{\PsiEJ}{\PsiBJ}}
 are not tensors,
 unlike the standard constitutive tensors
 {\xquad{\kappaDE}{\kappaDB}{\kappaHE}{\kappaHB}}
 in the traditional constitutive relations
 \eqref{DEDB_kappa} and \eqref{GH_kappa}.
We now specify the properties they must have
 in order for the fields upon which they act
 to be consistent with Maxwell's equations
 \eqref{GH_Max_NoMono} and \eqref{GH_Max_Faraday},
 and with charge conservation.

As a starting point,
 let us first focus on just the $\PsiErho\VAct{\VE}$ term
 in Eq. \eqref{M_New_Max_CR_rho}.
Its simplest expression is
 as the sum of a linear term and a first derivative,
 i.e. in a coordinate basis it is
~
\begin{align}
  \PsiErho\VAct{\VE}
&=
  (\PsiErho)^i\ \VE_i + (\PsiErho)^{0j}\ \dtime\VE_j +
  (\PsiErho)^{ij}\ \partial_i\VE_j
,
\label{M_Phi_components}
\end{align}
 where $i,j \in \left\{ 1,2,3 \right\}$
 and we have used implicit summation over repeated indices.
However,
 as we show in appendix \ref{ch_Proofs},
 \eqref{M_Phi_components} implies
 the following three coordinate-free linearity relations:
\begin{align}
  \PsiErho\VAct{{f}^2\VE}
&=
  2{f}\,\PsiErho\VAct{{f}\VE}
 -
  {f}^2\PsiErho\VAct{\VE}\,,
\label{M_MHOOF_Phi}\\
  \PsiErho\VAct{\VE_1+\VE_2}
&=
  \PsiErho\VAct{\VE_1}+\PsiErho\VAct{\VE_2}
,
\label{GH_Psi_+-linear}
\\
  \PsiErho\VAct{\lambda\,\VE}
&=
  \lambda\,\PsiErho\VAct{\VE}\,,
\label{GH_Psi_R-linear}
\end{align}
 where $f$ is a scalar field,
 and $\lambda$ is a real constant.
The converse is also true;
 Eqs. \eqref{M_MHOOF_Phi}--\eqref{GH_Psi_R-linear}
 together imply Eq. \eqref{M_Phi_components}.
We refer to operators {\xquad{\PsiErho}{\PsiBrho}{\PsiEJ}{\PsiBJ}}
satisfying Eq. \eqref{M_MHOOF_Phi}
 as first order operators,
 and regard Eqs.  \eqref{M_MHOOF_Phi}--\eqref{GH_Psi_R-linear}
 as collectively expressing their linearity.

We now postulate that all the CMCR operators
 {\xquad{\PsiErho}{\PsiBrho}{\PsiEJ}{\PsiBJ}}
 appearing in Eqs. \eqref{M_New_Max_CR_rho}
 and \eqref{M_New_Max_CR_J} are first order operators,
 and describe the consequences.

Without applying any further constraints,
 $\PsiErho$ and $\PsiBrho$ each have 15 components;
 while $\PsiEJ$ and $\PsiBJ$,
 which map vectors to vectors,
 each have 45 components.
This gives a grand total of $120$ components,
 but the number is
 reduced by the demand that the fields satisfy charge conservation.

%
% - - - - - - - - - - - - - - - - - - - - - - - - - - - - - - - - - - - -
%\subsubsection*{Charge conservation}\label{S-CR-properties-Cn}

Local conservation dictates that
 physical electromagnetic fields {\xpair{\VE}{\VB}},
 and sources {\xpair{\eCharge}{\eCurrent}}
 appearing in \eqref{M_New_Max_CR_rho},\eqref{M_New_Max_CR_J}
 obey % the local conservation law
%[
\begin{align}
  \dtime{\rho} + \nabla\cdot\VJ = 0
.
\label{EMCR_conservation}
\end{align}
%]
Since one can always solve
 equations \eqref{GH_Max_NoMono} and \eqref{GH_Max_Faraday}
 locally via the use of a potential {\xpair{\emScalar}{\emVector}},
 with \eqref{M_New_Max_CR_rho} and \eqref{M_New_Max_CR_J}, 
  we can re-express \eqref{EMCR_conservation} as
%[
\begin{align}
  \dTime{}
  \left[
    \PsiErho (-\nabla\emScalar-\dtime{\emVector})
  \right]
 +
  \dTime{}
  \left[
    \PsiBrho(\nabla\times\emVector)
  \right] \quad\quad\quad &
\nonumber
\\
%\quad\quad
 +
  \nabla \cdot
  \left[
    \PsiEJ(
     -\nabla\emScalar-\dtime{\emVector})
  \right]
 +
  \nabla \cdot
  \left[
    \PsiBJ(\nabla\times\emVector)
  \right]
&=
0
,
\label{M_New_Max_Consev}
\end{align}
%]
for all {\xpair{\emScalar}{\emVector}}.

It is not necessary to consider \eqref{M_New_Max_Consev} independently
 in the standard approach to the EMCR
 because there it is a guaranteed
 \emph{consequence} of Maxwell's equations
 \eqref{GH_Max_Gibbs} and \eqref{GH_Max_Amp}.
The constraint \eqref{M_New_Max_Consev} relates
 the 120 components of {\xquad{\PsiErho}{\PsiBrho}{\PsiEJ}{\PsiBJ}}
 to each other,
 and reduces the number
 of independent components from 120 to 55.
Although this represents a significant reduction,
 the number of independent parameters in our theory
 is still larger than the 35 
 (or 36, if $\axipot$ is also counted)
 independent components
 needed in the standard constitutive tensor approach.

%
% ----------------------------------------------------------------------
\subsection{Axionic response terms}
\label{sch_Description}

Of these 20 new parameters
 4 describe
 the axion-like constitutive response of the material.
Unlike the standard EM axion,
 these responses are not due to a coupling with an axion particle field,
 and
 couple to the Maxwell fields
 just as ordinary constitutive properties do,
 so we can now imagine media with
 a combination of ordinary and axionic properties.

These {\pAxionResponse}s
  do not involve derivatives of the electromagnetic field,
 and so correspond to the first term
 on the right hand side of \eqref{M_Phi_components}.
We can therefore investigate the {\pAxionResponse}
 by replacing \eqref{M_DH_axion} with
%[
\begin{align}
{\eCharge}_{\axion} = (\PsiErho)^i\,\VE_i + (\PsiBrho)^i\,\VB_i
,
\label{GH_rho_ax}
\end{align}
%]
where
 ${\eCharge}_{\axion}$ refers to that part of the total charge
 which relates to the {\pAxionResponse}.
If we now apply the constraint \eqref{M_New_Max_Consev},
 we find it demands $(\PsiErho)^i=0$,
 while the three components $(\PsiBrho)^i$
 are arbitrary fields%
  \footnote{
   For homogeneous static media,
   the argument is as follows:
   At each point and moment in time,
   the various derivatives of $\emScalar$ and $\emVector$ are all independent.
    Thus by comparing \eqref{M_Phi_components}
    and \eqref{M_New_Max_Consev} we see that $\dTime{}
  \left(
    \PsiErho (-\nabla\emScalar-\dtime{\emVector})
  \right)$ will generate the term
    $(\PsiErho)^i\dTime{}^2{A}_i$,
    the only such term in Eq. \eqref{M_New_Max_Consev}.
   Hence $(\PsiErho)^i=0$. By contrast the term $\nabla \cdot
  \left[
    \PsiEJ(
     -\nabla\emScalar-\dtime{\emVector})
  \right]$ contains terms $\partial_j\dTime^2{A}_i$.
   For full derivation see appendix \ref{ch_Proofs}.
  }.

Now let some vector $\axivav$ have components $(\axivav)^i = (\PsiBrho)^i$,
 so that \eqref{GH_rho_ax} becomes ${\eCharge}_{\axion}=\axivav\cdot\VB$.
Defining ${\eCurrent}_{\axion}$ in the same manner
 leaves us one more free component of $\PsiEJ$ and $\PsiBJ$,
 which we denote $\axivat$.
This means that the CMCR equations for just the {\pAxionResponse} are
 (cf. \eqref{FOO_CR_coord_alt})
%[
\begin{align}
  \axivav\cdot \VB
=
  {\eCharge}_{\axion}
,
\qquadand
  -\axivav\times\VE - \axivat \VB
=
  {\eCurrent}_{\axion}
,
\label{GH_new_axion}
\end{align}
%]

As an example,
 we can model a medium with an {\pAxionResponse}
 together with a simple constant permittivity $\EMepsilon$
 and permeability $\EMmu$
 as per \eqref{GH_Max_NoMono},\eqref{GH_Max_Faraday},
 and so replace
 \eqref{GH_Max_Gibbs},\eqref{GH_Max_Amp},\eqref{GH_kappa}
 for this type of medium with the relations
%[
\begin{align}
  \EMepsilon \nabla \cdot \VE
 -
  \axivav \cdot \VB
&=
  {\eCharge},
\label{M_New_Max_axion_charge}
\\
  \EMmu^{-1} \nabla \times \VB
 -
  \EMepsilon \dtime{\VE}
 +
  \axivav \times \VE
 +
  \axivat \, \VB
&=
 {\eCurrent}
.
\label{M_New_Max_axion_current}
\end{align}
%]
 where \xpair{\axivav}{\axivat} need not be constants.
We call this
 a local {\pAxionREA},
 and the special case
 when $\EMmu=\EMmu_0$ and $\EMepsilon=\EMepsilon_0$
 is a ``vacuum-like'' {\pAxionREA}.
We examine the behaviour of electromagnetic fields
 in such media in the examples
 of section \ref {ch_Examples}.

\ResubOne{It is worth noting that we cannot just pick any 
 \xpair{\axivav}{\axivat}
 that we would like --
 we need the result to be consistent with the constraints
 given above in \eqref{M_New_Max_Consev}.
It is easy to see that 
 for the local axionic material
 \eqref{M_New_Max_axion_charge}, 
 \eqref{M_New_Max_axion_current}
 conservation of charge gives 
\begin{align}
  0
&=
  \dtime{\rho} + \nabla\cdot\VJ
=
  \left( \nabla \times \axivav \right)
 \cdot
  \VE
 +
  \left( \nabla \axivat - \dtime{\axivav} \right)
 \cdot
  \VB
.
\label{EMCR_axi_proof}
\end{align}
 If we further assume that \xpair{\VE}{\VB}
 are unconstrained and could take any form,
 such as in a bulk material,
 then \xpair{\axivav}{\axivat} must also satisfy 
%[
\begin{align}
  \nabla\times\axivav = \Vzero
\qquadand
  \nabla\axivat - \dtime{\axivav} = \Vzero
.
\label{EMCR_axi_constraints}
\end{align}
However, 
 it is important to emphasise that 
 \eqref{EMCR_axi_constraints} is too restrictive for use in many situations, 
 e.g. such as those involving symmetries, 
 where \xpair{\VE}{\VB}
 may only have specific orientations with non-zero components.
 (See e.g. the examples of \ref{sch_cylinder_bdd} or \ref{S-axion-example}).}
%]
%The proof of \eqref{EMCR_axi_constraints} is given
% in appendix \ref{ch_Proofs},
% but it is easy to see for the local axionic material
% \eqref{M_New_Max_axion_charge},
% \eqref{M_New_Max_axion_current}
% the conservation of charge equation gives
%%[
%\begin{align}
%  0
%&=
%  \dtime{\rho} + \nabla\cdot\VJ
%=
%  \left( \nabla \times \axivav \right)
% \cdot
%  \VE
% +
%  \left( \nabla \axivat - \dtime{\axivav} \right)
% \cdot
%  \VB
%.
%\label{EMCR_axi_proof}
%\end{align}
%%]
Together \eqref{EMCR_axi_constraints}
 imply that the {\pAxionResponse} can be derived (locally)
 from an axionic scalar potential $\axipot(x,y,z,t)$ via
~
\begin{align}
  \axivav = \nabla\, \axipot
,
\quad
\textrm{and}
\quad
  \axivat = \dTime \,\axipot
.
\label{eqn-axion-chargeconstraintpotential}
\end{align}
This potential need not be given any specific physical meaning,
 since its existence is simply a calculational device
 to ensure constructions of \xpair{\axivav}{\axivat}
 stay consistent with the necessary constraints.
However,
 it can be interpreted as a specification of the properties
 required for a necessarily inhomogeneous medium,
 described by the
 traditional ``$\emKappa$'' tensorial formulation,
 to match a medium with an {\pAxionResponse}
 of the kind described here,
 i.e. by comparing
  \eqref{Intro_Axion_current} and \eqref{GH_new_axion}.

There are however
 two cases where it is not possible
 to simply replace {\xpair{\axivav}{\axivat}} with $\axipot$.
One case is when we make the natural demand
 that the axionic terms {\xpair{\axivav}{\axivat}} be homogeneous,
 which leads to
 an inhomogeneous $\axipot$,
 i.e.
%[
\begin{align}
  \axipot
&=
  \left(
    t \axivat + x \axivax + y \axivay + z \axivaz
  \right)
.
\label{EMCR_kax_hom}
\end{align}
%]
Here $t$ is time,
 {\xtrio{x}{y}{z}} are the usual Cartesian coordinates,
 and {\xquad{\axivax}{\axivay}{\axivaz}{\axivat}}
 are the 4 axionic material constants.
If we do make
 $\axipot$ constant then $\axivav=\Vzero$ and $\axivat=0$.
Thus, 
 for a block of material in which $\axipot$
 is constant,
 the traditional axionic terms $\axivav$
 will only appear on the surface of the material \cite{Obukhov-H-2005pla}.
The other case, 
 where the existence of $\axipot$ is prevented,
 is due to topological considerations,
 and is discussed below in section \ref{S-axion-topological}.

%
% ----------------------------------------------------------------------
\subsection{Non-axionic extension terms}
\label{sch_ccrestrictions}

   \def\VNullE{{\check{\VE}}}
   \def\VNullB{{\check{\VB}}}
   \def\VNullJ{{\check{\VJ}}}
   \def\Nullrho{{\check{\rho}}}

The above prediction of 20 extra constitutive parameters
 sounded extremely promising,
 suggesting many new possibilities for novel electromagnetic media.
Since we have just seen that 4 of those terms
 have response similar (but not the same) as known axion-like behaviour;
 this leaves us another 16 that require further consideration.

In particular
 we need to ensure consistency with section \ref{ch_GH},
 where
 as far as homogeneous constitutive relations are concerned,
 it was necessary to ignore the axionic contribution %{\pAxionresponse}
 since it did not relate
 the electromagnetic field to the current.
That is,
 any contribution to {\xpair{\VD}{\VH}} coming from $\axipot$
 in \eqref{M_DH_axion} would vanish
 when inserted into Maxwell's equations.
By the same argument we should ignore any components of 
 \xquad{\PsiErho}{\PsiBrho}{\PsiEJ}{\PsiBJ}
 which do not relate the electromagnetic field to the current.
Thus we say that,
 similar to a gauge freedom,
 we can replace
   %[
   \begin{equation}
   \begin{aligned}
   \PsiErho &\rightarrow \PsiErho + \NullPsiErho,\quad &
   \PsiBrho &\rightarrow \PsiBrho + \NullPsiBrho,
   \\
   \PsiEJ   &\rightarrow \PsiEJ + \NullPsiEJ,\quad &
   \PsiBJ   &\rightarrow \PsiBJ+ \NullPsiBJ
   ,
   \end{aligned}
   \label{ccrestrictions_Null_invar}
   \end{equation}
   %]
where for any valid electromagnetic fields
   %[
   \begin{equation}
   \begin{aligned}
   \NullPsiErho\VAct{\VE} &= 0
   ,\quad&
   \NullPsiBrho\VAct{\VB} &= 0
   ,
   \\
   \NullPsiEJ\VAct{\VE} &= \Vzero
   ,\quad &
   \NullPsiBJ\VAct{\VB} &= \Vzero
   ,
   \end{aligned}
   \label{ccrestrictions_Null_def}
   \end{equation}
   %]
 i.e. satisfying \eqref{GH_Max_NoMono}, \eqref{GH_Max_Faraday}.
Imposing this, % freedom,
 we show in appendix \ref{ch_Proofs} that this reduces
 the number of ``physical'' components
 of {\xquad{\PsiErho}{\PsiBrho}{\PsiEJ}{\PsiBJ}}
 by 16.
Thus we are left with 36+4 components for the CMCR,
 and so our main achievement is to have formulated
 a more general kind of {\pAxionResponse}.

At this point we could simply regard \eqref{ccrestrictions_Null_def}
 as a motivation for taking
 the apparently obvious and sufficient step
 of setting all the $\NullPsi$'s terms to zero,
 and so take no further interest in those 16 additional parameters.
However,
 in order to consider alternative scenarios
 where they do have a potential physical meaning,
 we now ask the following question:
 ``How,
   for the constitutive operators\footnote{We
     use the symbol $\NullPsi$ to refer to all four operators
  {\xquad{\NullPsiErho}{\NullPsiBrho}{\NullPsiEJ}{\NullPsiBJ}}}
  {\xquad{\NullPsiErho}{\NullPsiBrho}{\NullPsiEJ}{\NullPsiBJ}},
  might we determine any of their valid non-zero values?''
There are two possibilities,
 which we outline briefly below,
 both of which add to the standard 35 terms,
 and the 4 {\pAxionResponse} terms,
 to permit a total of {55} constitutive parameters.

%
% - - - - - - - - - - - - - - - - - - - - - - - - - - - - - - - -
\subsubsection{Measurable excitation fields: two versions of {\xpair{\VD}{\VH}}}

The first  option is to
 imagine that we \emph{do} in fact have
 some new kind of experimental apparatus
 that enables us to directly measure aspects of {\xpair{\VD}{\VH}}.
Although
 somewhat in opposition to the starting motivation
 for our generalised CMCR,
 it is nevertheless an interesting possibility.

We can notice immediately that each component of
 these fields appears \emph{twice}
 in Maxwell's equations;
 and that means that there can likewise be
 two distinct ways of measuring it --
 e.g.
 we might measure a $D_x$ based on charge from \eqref{GH_Max_Gibbs},
 or a $D_x$ based on current from \eqref{GH_Max_Amp}.

Normally we would expect such measurements to give the same outcomes.
However,
 \emph{if}
 for distinct measurement approaches on
 what are ordinarily seen as the same field components,
 we get \emph{different} results,
 then we can conclude that
 the $\NullPsi$ constitutive operators
 are non-zero;
 although they must be such that
 they still ensure that \eqref{ccrestrictions_Null_def} holds.
It is the differences in these measurements
 that are the key to determining the $\NullPsi$ parameters.

Since this proposal not only retains a role for {\xpair{\VD}{\VH}},
 but even demands an \emph{extra} pair of excitation fields
 {\xpair{\VD'}{\VH'}},
 we also consider a second option as described next.

%
% - - - - - - - - - - - - - - - - - - - - - - - - - - - - - - - -
\subsubsection{Auxiliary fields and charges}

An alternative scheme for determining the $\NullPsi$'s parameters,
 and one more in keeping with our premise
 of not relying on {\xpair{\VD}{\VH}},
 is to posit the existence of additional fields
 {\xpair{\VNullE}{\VNullB}}.
These fields would
 also require the existence of their
 related charges and currents {\xpair{\Nullrho}{\VNullJ}}.

Assuming we can measure these new physical properties,
 i.e. the auxiliary fields or sources,
 then we could determine the relevant $\NullPsi$ constitutive parameters.
This is because
 they need to satisfy
   %[
\begin{equation}
   \begin{aligned}
   \NullPsiErho\VAct{\VNullE} &= \Nullrho
   ,\quad&
   \NullPsiBrho\VAct{\VNullB} &= \Nullrho
,
   \\
   \NullPsiEJ\VAct{\VNullE} &= \VNullJ
   ,\quad &
   \NullPsiBJ\VAct{\VNullB} &= \VNullJ
   .
   \end{aligned}
   \label{ccrestrictions_Null_alt}
\end{equation}
   %]

It is important to note that whatever other dynamical equations
 that these auxiliary fields {\xpair{\VNullE}{\VNullB}} might follow,
 they cannot be the same as
 the standard Maxwellian \eqref{GH_Max_NoMono}, \eqref{GH_Max_Faraday}.
Note that here the sources $\Nullrho$ and $\VNullJ$ might be
 the ordinary charge and current $\eCharge$ and $\VJ$,
 and not new types of sources.
This is
 in contrast to the necessarily new fields $\VNullE$ and $\VNullB$.

To reiterate:
 If there is a medium for which the CMCR
  \eqref{M_New_Max_CR_rho}, \eqref{M_New_Max_CR_J}
 are valid
  and we can measure the $\NullPsi$'s to be non-zero
 using \eqref{ccrestrictions_Null_alt};
 then such a medium cannot be
   modelled via the standard Maxwell
 \eqref{GH_Max_NoMono} -
 %\eqref{GH_Max_Faraday},
 %\eqref{GH_Max_Gibbs},
 \eqref{GH_Max_Amp}.

%
% ======================================================================
\section{Topology and the CMCR}\label{S-axion-topological}

The differences between the standard EMCR
 and our generalised CMCR are particularly marked
 when considered in the context of
 interesting topologies.
It has already been shown that a non-trivial topology
 can have some remarkable consequences --
 notably
 it has already been demonstrated \cite{Gratus-KM-2019foop-nocharge}
 that that if {\xpair{\VD}{\VH}} are
 not considered to be true physical fields,
 charge need not be globally conserved;
 e.g. if a black hole forms and then evaporates.
Here,
 however,
 we can give rather less exotic examples
 where topology,
 in combination with
 the existence of the {\pAxionResponse},
 can lead to new phenomena.
One,
 in principle,
 could be built in the laboratory,
 whilst the other can be used for computer modelling of
 periodic materials (see \ref{sch_torus}).

\def\backup{\hspace{-0.33em}}

\begin{figure}
\centering
%\resizebox{0.70\columnwidth}{!}{
% \input{fig01-perfectcyl.tikz}
%}
\includegraphics[width=0.70\columnwidth]{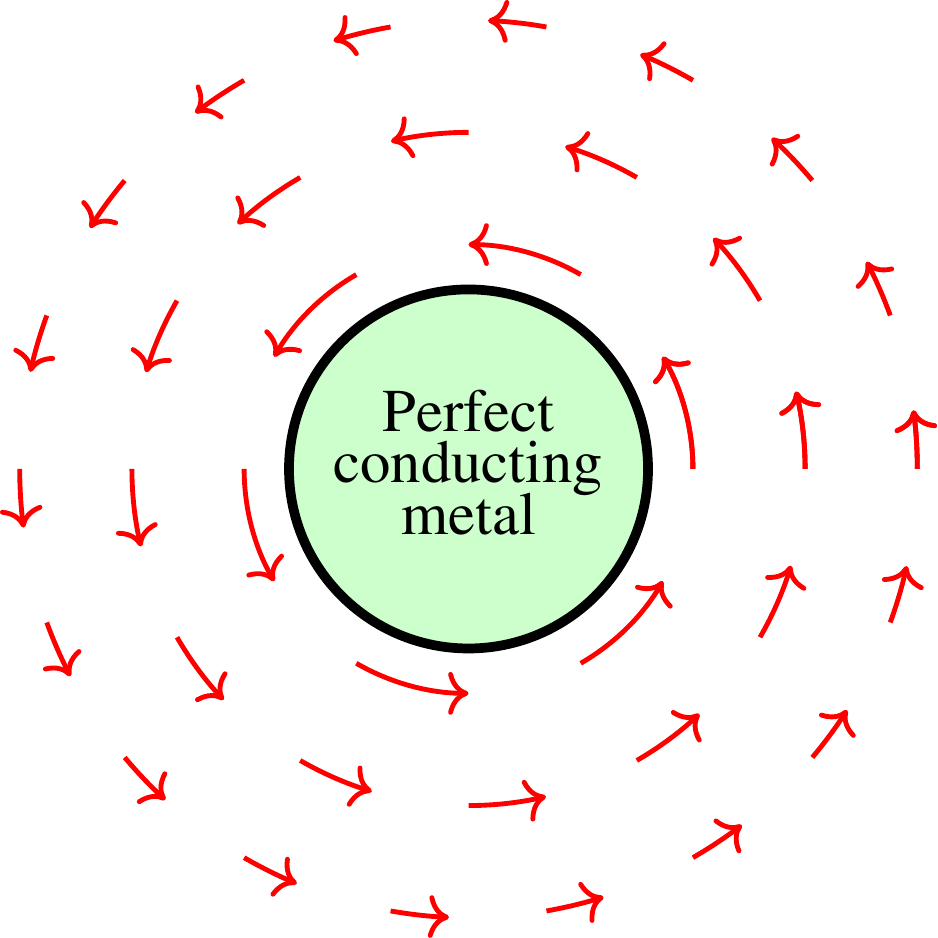}
\caption{An {\pAxionResponse} 
 $\axivav = \axivaa \AxisTheta$ (arrows)
 outside a charged conducting cylinder,
 shown in the cross section of
 the plane perpendicular to the cylinder.}
\label{fig_example-cylinder}
\end{figure}

\begin{figure}
\centering
%\resizebox{0.80\columnwidth}{!}{
% \input{fig02-axiondrive.tikz}
%}
\includegraphics[width=0.80\columnwidth]{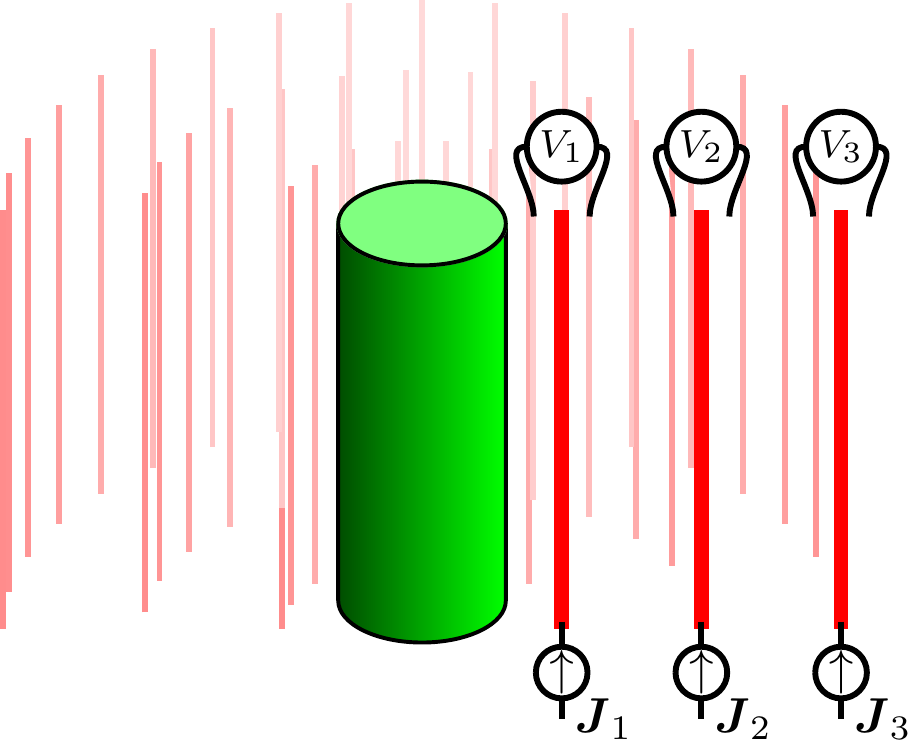}
\caption{Emulating the $-\Vzeta\times\VE$ term
 using a conducting metal cylinder (green)
 surrounded by a radial array of wires (pink),
 with a few wires visually emphasised in order to clarify the setup.
A series of voltmeters measure the radial electric fields
 (e.g. $V_1$, $V_2$, ...) near each wire,
 and the resulting information is used
 to control a current source that drives currents along those wires
 (e.g. $\VJ_1 \propto V_1/r$, $\VJ_2 \propto V_2/r$, ...).
As a result these actively monitored and driven wires
 will act as a metamaterial,
 modifying the electromagnetic field
 as if a constitutive {\pAxionResponse} $\Vzeta$ were present.
As an alternative,
 one might imagine replacing the array of wires
 with an array of controllable high energy electron beams,
 since at sufficiently high energy
 the electrons will not be deflected significantly
 by a (relatively weak) background electromagnetic field.}
\label{fig_example-cylinder_emulate}
\end{figure}

%
% ----------------------------------------------------------------------
\subsection{Outside a conducting cylinder}
\label{sch_cylinder_bdd}

An example that
 could in principle be built
 consists of an infinite conducting metal cylinder,
 which enables us to avoid difficulties at $r=0$.
The cylinder is charged
 in order to create a \ResubOne{purely} radial electric field ($\pE_r$)
 \ResubOne{with no axial component ($\pE_z=0$}),
 and then surrounded by an array of parallel wires used
 to implement the {\pAxionResponse}.
With the cylinder and wires being oriented parallel to
 the $z$-axis,
 then for a cylinder of radius $r=r_0$,
 \ResubOne{in this static case}
 we can set
%[
\begin{align}
\axivav = \frac{Z_0}{r}\ \AxisTheta
\qquadand
\axivat=0
\label{Top_cyl_zeta}
\end{align}
%]
where $Z_0$ is a constant, 
 as shown in figure \ref{fig_example-cylinder}.
Here {\xtrio{r}{\theta}{z}} are the cylindrical coordinates
 and {\xtrio{\AxisR}{\AxisTheta}{\AxisZ}}
 the corresponding orthonormal vectors.
However,
 this does not satisfy \eqref{EMCR_axi_constraints}
 since
%[
\begin{align}
 \nabla \times \axivav
=
 \frac{1}{r_0}\, \delta(r-r_0) \,\AxisZ
,
\label{Top_cyl_curl_zeta}
\end{align}
%]
 but of course
 \eqref{EMCR_axi_constraints} is required only inside bulk materials.
Fortunately,
 on the boundary of the cylinder,
 $\VE$ is along $\AxisR$ so \eqref{EMCR_axi_proof},
 and hence the conservation of charge is satisfied after all.

A crucial point that separates  our CMCR from
 the standard constitutive relations
 is that in this situation
 $\axivav$ \emph{cannot} be the gradient of any field
 as in \eqref{eqn-axion-chargeconstraintpotential};
 thus the standard constitutive relations cannot describe this.
Although locally we can set $\axipot = Z_0 \theta$,
 globally $\theta$ is not single valued and continuous;
 thus it would be impossible to model such a material
 using the traditional Maxwell's equations,
 no matter what constitutive relations were used.

However,
 with our CMCR
 we \emph{can} emulate the {\pAxionResponse} \eqref{Top_cyl_zeta}.
This is done by first measuring the radial electric field
 at each point,
 and then using that information to specify a current source
 $-\Vzeta\times\VE$ along $z$-directed wires.
As
 shown in figure \ref{fig_example-cylinder_emulate}, 
 a radially directed electric field could --
 by means of an active measurement and current generation process --
 give rise to the necessary axial current,
 as per \eqref{M_New_Max_axion_current}.
\ResubOne{Naturally,
 if the charge on the cylinder varies over time,
 such variation will need to be on much slower timescales
 than the reaction time
 of the active measurement and current generation processes
 in order to remain causal \cite{Kinsler-2010pra-lfiadc}.}
An alternative method of emulating
 an {\pAxionResponse} dynamically,
 using a mechanical substructure, 
 is suggested in section \ref{S-other-metamech}.

For the angular {\pAxionResponse} discussed here,
 we can solve the local axionic media equations
 \eqref{M_New_Max_axion_charge},
 \eqref{M_New_Max_axion_current}
 with a static radial solution given by
%[
\begin{equation}
\begin{aligned}
  \VE(r)
&=
  \left( G_1 r^{-1+\alpha} + G_2 r^{-1-\alpha} \right) \,
  \AxisR
,
\\
\text{and}\qquad
  \VB(r)
&=
  \alpha Z \EMepsilon^{-1}
  \left( G_1 r^{-1+\alpha} - G_2 r^{-1-\alpha} \right) \,
  \AxisTheta
,
\end{aligned}
\label{Top_cyl_soln}
\end{equation}
%]
 where $\alpha=Z_0^2\, \EMmu\,\EMepsilon^{-1}$.

%
% ----------------------------------------------------------------------
\subsection{Toroidal universes and periodic lattices,
 with charge.}
\label{sch_torus}

The toroidal universe imagined here is one in which
 the spatial $x$, $y$ and $z$ coordinates
 are periodic,
 with $x\to x + L_x$,
 $y\to y + L_y$ and
 $z\to z + L_z$.
This situation is also compatible with
 an infinitely periodic system, 
 whose physical properties are periodic, 
 even if the coordinates themselves are not.
In such a toroidal universe 
 the standard Maxwell's equation \eqref{GH_Max_Gibbs}
 in concert with the divergence theorem implies
 that the total charge is zero:
%[
\begin{equation}
\begin{aligned}
  \int_0^{L_x}
   \backup
   dx
&
  \int_0^{L_y}
   \backup
   dy
  \int_0^{L_z}
   \backup
   dz
 ~~
  \rho
=
  \iiint_V \rho\,  dx \, dy \, dz
  \\
&=
  \iiint_V
  \left( \nabla \cdot \VD \right)
   ~~
  dx \,
  dy \,
  dz
=
  \iint_{\partial V}
  \VD
  \cdot d\boldsymbol{S}
=
0
\end{aligned}
\label{axion-topological_int_zero}
\end{equation}
%]
 where $V$ is the 3-torus and $\partial V$ is its boundary.
However,
 since a torus does not possess a boundary
 (i.e. $\partial V=\emptyset$),
 any integral over it is zero --
 i.e. there can be no net charge on the torus.
In the periodic counterpart to \eqref{axion-topological_int_zero}, 
 the torus maps onto each cell
 in the periodic lattice, 
 and since the contributions from opposite cell boundaries
 are equal but have opposite orientations, 
 they exactly cancel,
 and again no net charge is ensured.

By contrast,
 our generalised CMCR
 gives a different substitution for the charge,
 i.e. according to \eqref{M_New_Max_CR_rho}.
The total charge is then given by
%[
\begin{equation}
\begin{aligned}
  \int_0^{L_x}
   \backup
   dx
&
  \int_0^{L_y}
   \backup
   dy
  \int_0^{L_z}
   \backup
   dz
   ~~
   \rho
\\
&=
  \int_0^{L_x}
   \backup
   dx
  \int_0^{L_y}
   \backup
   dy
  \int_0^{L_z}
   \backup
   dz
   ~~
   \left(
    \PsiErho \VAct{\VE} + \PsiBrho\VAct{\VB}
   \right)
,
\end{aligned}
\label{axion-topological_int_nonzero}
\end{equation}
%]
 which depends on $\PsiErho$ and $\PsiBrho$,
 and can therefore be non-zero.

\newcommand{\pNfac}[2]{\frac{#1 \pi}{2 L_#2}}

\def\pnx{n}
\def\pny{m}
\def\pnm{\pnx \pny}

The result has practical implications when considering
 any periodic system of size {\xtrio{L_x}{L_y}{L_z}},
 and determining its Floquet modes,
 for example using a numerical electromagnetic solver.
The approach using the standard EMCR
 yeilds \eqref{axion-topological_int_zero},
 which insists that
 the total charge on the lattice is zero.
However,
 if the charge is \emph{not} zero,
 then we must either abandon
 (a) the claim that the system is periodic,
 or
 (b) the concept of a meaningful $\VD$ in the lattice,
  and choose to use the CMCR proposed here.

We now show that a periodic solution
 with net free charge is possible,
 in the following static case based on our CMCR.
Consider a toroidal space,
 or an equivalent infinitely periodic one
 where the coordinates $x$,
 $y$, 
 and $z$
 range over (or are periodic on)
 lengths $L_x$, $L_y$, and $L_z$.
Assume that the free charge density has no $z$ dependence
 and is $\eCharge(x,y)$, 
 there is no free current density so that $\VJ=\Vzero$,
 that the {\pAxionResponse} 
 consists solely of a homogeneous $\axivaz$ component.
In this situation
 the electric field components that exist
 are ${\pE}_x$ and ${\pE}_y$, 
 so that $\VE = \AxisX {\pE}_x(x,y) + \AxisY  {\pE}_y(x,y)$.
The only magnetic field contribution will be generated
 by the {\pAxionResponse} $\axivaz$, 
 and so consist only of ${\pB}_z$, 
 so that $\VB = \AxisZ {\pB}_z(x,y)$.

The source free axionic vacuum equations
  \eqref{M_New_Max_axion_charge} and
  \eqref{eqn-axion-chargeconstraintpotential}
 then reduce to 
 just three non-zero contributions
~
\begin{align}
  \partial_x {\pE}_x + \partial_y {\pE}_y - \axivaz {\pB}_z &= \eCharge
,
\label{eqn-axion-topological-rho}
\\
 \partial_y {\pB}_z + \axivaz {\pE}_y &= 0
,
\label{eqn-axion-topological-Jx}
\\
 -\partial_x {\pB}_z - \axivaz {\pE}_x &= 0
\label{eqn-axion-topological-Jy}
.
\end{align}
For non-zero $\axivaz$, 
 \eqref{eqn-axion-topological-Jx} and \eqref{eqn-axion-topological-Jy}
 can be substituted into \eqref{eqn-axion-topological-rho}
 to give
~
\begin{align}
 -
  %\frac{1}{\axivaz}
  \partial_x^2 {\pB}_z 
 -
  %\frac{1}{\axivaz}
  \partial_y^2 {\pB}_z 
 -
  \axivaz^2
  {\pB}_z
&=
  \axivaz
  {\eCharge}
.
\label{eqn-axion-topological-tode}
\end{align}

A periodic $\eCharge(x,y)$
 can be written as a Fourier series where 
~
\begin{align}
  \eCharge(x,y)
 &=
  \sum_{\pnm}
      {\eCharge}_{\pnm}
      \cos\left( \pNfac{\pnx}{x} x \right)
      \cos\left( \pNfac{\pny}{y} y \right)
,
\end{align}
 where the sums are over $\pnx,\pny \in \{0,1,2,\ldots,\infty\}$
 and the coordinate range is centred about the origin.

Since the system is periodic, 
 ${\pB}_z$ is also periodic, 
 and can also be written as a Fourier series
~
\begin{align}
  {\pB}_z(x,y)
 &=
  \sum_{\pnm}
      b_{\pnm}
      \cos\left( \pNfac{\pnx}{x} x \right)
      \cos\left( \pNfac{\pny}{y} y \right)
.
\end{align}
Now we can use 
 \eqref{eqn-axion-topological-tode}
 to relate the ${\eCharge}_{\pnm}$ coefficients 
 to the $b_{\pnm}$, 
 i.e.
~
\begin{align}
  \left[
    \left( \pNfac{\pnx}{x} \right)^2
   +
    \left( \pNfac{\pny}{y} \right)^2
   -
    \axivaz^2
  \right]
  b_{\pnm}
&=
  \axivaz
  {\eCharge}_{\pnm}
&
,
\end{align}
and this allows us to easily determine
 each of the $b_{\pnm}$ from 
 any given $\eCharge(x,y)$.
Further, 
 since any electric field component ${\pE}_i(x,y)$
 must also be periodic, 
 with Fourier components $e_{i,\pnm}$, 
 then we also have
 from \eqref{eqn-axion-topological-Jx}
  and \eqref{eqn-axion-topological-Jy}
~
\begin{align}
  e_{x,\pnm}
&=
  \frac{1}{\axivaz}
    \pNfac{\pnx}{x}
  b_{\pnm}
,
\quad
\textrm{and}
\quad
  e_{y,\pnm}
&=
  \frac{1}{\axivaz}
    \pNfac{\pny}{y}
  b_{\pnm}
.
\end{align}

We also need to satisfy both of 
 \eqref{GH_Max_NoMono} and \eqref{GH_Max_Faraday}.
We already have $\grad \cdot \VB=0$
 because ${\pB}_z$ depends only on $x,y$.
This leaves the condition $\grad \times \VE = 0$, 
 i.e. that
~
\begin{align}
%  \partial_y {\pE}_x - \partial_x {\pE}_y = 0
%~{\Leftrightarrow}
  \pNfac{\pny}{y} e_{x,\pnm}
&=
  \pNfac{\pnx}{x} e_{y,\pnm}
,
\end{align}
 which can be converted into a condition
 not only on $b_{\pnm}$ 
 but also the charge distribution parameters $\eCharge_{\pnm}$;
 fortunately,
 a back substitution reveals that 
 this is already satisfied.

Note that the interesting case here
 is when the net charge on the system is non-zero, 
 which occurs solely when ${\eCharge}_{00}$ is non-zero.
Indeed, 
 the total charge on the torus --
 or one element in the periodic lattice --
 is simply $\eCharge_{\textrm{total}} = {\eCharge}_{00} L_x L_y L_z$.
Here
 the charge density, 
 and hence the fields,
 are constant, 
 has a field solution trivially obtained from 
 \eqref{eqn-axion-topological-rho}
 where $\VE=\Vzero$
 and $\VB= - \AxisZ \left(\eCharge/\axivaz\right)$.
Thus although there is a charge on the torus, 
 there is no electric field from that charge; 
 there is only an axionically-induced magnetic field
 whose field lines form (topologically allowed) loops.

In the more general solutions,
 we can see that the additional coupling between
 the electric and magnetic fields permitted by the presence
 of the {\pAxionResponse}
 allows a finite charge to be supported in the system.
Consider each effect in turn:
 the free charge creates an electric field, 
 then 
 that electric field causes the {\pAxionResponse}
 to generate a polarization current, 
 then this polarization current in turn creates a magnetic field.
Finally, 
 this magnetic field causes the {\pAxionResponse}
 to (also) create a polarization charge; 
 and we find that this exactly counteracts the free charge.

%
% ======================================================================
\section{Examples}
\label{ch_Examples}

In this section
 we consider situations
 involving conventional media
 with additional constitutive properties
 that match the {\pAxionResponse}s described above.
We will call these {\pAxionREA}s,
 and it is important to note
 that they have a vacuum contribution
 (or a conventional and homogeneous $\EMepsilon$ and $\EMmu$)
 in their constitutive properties
 as well as the added {\pAxionResponse}.
In what follows,
 remember that these are constitutive properties,
 and are not the result of coupling Maxwell's equations
 to an external axion particle field.

%
% ----------------------------------------------------------------------
\subsection{Homogeneous case: longitudinal waves}\label{ch_axion-longitudinal}

The propagation of electromagnetic fields
 in media with an axion-like response has been an area of interest
 for some time \cite{Itin-2008grg,Visinelli-2013mpla},
 particularly with regard to its symmetry properties
 and axionic dispersion relations.
A starting point of a homogeneous {\pAxionREA}
 with all 4 constitutive terms being non-zero
 contains a large number of interesting cross-couplings
 between the $\VE$ and $\VB$ fields,
 leading to a range of new behaviours.
Here,
 however
 we will highlight one interesting feature --
 namely that propagating EM fields
 in a
 homogeneous medium, 
 with isotropic $\EMepsilon$ and $\EMmu$,
 need no longer be purely transverse.

\begin{figure}
\includegraphics[angle=-0,width=0.80\columnwidth]{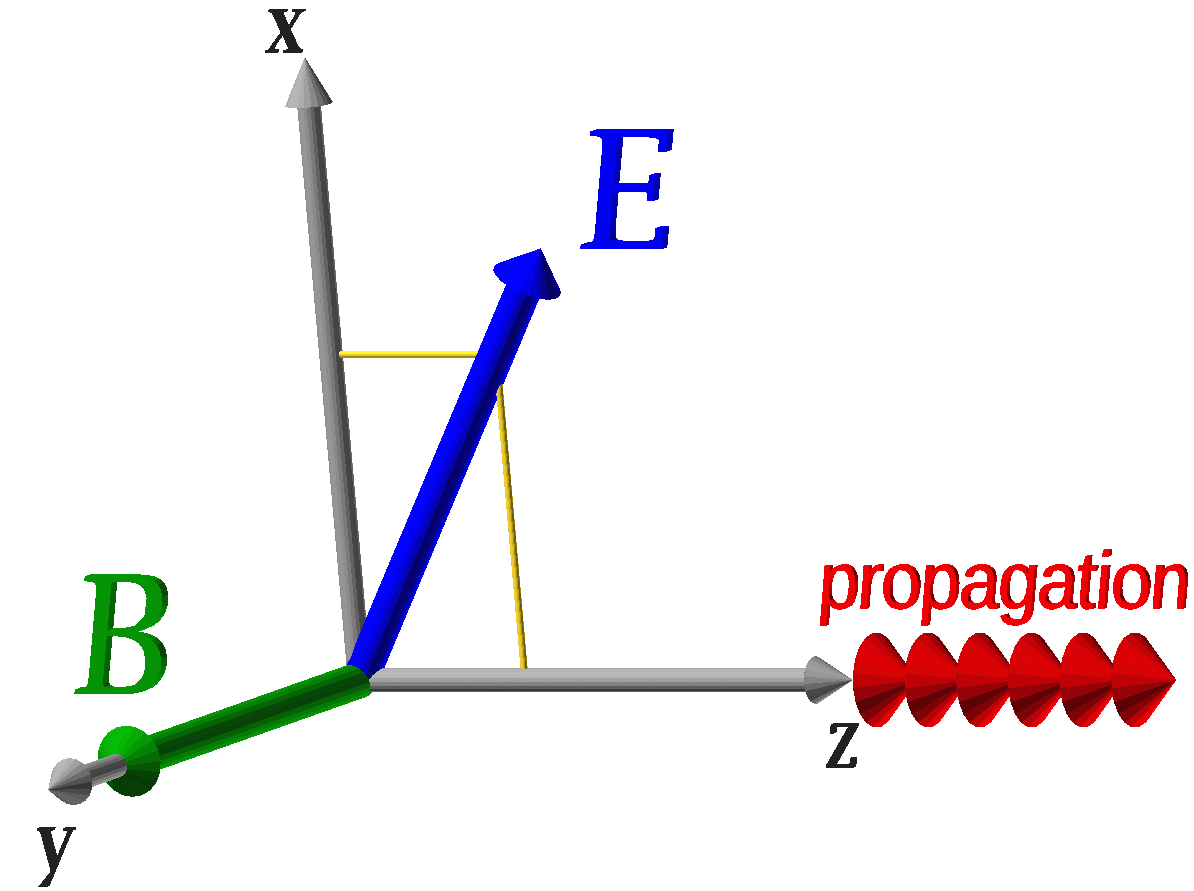}
\caption{An EM wave propagating in the $\AxisZ$ direction
 has its electric field vector $\VE$ (thick blue arrow) rotated forwards
 in the ${\AxisX}{-}{\AxisZ}$ plane
 by an axionic perturbation to the propagation medium;
 in the case shown the magnetic field vector $\VB$ (thick green arrow)
 remains transverse.
The propagation direction $\Vk$ is indicated with
 multiple arrowheads (red) for clarity.
The thin arrows indicate the coordinate axes.
}
\label{fig_example-longwave}
\end{figure}

Starting with Maxwell's equations \eqref{GH_Max_NoMono},
  \eqref{GH_Max_Faraday} and the source free axionic vacuum equations
  \eqref{M_New_Max_axion_charge} and
  \eqref{eqn-axion-chargeconstraintpotential} we have 
%[
\begin{equation}
\begin{gathered}
  \nabla \cdot  \VB            = 0,\qquad
  \nabla \times \VE +           \dtime{\VB} = \Vzero,\\
  \EMepsilon_0\nabla \cdot \VE  +  \axivav \cdot \VB = 0,\\
  \EMmu^{-1}_0\nabla \times \VB  -  \EMepsilon_0 \dtime{\VE}
   - \axivav\times\VE - \axivat \VB =  \Vzero
.
\end{gathered}
\label{axion-longitudinal_Vac_Max}
\end{equation}
%]
 We can show that the existence of the {\pAxionResponse} terms enables
 the propagation of EM waves with a \emph{longitudinal} component,
 as depicted on figure \ref{fig_example-longwave}.
Although achievable with the aid of artificial functional materials
 \cite{Gratus-KLB-2017apa-malaga,Boyd-GKL-2018oe-tbwire,Boyd-GKLS-2018as-mdpi},
 or in an ordinary anisotropic (birefringent) medium,
 here we can do this with simple homogeneous constitutive properties.

For example for a wave travelling in the $\AxisZ$-direction
 satisfying the dispersion relations
%[
\begin{align}
  \omega^2
 +
  \axivay^2
  \EMepsilon_0^{-2}
 -
  c^2 k^2
&=
  0
,
\label{axion-longitudinal_dispersion_rel}
\end{align}
%]
and compatible with constraints on the {\pAxionResponse}
%[
\begin{align}
  \omega
  \axivaz
 +
  k
  \axivat
&=
  0
,
\qquadand
  \axivax
=
  0
.
\label{axion-longitudinal_dispersion_rel_real}
\end{align}
%]
By direct substitution we see the propagating electromagnetic field given by
%[
\begin{align}
  \VE
&=
  {\pE}_0
  \omega
  \cos \left( \omega t -  k z \right)
  \AxisX
 -
  {\pE}_0
  \frac{\axivay}{\EMepsilon_0}
  \sin \left(\omega t -  k z \right)
  \AxisZ
,
\label{axion-longitudinal_Esol_real}
\\
  \VB
&=
  {\pE}_0
  k
  \cos \left( \omega t -  k z \right)
  \AxisY
.
\label{axion-longitudinal_Bsol_real}
\end{align}
%]
is a solution to Maxwell's equations
 in a vacuum {\pAxionREA}
 \eqref{axion-longitudinal_Vac_Max}.
Note here how the homogeneous $\axivay$ {\pAxionREA}
 still supports the propagation whilst having
 rotated the
 electric field orientation forward,
 away from a purely transverse direction.

%
% ----------------------------------------------------------------------
\subsection{Inhomogeneous case: Static solutions}\label{S-axion-example}

\def\aradius{R}
\def\cradius{R_c}
\def\wradius{R_w}

\def\aarea{A}
\def\warea{\aarea_w}
\def\carea{\aarea_c}

\def\cthick{T}
\def\alength{L}

\def\pHstep{\mathfrak{H}}

\begin{figure}
\includegraphics[angle=-0,width=0.7\columnwidth]{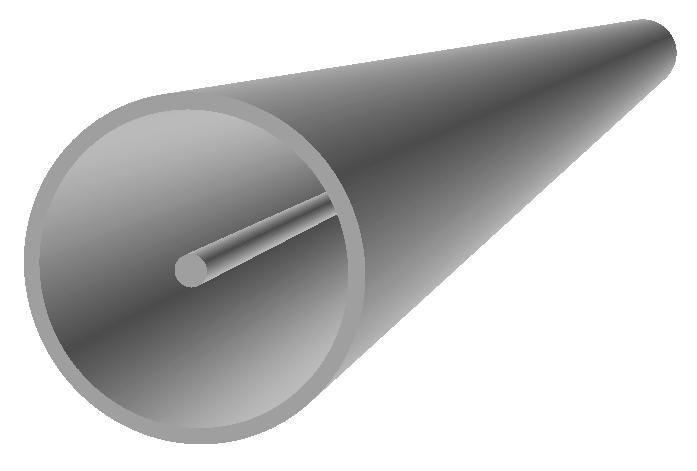}
\caption{The wire and shell system
 represented by \eqref{eqn-example-modulationfn}
 with a wire of radius $\wradius=1$,
 and a shell centred at $\cradius=10$ with thickness $\cthick=1$.
The wire and cylindrical shell lie along the $z$ axis.
}
\label{fig-wireshell}
\end{figure}

Here we consider two static cases
 based on cylindrical symmetry.
These are based primarily on a combination of
 a thin cylindrical shell
 which has radius $\cradius$
 and thickness $\cthick$,
 and a thin wire
 with radius $\wradius$,
 as depicted on figure \ref{fig-wireshell}.
They are made of 
 a vacuum augmented with
 an {\pAxionResponse} of the kind described above.
Although we would like to treat each of the
 4 axionic components separately,
 it turns out that only the $z$-directed ($\axivaz$)
 and the $t$-directed ($\axivat$) are
 sufficiently interesting
 whilst still allowing a straightforward discussion
 --
 the case where radial $\axivar$ is non-zero is very simple,
 the case where angular $\axivaa$ is non-zero is rather complicated.

The cylindrical symmetry here means that
 the modulation function for the axionic % {\pAxionResponse} dependent
 wire and shell properties depends only on $r$,
 and is a sum of offset step functions ${\pHstep}$.
For a wire of radius $\wradius$ % = \sqrt(\warea/\pi)$
 the modulation (density) function $W(r)$ is
~
\begin{align}
  W(r)
&=
  W_w(r)
+
  W_c(r)
\nonumber
\\
&=
  \tfrac{1}{\pi \wradius^2}
  {\pHstep} ( \wradius - r )
\nonumber
\\
&\quad
 +
  \tfrac{1}{\pi \wradius^2} %{\cthick}
  \left[
    {\pHstep} ( r - \cradius + \tfrac{\cthick}{2} )
   -
    {\pHstep} ( r - \cradius - \tfrac{\cthick}{2} )
  \right]
.
\label{eqn-example-modulationfn}
\end{align}
Here this modulation function
 has been normalised to compensate
 for the effect of the wire's cross-sectional area.
In the examples below,
 there are only the fields $\VE$ and $\VB$,
 and two types of sources.
There is
 free (source) charge ${\pS}$
 and
 free (source) current ${\eCurrent}$;
 in addition there are
 polarization (charge and current) sources.
The polarization sources are those induced %excited (brought into existence)
 in the medium
 by the presence of an $\VE$ or $\VB$ field
 acting on the {\pAxionResponse}.

%
% - - - - - - - - - - - - - - - - - - - - - - - - - - - - - - - - - - - -
\subsubsection{Free charge and axial {\pAxionResponse} $\axivaz$}\label{S-axion-exampleQ}

Here we take the wire and cylindrical shell
 to consist of an {\pAxionREA}
 with only $\axivaz(r)$ being non-zero.
This can be derived from a $\axipot(r,\theta,z,t)$
 with an appropriate $r$-dependent modulation $W(r)$
 of a linear increase along the axis $z$;
~
\begin{align}
  \axipot(r,z) = \axivaz W(r) z
,
\end{align}
 where $W(r)$ is zero everywhere but in the wire and shell.
However,
 somewhat inconveniently,
 this $\axipot(r,z)$ also has an $r$ derivative,
 and so along with our desired $\axivaz$ properties
 we also get a non-zero and $z$-dependent $\axivar$ on the surfaces
 of the wire and shell.

If we write just the
 combinations that are potentially non-zero,
 then with
 $\VB = {\pB}_r \AxisR + {\pB}_\theta \AxisTheta + {\pB}_z \AxisZ$
 and
 $\VE = {\pE}_r \AxisR + {\pE}_\theta \AxisTheta + {\pE}_z \AxisZ$,
 the static constitutive relations in the medium
 are
~
\begin{align}
  \EMepsilon\nabla\cdot\VE
 -
  W(r)
  \axivar {\pB}_r
 -
  W(r)
  \axivaz {\pB}_z
&=
  W(r)
  {\pS}   %{\eCharge}
\\
  \frac{1}{\EMmu}
  \nabla\times\VB
% -
%  \EMepsilon\dtime{\VE}
 +
  W(r)
  \left( \axivar \AxisR + \axivaz \AxisZ \right)
  \times
  \VE % \left( {\pE}_r \AxisR {\pE}_\theta, {\pE}_z \right)
&=
 W(r)
 {\eCurrent}
.
\label{eqn-axion-exampleS-zetaz-V}
\end{align}
Note the nature of the $\axivar$ surface terms
 in these two equations,
 and,
 in particular,
 that
 (a) if the magnetic field has no radial component,
 and
 (b) if the electric field remains purely radial,
 they will have no effect.
Because of our construction,
 the radial symmetry guarantees both
 a radial electric field,
 and an enforced non-radial magnetic field.
This means the $\axivar$ surface terms
 play no role in the following calculation;
 but if the symmetry was broken and they did,
 to first order
 they would induce surface charges and currents on the wire and shell.

These symmetry restrictions reduce the above equations to simpler ones; 
and also 
 mean that the other Maxwell equations, 
 namely \eqref{GH_Max_NoMono} and \eqref{GH_Max_Faraday}
 are automatically satisfied.
The first equation is for the radial electric field ${\pE}_r$
 given the presence of the free charge line density ${\pS}$
 and an axial magnetic field ${\pB}_z$:
~
\begin{align}
  \EMepsilon r^{-1}
  \partial_r \left[ r {\pE}_r \right]
% -
%  \axivaz {\pB}_z
&=
  W_w(r) {\pS} + W(r) \axivaz {\pB}_z
,
%& \quad [C/m^3]
\label{eqn-axion-exampleS-zetaz-simple-Er}
\end{align}
where the second RHS term can be interpreted as
 a polarization charge density
 induced by the action of the free fields
 on the {\pAxionResponse} of the medium.
The second equation is for the axial magnetic field ${\pB}_z$
 under the influence of the radial electric field ${\pE}_r$:
~
\begin{align}
  \EMmu^{-1}
  \partial_r {\pB}_z
&=
 -
  W(r)
  \axivaz
  {\pE}_r
%=
% {\pj}_\theta
%&
%\quad [C/sm^2]
,
\label{eqn-axion-exampleS-zetaz-simple-Bz}
\end{align}
 where the RHS can be interpreted as % ${\pj}_\theta$
 a polarization current density
 induced by the action of the free fields
 on the {\pAxionResponse}.

These two equations can be combined,
 %and $r$ scaled appropriately,
 resulting in inhomogeneous Bessel's equations
 of order $\nu=0$ for ${\pB}_z$ 
 and order $\nu=1$ for ${\pE}_r$.
With $\bar{r} = W \axivaz r$
 and a suitably scaled charge density $\bar{\pS}$,
 we have
~
\begin{align}
  \bar{r}^2
    \partial_{\bar{r}}^2
    {\pB}_z
   +
    \bar{r}
    \partial_{\bar{r}}
    {\pB}_z
 +
  \bar{r}^2
  {\pB}_z
&=
 -
  \bar{r}^2
    \bar{\pS}  % {\pS} \axivaz^{-1}
\label{eqn-axion-exampleS-zetaz-Bessel-Bz}
,
\\
  \bar{r}^2
    \partial_{\bar{r}}^2
    {\pE}_r
 +
  \bar{r}
    \partial_{\bar{r}}
    {\pE}_r
 +
  \left[
    \bar{r}^2
   -
    1
  \right]
    {\pE}_r
&=
  0
.
\label{eqn-axion-exampleS-zetaz-Bessel-Er}
\end{align}
Note in particular that such behaviour emphasises again
 the difference between our CMCR response
 and the standard tensorial EMCR one.
Standard EMCRs only allow coupling to axionic effects
 to occur at boundaries,
 but here we see the effects of the {\pAxionResponse}
 present in a homogeneous system
 (i.e. inside the wire and/or shell).

\begin{figure}[h!]
%\resizebox{0.9\columnwidth}{!}{
\includegraphics[angle=-90,width=0.90\columnwidth]{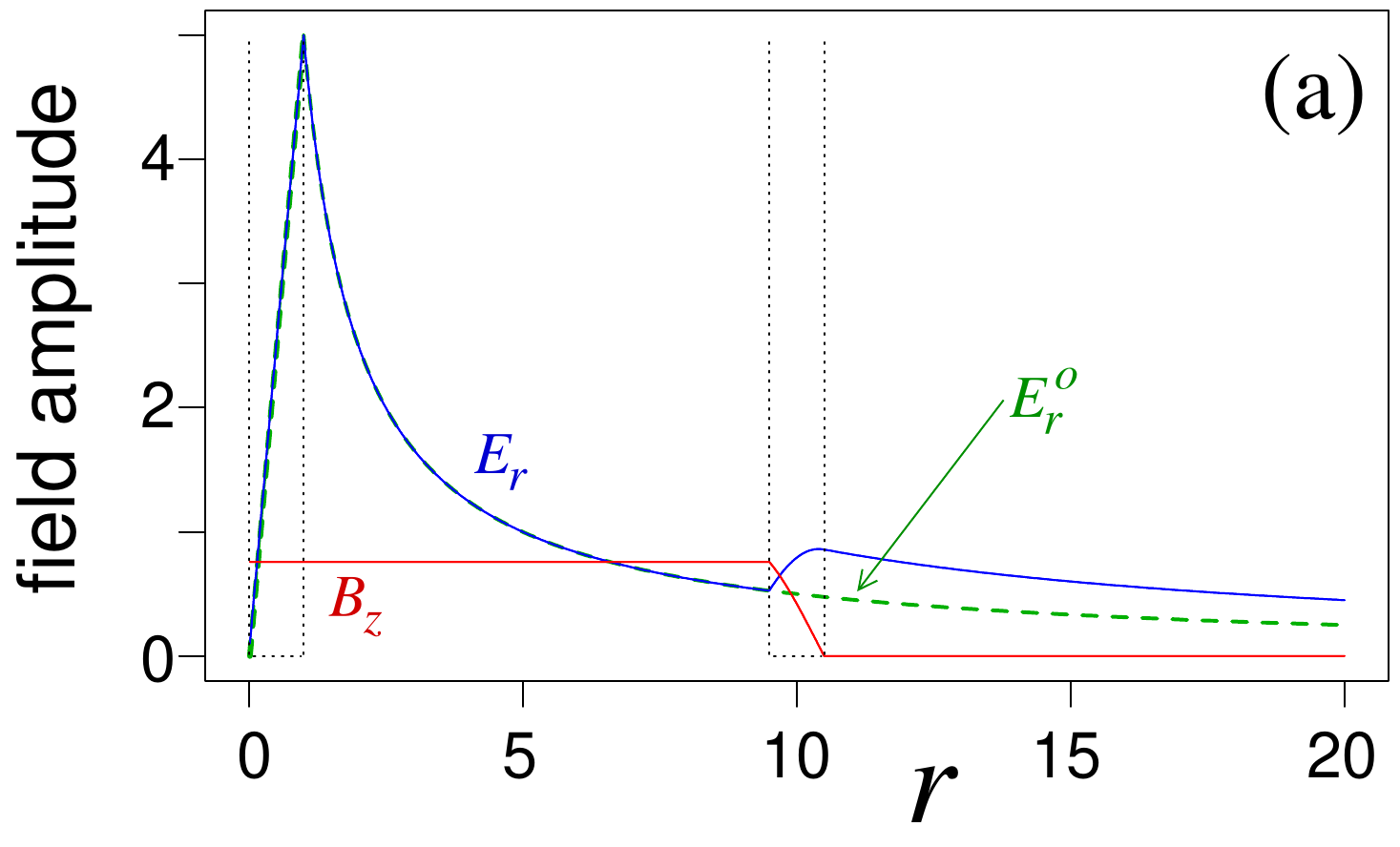}
\includegraphics[angle=-90,width=0.90\columnwidth]{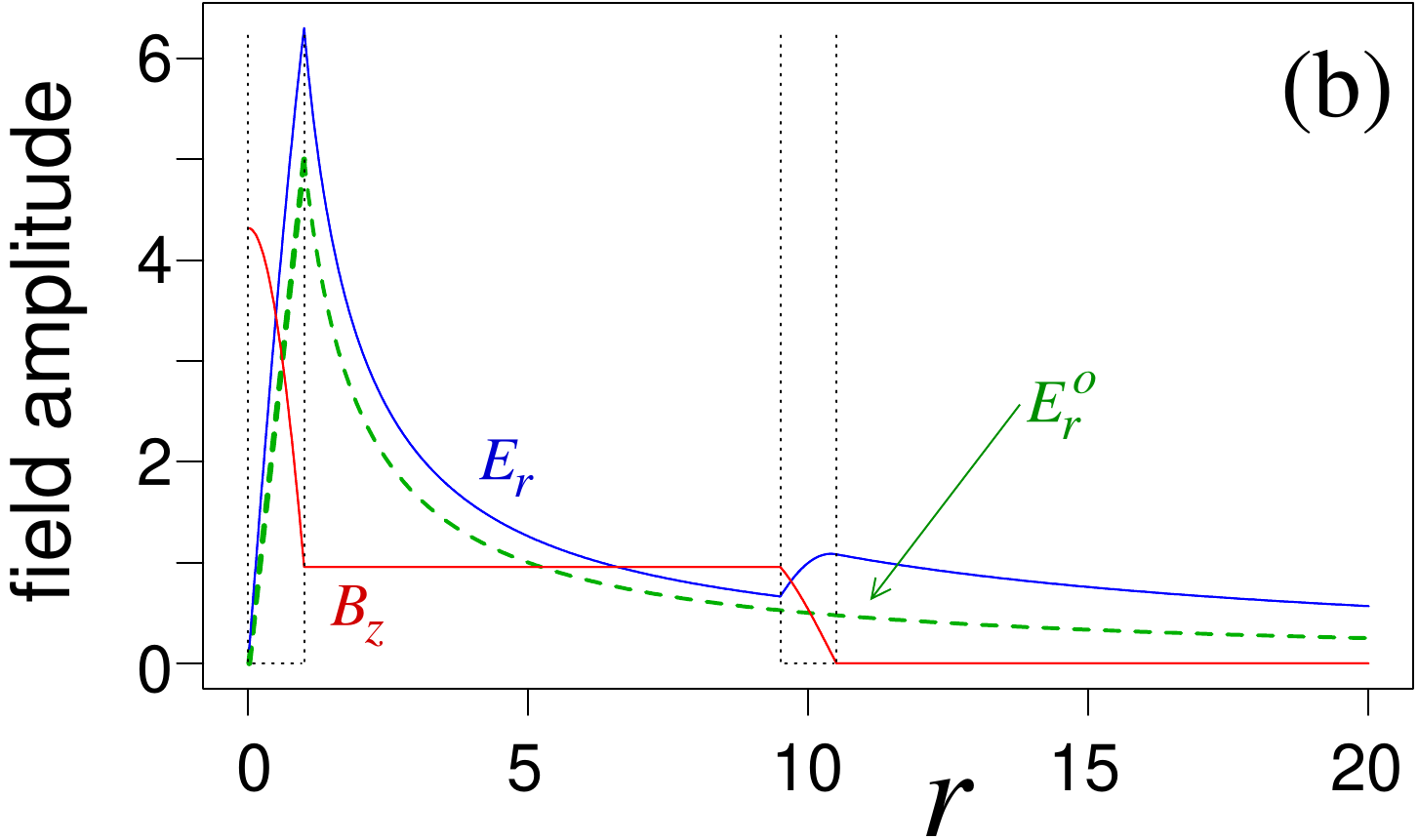}
%}
\caption{The fields in a wire and shell system
 in scaled units with $\EMepsilon=1, \EMmu=1$,
 and $\axivaz=1$,
 with line charge density ${\pS}=1$ on the wire.
The wire has a radius $\wradius=1$
 and the shell is centred at $\cradius=10$
 with thickness $\cthick=1$.
The electric field ${\pE}_r$ is in blue,
 and the magnetic field ${\pB}_z$ is in red.
The green dashed curve is the radial electric field ${\pE}_r^o$
 obtained for the same charge density distribution
 in non-axionic materials.
The upper panel (a) shows an alternate simpler case where the wire is charged
 but has no {\pAxionResponse}
 (i.e. its modulation function differs from \eqref{eqn-example-modulationfn}),
 whereas the lower panel (b) shows results for
 a both charged and axionic wire
 (and is consistent with \eqref{eqn-example-modulationfn}).
}
\label{fig-numerfig-E01-axivaz-A}
\end{figure}

We now numerically solve
 \eqref{eqn-axion-exampleS-zetaz-simple-Er}
 and
 \eqref{eqn-axion-exampleS-zetaz-simple-Bz}
 in order to give a qualitative
 sense of the axionic effects.
Solution is straightforward,
 the only constraint being that we must pick an initial
 ${\pB}_z$ value at $r=0$ that results in ${\pB}_z=0$
 outside the shell.
We achieve this using a simple iterative process for zero-finding 
 that converges to the right answer.

\begin{figure}
%\resizebox{0.9\columnwidth}{!}{
\includegraphics[angle=-90,width=0.90\columnwidth]{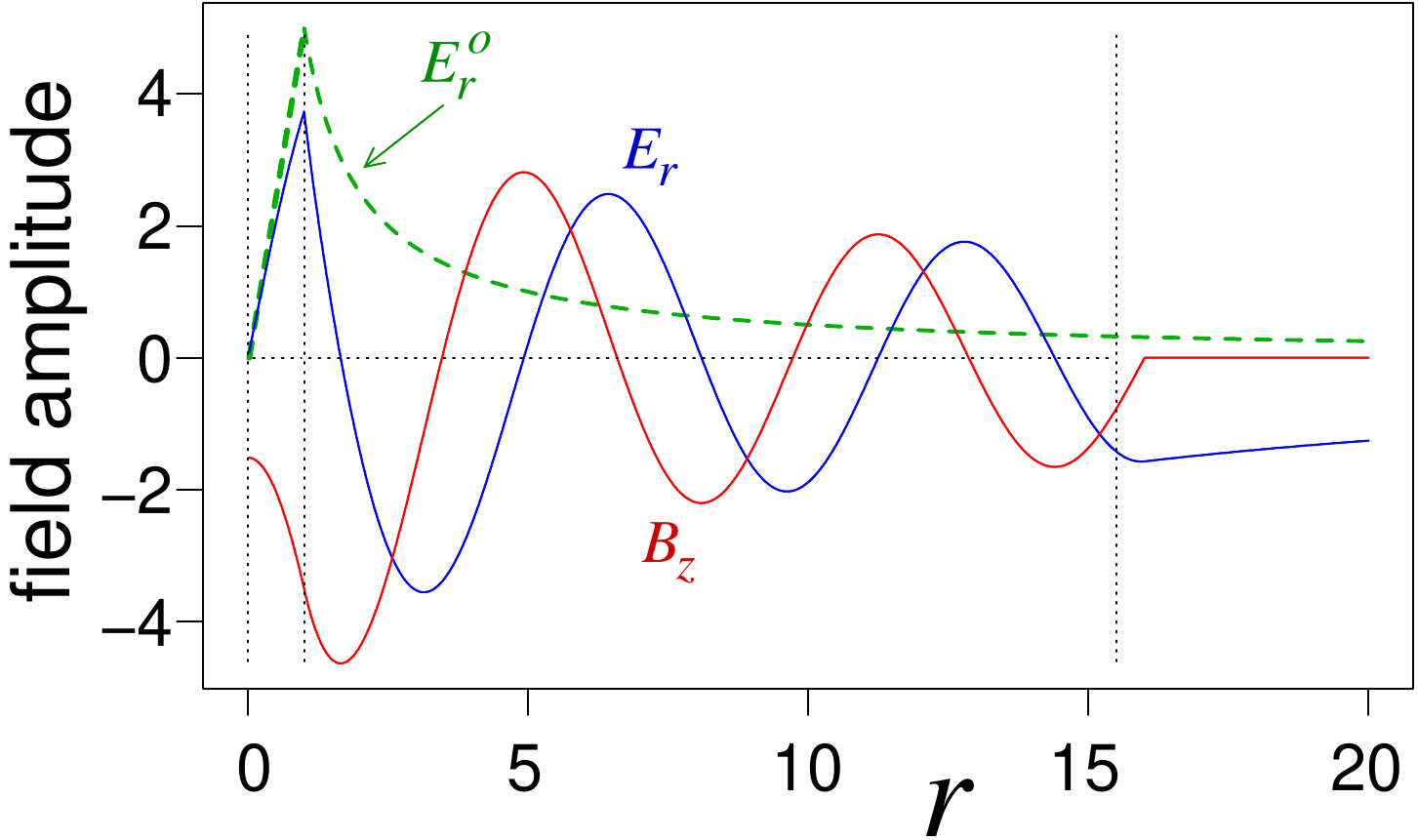}
%}
\caption{The fields in a wire embedded inside a
 thick cylinder,
 in scaled units with $\EMepsilon=1, \EMmu=1$,
 and $\axivaz=1$,
 with line charge density ${\pS}=1$ on the wire.
The wire has a radius $\wradius=1$
 and the shell is centred at $\cradius=8.5$
 with thickness $\cthick=15$.
The electric field ${\pE}_r$ is in blue,
 and the magnetic field ${\pB}_z$ is in red.
The green dashed curve is the radial electric field ${\pE}_r^o$
 obtained for the same charge density distribution
 in non-axionic materials.
The Bessel-like nature of the fields
 under the influence of this axionic material
 can be clearly seen.
Notably,
 even though the oppositely-signed ${\pB}_z$ inside the wire
 acts to suppress the effective charge density there,
 the field oscillations here
 enhance the electric field strength outside the shell,
 whilst also swapping its sign.
}
\label{fig-numerfig-E01-axivaz-B}
\end{figure}

A typical result
 is shown on figure \ref{fig-numerfig-E01-axivaz-A}.
Here the axionic effect is fairly strong,
 which enables the character of the modifications
 from the non-axionic result to be easily seen.
In  figure \ref{fig-numerfig-E01-axivaz-A}(a),
 where there is no {\pAxionResponse} in the wire,
 we see that the ${\pE}_r$ matches that for
 an ordinary charged rod until it reaches the shell.
In the shell the ${\pE}_r$ induces a circulating polarization current,
 which generates the constant ${\pB}_z$ inside the shell and wire.
As the ${\pB}_z$ falls to zero across the shell,
 it induces a polarization charge,
 which enhances the electric field.

In  figure \ref{fig-numerfig-E01-axivaz-A}(b),
 the {\pAxionResponse} in the wire leads to
 an extra inverted parabolic behaviour for ${\pB}_z$,
 and this increased ${\pB}_z$ induces a polarization charge
 which enhances the electric field.
However,
 note that increasing $\axivaz$ further can push the results
 into a regime where axionic effects
 triggered by the charge density dominate,
 and the electric field can even change sign.

With sufficient distance in which to accumulate these
 {\pAxionResponse} effects further,
 we see more complicated behaviour.
In figure \ref{fig-numerfig-E01-axivaz-B}
 we see how if the shell is thick and there is no gap between it
 and the wire,
 an oscillatory Bessel function behaviour manifests itself,
 in line with \eqref{eqn-axion-exampleS-zetaz-Bessel-Bz}
 and \eqref{eqn-axion-exampleS-zetaz-Bessel-Er}.

%
% - - - - - - - - - - - - - - - - - - - - - - - - - - - - - - - - - - - -
\subsubsection{Free current and time-directed {\pAxionResponse}: $\axivat$}\label{S-axion-exampleJ}

Here we take the wire and cylindrical shell
 to consist of an {\pAxionREA}
 with only $\axivat(r)$ being non-zero{; 
  note that despite the different physics, 
  this derivation follows very similar steps 
  to the previous one.
This can be derived from a $\axipot(r,\theta,z,t)$
 with an appropriate $r$-dependent modulation $W(r)$
 of a linear increase in time $t$;
~
\begin{align}
  \axipot(r,t) = \axivat W(r) t
,
\end{align}
 where $W(r)$ is zero everywhere but in the wire and shell.
However,
 somewhat inconveniently,
 this $\axipot(r,t)$ does have an $r$ derivative,
 and so along with our desired $\axivat$ properties
 we also obtain a non-zero and $t$-dependent  $\axivar$ on the surfaces
 of the wire and shell.

If we write down just the 
 potentially non-zero combinations,
 the static constitutive relations in the medium
 are
~
\begin{align}
  \EMepsilon\nabla\cdot\VE
 -
  W(r)
  \axivar \cdot {\pB}_r
&=
  W(r)
  {\pS}
,
\label{eqn-axion-examplerho-zetat}
\\
  \frac{1}{\EMmu}
  \nabla\times\VB
 %-
 % \EMepsilon\dtime{\VE}
 +
  W(r)
  \left( \axivar \AxisR \right)
  \times
  \VE % \left( 0, {\pE}_\theta, {\pE}_z \right)
& %\quad &
\nonumber
\\
 +
  W(r)
  \axivat \VB
&=
  W(r)
 {\eCurrent}
.
\label{eqn-axion-exampleJ-zetat}
\end{align}
Note the nature of the $\axivar$ surface terms
 in these two equations,
 and in particular that if the electric field remains purely radial,
 they will have no effect.
Because of our construction,
 there is no electric behaviour in this system,
 and even if one somehow appeared,
 the radial symmetry would guarantee a concommitantly radial electric field.
This means that the $\axivar$ surface terms
 play no role in the following calculation;
 but if the symmetry was broken and they did,
 they would induce surface charges and currents on the wire and shell.

These symmetry restrictions reduce the above equations to simpler ones;
 and also 
 mean that the other Maxwell equations, 
 namely \eqref{GH_Max_NoMono} and \eqref{GH_Max_Faraday}
 are automatically satisfied.
The first equation is for the angular magnetic field ${\pB}_\theta$
 resulting from the current density ${\pJ}_z$
 and any axial magnetic field ${\pB}_z$:
~
\begin{align}
  \EMmu^{-1}
  r^{-1} \partial_r \left[ r {\pB}_\theta \right]
&=
  W(r)
 {\pJ}_z
 -
  W(r)
  \axivat
  {\pB}_z
%=
%  W(r)
%  {\pJ}_z
% +
%  W(r)
%  {\pj}_z
,
\label{eqn-axion-exampleJ-zetat-simple-Bang}
\end{align}
 where the second RHS term can be interpreted
 as a polarization current density
 induced by the action of the axial magnetic field
 on the {\pAxionResponse} of the medium.
The second equation is for the axial magnetic field ${\pB}_z$
 under the influence of the angular magnetic field ${\pB}_\theta$:
~
\begin{align}
  \EMmu^{-1}
  \partial_r {\pB}_z
&=
 -
  W(r)
  \axivat
  {\pB}_\theta
%=
%  W(r)
% {\pj}_\theta
,
\label{eqn-axion-exampleJ-zetat-simple-Bz}
\end{align}
 where the RHS term can be interpreted
 as a polarization current density
 induced by the action of the angular magnetic field
 on the {\pAxionResponse}. % of the medium.

These two equations can be combined,
 resulting in inhomogeneous Bessel's equations
 of orders $\nu=0$ and $\nu=1$
 for ${\pB}_z$ and $ {\pB}_{\theta}$ respectively.
With $\bar{r} = W \axivat r$
 and a suitably scaled current  $\bar{\pJ}_z$,
 we have
~
\begin{align}
  {\bar{r}}^2
  \partial_{\bar{r}}^2
  {\pB}_z
 +
  {\bar{r}}
  \partial_{\bar{r}}
  {\pB}_z
 +
  {\bar{r}}^2
  {\pB}_z
&=
  {\bar{r}}^2
  \bar{{\pJ}}_z
,
\\
  {\bar{r}}^2
  \partial_{\bar{r}}^2
    {\pB}_{\theta}
 +
  {\bar{r}}
  \partial_{\bar{r}}
    {\pB}_{\theta}
 +
  \left[
    {\bar{r}}^2
   -
    1
  \right]
    {\pB}_{\theta}
&=
  0
% constant J
%  {\bar{r}}^2
%   \partial_r
%  \bar{\pJ}_z
.
\end{align}

Note that solutions for this system
 are mathematically identical to those presented
 in the previous subsection,
 for an $\axivaz$ response.
This can be seen by inspection of
 \eqref{eqn-axion-exampleJ-zetat-simple-Bang}
 and \eqref{eqn-axion-exampleJ-zetat-simple-Bz},
 and comparison with
 \eqref{eqn-axion-exampleS-zetaz-simple-Er}
 and \eqref{eqn-axion-exampleS-zetaz-simple-Bz};
 the differences being solely the replacement
 of ${\pE}_r$ with ${\pB}_{\theta}$,
 of $\EMepsilon$ with $1/\EMmu$,
 a sign on the axionic term,
 and replacing the charge density $\pS$ with ${\pJ}_z$.

%
% ======================================================================
\section{Other Topics}\label{ch_other}

%
% ----------------------------------------------------------------------
\subsection{Metamechanical {\pAxionResponse}}\label{S-other-metamech}

An interesting question is whether or not
it is possible to design a metamaterial unit cell
 which can generate an {\pAxionResponse}
 of the kinds discussed here.
Broadly speaking,
 there are two sorts of {\pAxionResponse},
 those that generate charge (see \eqref{M_New_Max_axion_charge})
 and those that generate current (see \eqref{M_New_Max_axion_current}).
Since it is hard to generate a free charge from nowhere,
 or turn on and off some mechanism for isolating that charge,
 it is easiest to focus on current generating {\pAxionResponse}s.

Since the {\pAxionResponse}
 is outside the scope of standard electromagnetism,
 we will utilise concepts from the area
 of mechanical metamaterials \cite{Surjadi-GDLXFL-2019aem,Yu-LJW-2018pms},
 albeit ones driven by applied electromagnetic fields,
 and which --
 with the addition of moving charged elements,
 can generate currents.
It is the mechanical movement and couplings of
 the metamaterial component that create constitutive properties
 of the necessary orientation for an {\pAxionResponse}.

The metamaterial unit cell concept shown in
 figure \ref{fig_example-metamech}
 depicts a mechanism
 intended to create the {\pAxionResponse}
 where an electric field applied across the page
 (e.g. along $\AxisX$)
 generates a vertical current (oriented along $\AxisY$).
Whilst this suffices for selecting the
 appropriate {\pAxionResponse} orientations with respect to the field,
 this is not the complete picture.
As with the great majority
 of metamaterial schemes,
 this is (a) a dynamic response suitable for oscillating fields only,
 and (b) will typically only generate the desired response current
 with a phase offset to the driving field.
Further,
 it will also generate a side effect current (along $\AxisX$),
 and of itself generate a background dipolar/quadrupolar electric field.
Nevertheless,
 if the driving electric field is strong,
 and the detector charges $q_b$ are weak compared
 to the response ones $q_r$,
 and the mechanical system oscillates
 and is driven at the correct frequency,
 the system can function as an {\pAxionResponse}.

\begin{figure}
\includegraphics[angle=-0,width=0.95\columnwidth]{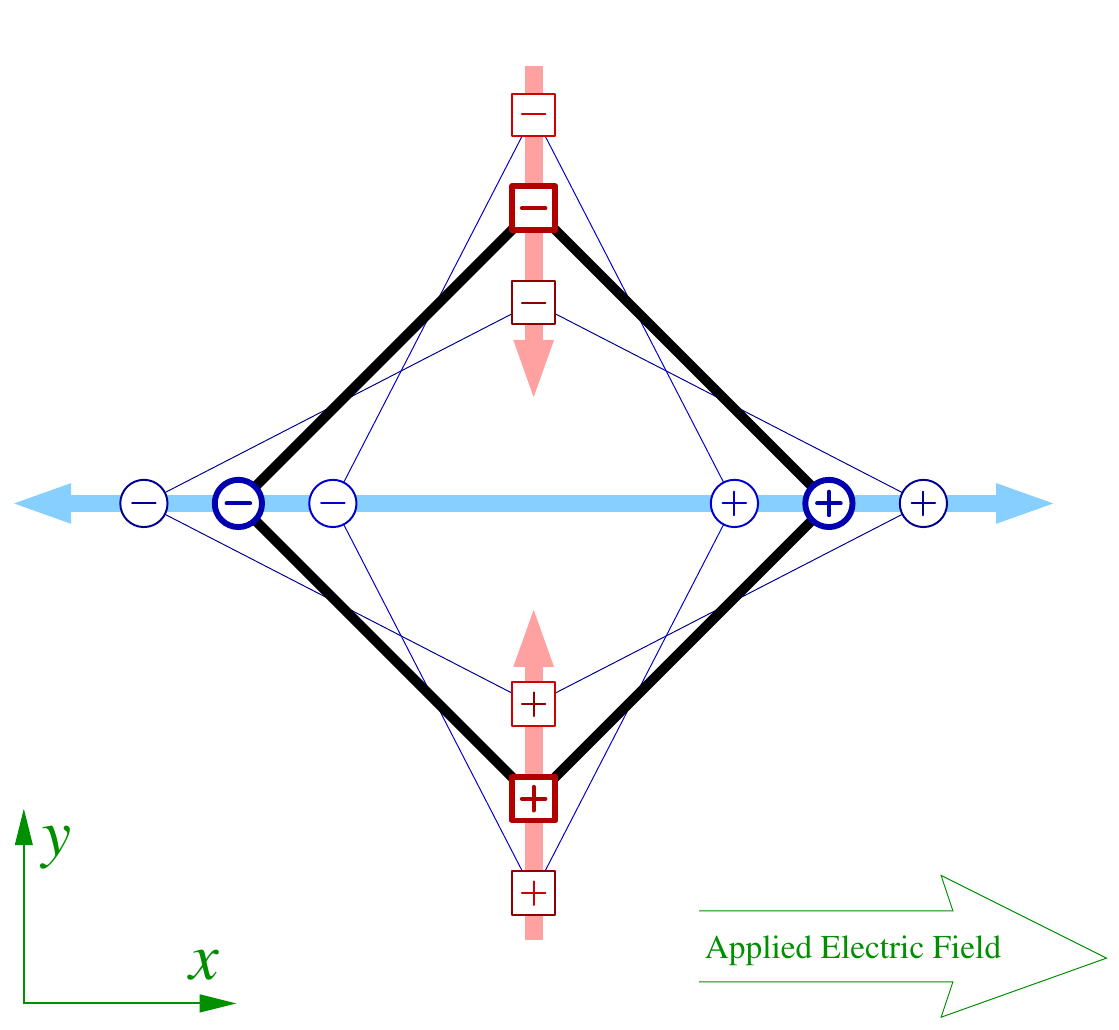}
\caption{Diagram
 of a current-generating {\pAxionResponse} $\axivaz \AxisZ$,
 based on a diamond shape that stretches ({or} shrinks) in $x$
 whilst simultaneously contracting (or expanding) in $y$.
When an oscillating electric field ${\pE}_x$ is applied,
 the ``detector'' charges (blue circles) are
 pulled apart (or pushed together) horizontally,
 so that the ``response'' charges (red squares)
 are moved together (apart) vertically.
The $y$-direction current produced by those moving response charges
 is the {\pAxionResponse}.
Not shown are the mechanisms
 that return the system to a default shape
 when the applied field is removed;
 for a construction with sufficiently flexible corners,
 this could be provided by the elastic properties of the material.
}
\label{fig_example-metamech}
\end{figure}

\def\pRforce{K}

For a minimal material providing such properties,
 we assume a model response dynamics,
 where $x$ is the charge displacement,
 $v$ is the speed of motion of any corner,
 $\pRforce$ the restoring force constant,
 and $m$ the effective mass of the structure,
 of
~
\begin{align}
  \dTime x
&=
  v
\label{eqn-example-metamech-velocity}
\\
  \dTime v
&=
 -
  \pRforce x
 -
  \gamma v
 -
  \frac{2q_b}{m} {\pE}_x
\label{eqn-example-metamech-force}
\\
\textrm{i.e.}\qquad
  \dTime^2 v
&=
 -
  \pRforce v
 -
  \gamma \dTime v
 -
  \frac{2q_b}{m} \dTime {\pE}_x
\\
\textrm{or}\qquad
  \dTime^2 {\pJ}_y
&=
 -
  \pRforce {\pJ}_y
 -
  2 q_r \gamma \dTime v
 -
  \frac{4 q_r q_b}{m} \dTime {\pE}_x
%\\
%\textrm{or}\qquad
%  \dTime^2 {\pJ}_y
%&=
% -
%  \pRforce {\pJ}_y
% -
%  2 q_r \gamma \dTime v
% -
%  \gamma {\axivaz} \times \dTime \VE
,
\end{align}
 since ${\pJ}_y = 2 q_r v$.
Here,
 $\gamma$ represents frictional losses in the mechanical oscillator.

In the quasistatic case with ${\pE}_x$ oscillating at a frequency $\omega_0$,
 and with negligible losses,
 we have 
~
\begin{align}
  \AxisY
  {\pJ}_y
&=
  \AxisZ
  \left[
    \frac{\omega_0}{\pRforce + \imath \gamma \omega_0 - \omega_0^2}
    \frac{4 q_r q_b}{m}
  \right]
 \times
  \AxisX
  {\pE}_x
,
\end{align}
 which means that
~
\begin{align}
  {\axivaz}
&=
  \frac{\omega_0}{\pRforce + \imath \gamma \omega_0 - \omega_0^2}
  \frac{4 q_r q_b}{m}
.
\label{eqn-example-metamech-dispersion}
\end{align}
\ResubOne{Note that
 \eqref{eqn-example-metamech-dispersion} is frequency dependent,
 and is the dispersion relation for the axionic response. 
As the response is derived from the causal dynamical model
 of \eqref{eqn-example-metamech-velocity}, 
 \eqref{eqn-example-metamech-force}
 it automatically satifies
 the Kramers-Kronig relations \cite{Kinsler-2011ejp};
 and in some suitably narrowband limit
 could be approximated as being dispersionless.}

Note that at the expense of additional complication,
 the unwanted current due to the detector charges
 could be (partly) cancelled
 by stacking the element of figure \ref{fig_example-metamech}
 with the complementary ``auxetic'' 
 one of figure \ref{fig_example-metamech-b}.
In the auxetic \cite{Yu-LJW-2018pms} cell,
 the shape expands simultaneously in $x$ and $y$;
 as a result,
 with (blue) detector charges of reversed sign (as shown),
 you can cancel the side-effect detector-charge
 currents with each other.

\begin{figure}
\includegraphics[angle=-0,width=0.95\columnwidth]{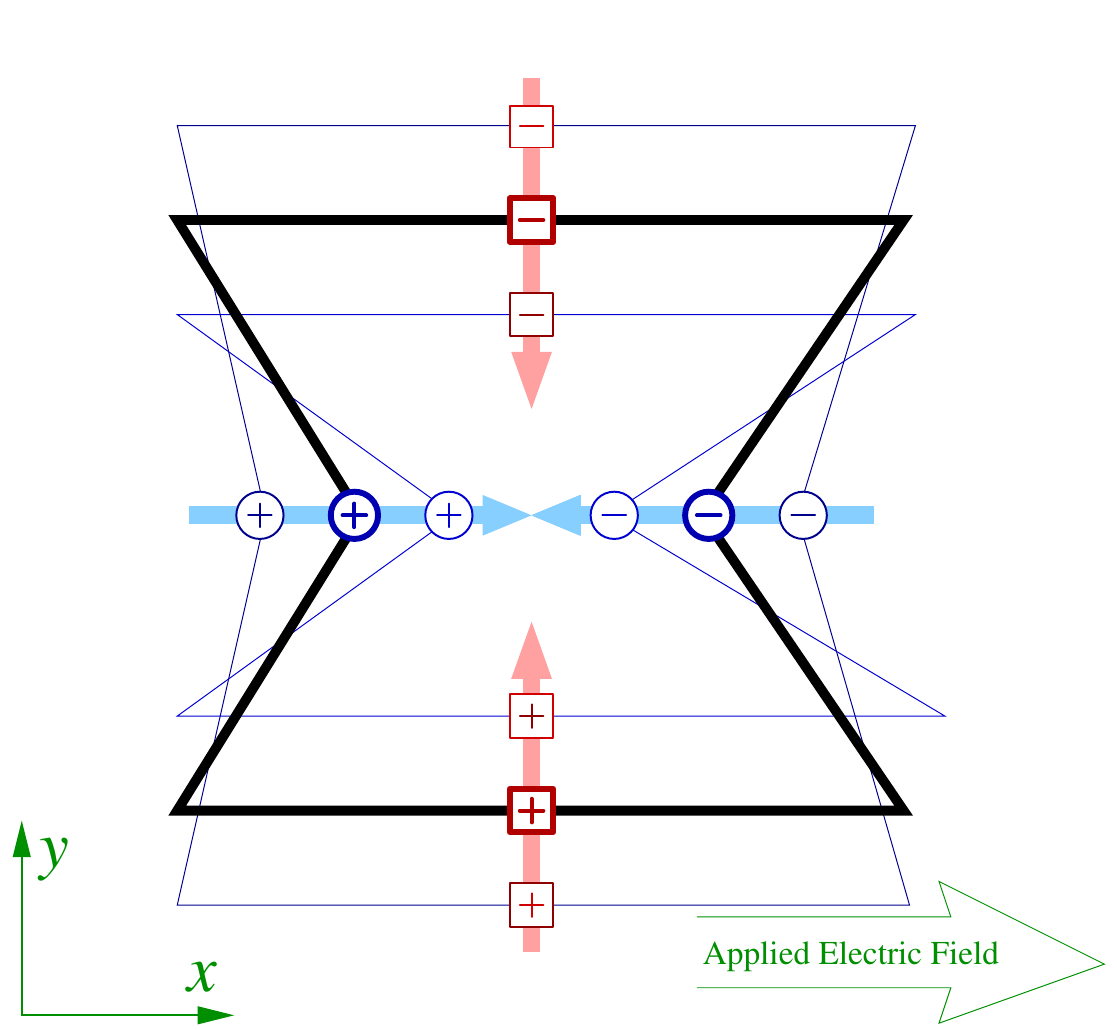}
\caption{Diagram
 of an auxetic current-generating {\pAxionResponse} $\axivaz \AxisZ$,
 based on a shape whose core stretches (or shrinks) in $x$
 whilst simultaneously expanding (or contracting) in $y$.
When an oscillating electric field ${\pE}_x$ is applied,
 the ``detector'' charges (blue circles) are
 pushed together (or pulled apart) horizontally,
 so that the ``response'' charges (red squares)
 are moved together (or apart)  vertically.
The $y$-direction current produced by those moving response charges
 is the {\pAxionResponse};
 notice that this response current
 is the same as in figure \ref{fig_example-metamech},
 whilst the current due to the detector charges is opposed.
Not shown are the mechanisms
 that return the system to a default shape
 when the applied field is removed;
 for a construction with sufficiently flexible corners,
 this could be provided by the elastic properties of the material.
}
\label{fig_example-metamech-b}
\end{figure}

By placing elements of this kind radially around a conducting cylinder,
 with $\AxisX$ replaced with the radial direction $\AxisR$,
 and $\AxisY$ replaced with the angular direction $\AxisTheta$,
 we could envisage constructing a dynamical counterpart
 to an active,
 driven scheme such as that
 shown in figure \ref{fig_example-cylinder};
 this is depicted in figure \ref{fig-metamech-radial}.

\begin{figure}
\centering
%\resizebox{0.80\columnwidth}{!}{
% \input{fig09-metamech-radial.tikz}
%}
\includegraphics[width=0.80\columnwidth]{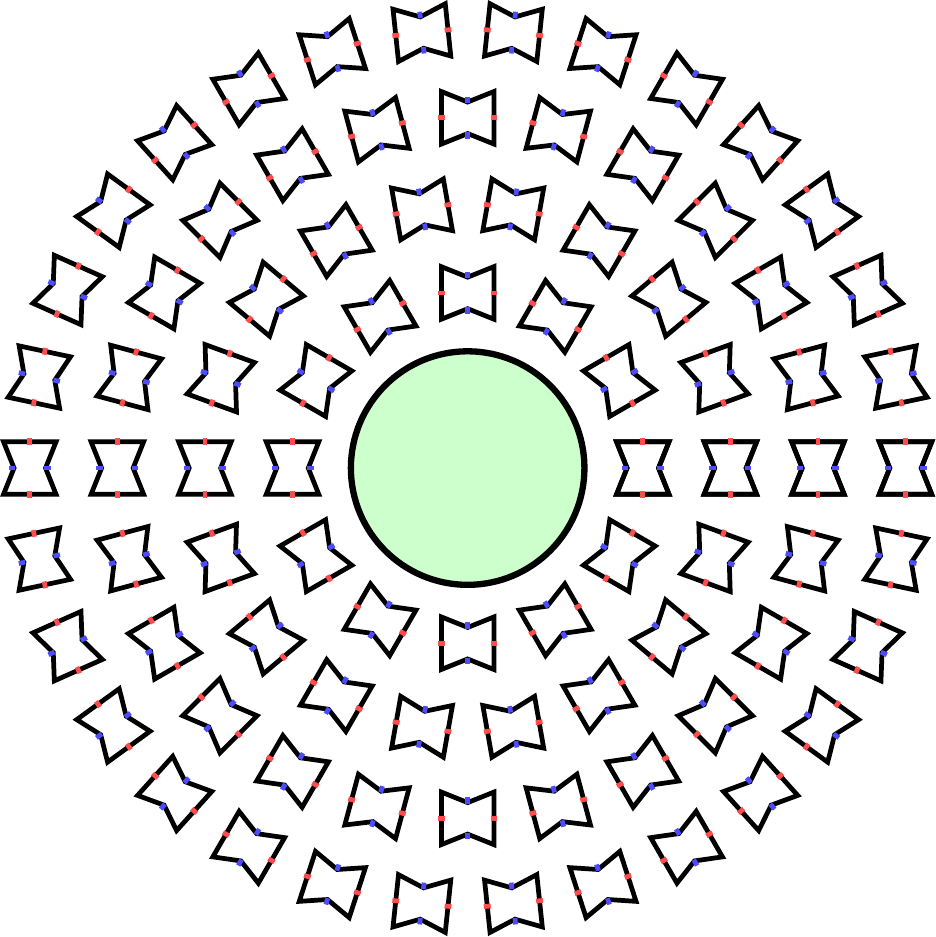}
\caption{Metamaterial {\pAxionResponse} cells
 arranged radially about a conducting cylinder,
 to provide an axial {\pAxionResponse} $\axivaz \AxisZ$.
This is
 rather like the concept of figure \ref{fig_example-cylinder},
 except that to achieve the $\axivaa \AxisTheta$ in that case
 the cells would need to be rotated about the radial axis
 and re-oriented into the $\AxisR \AxisZ$ plane.
The auxetic cells of figure \ref{fig_example-metamech-b} are used
 so that their proper orientation is clearer.
}
\label{fig-metamech-radial}
\end{figure}

%
% ----------------------------------------------------------------------
\subsection{Transformation Electromagnetics}\label{S-other-TO}

There is much
 current interest in the concepts and applications
 of transformation optics in both space and spacetime
 \cite{Kinsler-M-2014adp-scast,Baev-PASW-2015pr,Kinsler-M-2015pnfa-tofu,McCall-etal-2018jo-roadmapto},
 most notably involving various sorts of spatial and spacetime cloaking
 in free space
 \cite{Dolin-1961ivuzr,Pendry-SS-2006sci,McCall-FKB-2011jo,Gratus-KMT-2016njp-stdisp},
 against or on surfaces
 \cite{Li-P-2008prl,Kinsler-M-2014pra,MitchellThomas-MQHH-2013prl},
 and at distance \cite{Lai-CZC-2009prl}.

It is worth investigating how such techniques
 are implemented under our new CMCR scheme.
In transformation optics
 it is the optical metric where the cloaking or other transformation design
 is implemented;
 and it is important to note that that optical metric
 is distinct from the background physical (spacetime) metric
 \cite{Kinsler-M-2015raytail,Fathi-T-2016prd,Thompson-F-2015pra}.

For the constitutive model presented here,
 as detailed in our CMCR,
 \eqref{GH_Max_NoMono},
 \eqref{GH_Max_Faraday},
 \eqref{M_New_Max_CR_rho},
 \eqref{M_New_Max_CR_J},
 we see that the metric
 (whether of the background spacetime or optical signals)
 does not explicitly appear.
Instead its components and its derivatives
 are encoded within the components of the $\PsiAny$'s.

Since the operators {\xquad{\PsiErho}{\PsiBrho}{\PsiEJ}{\PsiBJ}}
 contain derivatives,
 one idea to perform a transformation optics
 is to promote the partial derivatives $\partial_i$
 in \eqref{M_Phi_components} to covariant derivatives $\nabla_i$.
However this will not work
 because the operators {\xquad{\PsiErho}{\PsiBrho}{\PsiEJ}{\PsiBJ}}
 are \emph{not} tensors.
Instead one should use the change of coordinates formula
 \eqref{GR_Change_coords_abc},
 \eqref{GR_Change_coords_abcd}
 given in appendix \ref{ch_GR}.
For example, 
 from this one can extract
 the components $(\PsiErho)^i$
 using \eqref{GR_comp_extract}.  
In \eqref{GR_Change_coords_abcd} we see that the non-axionic terms,
 i.e. those with two indices 
 such as $(\PsiErho)^{0j}$, 
 $(\PsiErho)^{ij}$, 
 transform like tensor densities,
 albeit ones that can mix the non-axionic terms
 when performing a spacetime transformation. 
By contrast,
 \eqref{GR_Change_coords_abc} says that
 the transformation of the axionic terms has two contributions:
 one is the standard tensor transformation,
 and the other introduces the non-axionic terms
 multiplied by the derivative of the Jacobian matrix. 
This means that a medium which does not have an axionic response
 can be transformed into one which does. 
This is analogous to the way the Christoffel symbols,
 which in Minkowski spacetime with Cartesian coordinates are zero,
 can be transformed into non-zero values.

%
% ----------------------------------------------------------------------
\subsection{Ohmic Resistance}\label{S-other-ohmic}

Ohm's law $\VJ=\sigma \VE$
 relates the current $\VJ$ through a conductor
 to the electric field $\VE$ experienced by that conductor, 
 with the proportionality between them 
 being due to the conductivity $\sigma$.
\ResubOne{Here we are in particular looking for
 a constitutive relation giving us $\eCharge$
 in terms of the electric field $\VE$. 
Since --
 by conservation of charge --
 the charge at a point can be given by the time integral of the current,
 such a constitutive relation is obtained
 by inserting Ohm's Law into this integral\footnote{This
    observation is often ignored, 
   either by assuming that $\nabla\cdot(\sigma \VE)=0$ so we can set $\rho=0$,
   or by working in frequency space so that \eqref{other_Ohm_charge}
   becomes
   $\eCharge = \left( \imath \omega \right)^{-1} \nabla \cdot (\sigma \VE) $.}.}
Thus we have 
%[
\begin{align}
  \eCharge(t)
&=
 -
  \int_{t_0}^{t} \nabla\cdot \left( \sigma \VE \right) \ dt'
 +
  {\eCharge}(t_0)
\label{other_Ohm_charge}
\end{align}
%]
Since the CMCR equations \eqref{M_New_Max_CR_rho} and \eqref{M_New_Max_CR_J}
 are a more general relationship
 between \xpair{\VE}{\VB} and \xpair{\eCharge}{\VJ}
 it is natural to ask if these avoid the problem
 of integral constitutive relations.

Consider a simple non-axionic isotropic homogeneous static medium with
 $\PsiEJ\VAct{\VE} = \EMepsilon \dtime{\VE} - \sigma \VE$ and
 $\PsiBJ\VAct{\VB} = \EMmu^{-1} \nabla \times \VB$ so that
 \eqref{M_New_Max_CR_J} becomes
%[
\begin{align}
  \EMmu^{-1}
 \nabla
 \times
  \VB
 -
  \EMepsilon \dtime{\VE}
&=
  \sigma
  \VE
 +
   {\eCurrent}
\label{M_ohmic_current}
\end{align}
%]
where $\EMepsilon$, $\EMmu$ and $\sigma$ are constants
 and $\VJ$ is the external current.
In order to be consistent with \eqref{EMCR_conservation},
 the other CMCR is given by
%[
\begin{align}
  \EMepsilon \nabla \cdot \VE
-
  \sigma
  \int \left( \nabla \cdot \VE \right) dt
&=
{\eCharge}\,.
\label{M_ohmic_charge}
\end{align}
%]
Since this expression contains an integral
 clearly there do not exist any $\PsiErho$ and $\PsiBrho$ such that
 \eqref{M_New_Max_CR_rho} becomes \eqref{M_ohmic_charge}.

However,
 we could decide to further extend the constitutive relations,
 so that the CMCR are also allowed to include
 second derivatives of {\xpair{\VE}{\VB}}
 as well as first derivatives of $\eCharge$.
In such a case,
 differentiating \eqref{M_ohmic_charge} with respect to time gives
%[
\begin{align}
  \EMepsilon
  \nabla
  \cdot
   \dtime{\VE}
 -
  \sigma
  \nabla
  \cdot
  \VE
&=
  {\dtime{\eCharge}}
.
\label{M_ohmic_charge_diff}
\end{align}
%]
This means that
 such a ``second order'' extension to the CMCR
 would allow us to include a conductivity model of Ohmic current
 in the constitutive relations,
 just the standard constitutive relations can be extended
 by the addition of the conductivity parameter $\sigma$.
In any such extension, 
 there would be very many more allowed constitutive parameters
 than are considered here in our first order CMCR model.

%
% ----------------------------------------------------------------------
\subsection{Poynting Vector}
\label{ch_AxionPoynting}

It is instructive to consider how these new {\pAxionResponse}s
 appear in the derivation
 of the electromagnetic {energy flux} equation,
 i.e. as applied to the Poynting vector \cite{RMC,Kinsler-FM-2009ejp}.
We have,
 using a standard vector identity
 and \eqref{M_New_Max_axion_current},
%\begin{align*}
%  \nabla \cdot \left( \VE \times \VB \right) &=
% -
%  \VE \cdot \left( \nabla \times \VB \right)  +
%  \VB \cdot \left( \nabla \times \VE \right)
%.
%\end{align*}
~
\begin{align*}
  \nabla \cdot &\left( \VE \times \VB \right) \\
&=
 -
  \VE \cdot \left( \nabla \times \VB \right)  +
  \VB \cdot \left( \nabla \times \VE \right)
\\
  &=
 -
  \VE
  \cdot
  \EMmu
  \left(
    \EMepsilon \dtime{\VE}
   -\axivav \times \VE
   -\axivat \VB
   +\VJ
  \right)
 {-}
  \VB
  \cdot
  \dtime{\VB}
\\
&=
 -
  {\tfrac12}\dTime
  \left(
    \EMepsilon \EMmu
    \VE \cdot \VE
   +
    \VB \cdot \VB
  \right)
 +
  \EMmu
  \VE
  \cdot
  \left( \axivav \times \VE \right)
\\&\qquad\qquad
 +
  \EMmu
  \axivat \VE \cdot \VB
 -
  \EMmu
  \VE \cdot \VJ
.
\end{align*}
%]
Hence
%[
\begin{equation}
\begin{aligned}
  &\dTime
  \left(
    {\tfrac12}\EMepsilon \VE \cdot \VE
   +
    {\tfrac12}\EMmu^{-1} \VB \cdot \VB
  \right)
+
\nabla \cdot \left( \VE \times \EMmu^{-1}\VB \right)
\\&\qquad =
 -
  \VE \cdot \VJ
 +
  \axivat \VE \cdot \VB
,
\end{aligned}
\label{AxionPoynting_Value}
\end{equation}
%]
where
 the only unconventional effect arises from the $\axivat$ term.

The conservation of momentum may also be derived as
%[
\begin{align}
  \partial_t \Vp
&=
  \nabla \cdot \TenStress
 -
  \left( \VE \cdot \VB \right) \, \Vzeta
 +
  \rho \, \VE
 +
  \VJ \times \VB
,
\label{Stress_conserv_law}
\end{align}
%]
where the (Minkowski) electromagnetic momentum is
%[
\begin{align}
  \Vp =\EMepsilon \, \VE \times \VB
,
  \label{Stress_def_p}
\end{align}
%]
and the electromagnetic stress tensor is
%[
\begin{align}
  \left( \TenStress \right)_{ij}
&=
  \EMepsilon
  \VE_i\VE_j
 +
  \EMmu^{-1}
  \VB_i
  \VB_j
\nonumber
\\
&\qquad\quad
 -
  \tfrac12
  \left(
    \EMepsilon \VE \cdot \VE
   +\EMmu^{-1} \VB \cdot \VB
  \right)
\delta_{ij}
.
\label{Stress_def_Stress}
\end{align}
%]
The divergence of the stress tensor turns out to be
~
\begin{align}
  \left(
    \nabla \cdot \TenStress
  \right)_j
&=
  \partial^i
  \left( \TenStress \right)_{ij}
\nonumber
\\
&=
  \partial_t
  \left( \EMepsilon \VB \times \VE \right)_j
 +
  \rho\VE_j
\nonumber
\\
&\qquad\quad
 +
  \left( \VB \times \VJ \right)_j
 +
  \Vzeta_j
  \left( \VB \cdot \VE \right)
.
\label{Stress_def_div_Stress}
\end{align}
%]
%
%\begin{align*}
%\partial^i
%\TenStress_{ij}
%&=
%\EMepsilon\nabla\cdot(\VE\VE_j)
%+
%\EMmu^{-1}\nabla\cdot(\VB\VB_j)
%-
%\tfrac12 \partial_j\big(\EMepsilon\VE\cdot\VE + \EMmu^{-1}\VB\cdot\VB\big)
%\\&=
%\EMepsilon(\nabla\cdot\VE)\VE_j
%+
%\EMmu^{-1}(\nabla\cdot\VB)\VB_j
%-
%\big(\EMepsilon\VE\times(\nabla\times\VE)\big)_j
%-
%\big(\EMmu^{-1}\VB\times(\nabla\times\VB)\big)_j
%\\&=
%(\rho+\zeta\cdot\VB)\VE_j
%+
%\epsilon  \big(\VE\times \partial_t \VB\big)_j
%-
%\big(\VB\times(\EMepsilon\partial_t\VE-\Vzeta\times\VE-\zeta_t\VB)+\VJ\big)_j
%\\&=
%\EMepsilon  \big(\VE\times \partial_t \VB\big)_j
%-
%\EMepsilon \big(\VB\times\partial_t \VE\big)_j
%+
%\rho\VE_j+(\zeta\cdot\VB)\VE_j
%+
%\big(\VB\times(\Vzeta\times\VE)\big)_j
%+
%(\VB\times\VJ)_j
%\\&=
%\partial_t \big(\EMepsilon\VB\times\VE\big)_j
%+
%\rho\VE_j+(\zeta\cdot\VB)\VE_j
%+
%\Vzeta_j (\VB\cdot\VE)
%-
%\VE_j (\VB\cdot\Vzeta)
%+
%(\VB\times\VJ)_j
%\\&=
%\partial_t \big(\EMepsilon\VB\times\VE\big)_j
%+
%\rho\VE_j
%+
%\Vzeta_j (\VB\cdot\VE)
%+
%(\VB\times\VJ)_j
%\end{align*}
%%]
%}
Here,
 as throughout this paper,
 we do not wish to consider the axionic terms \xpair{\Vzeta}{\axivat}
 as due to an axion particle field
 as in \eqref{IN_Axio_Lagrangian},
 but instead as background constitutive relations in the medium.
Setting the external
 current to zero we see from \eqref{AxionPoynting_Value}
 that energy is conserved only if $\axivat=0$.
Similarly,
 a component of the $\Vp_i$ momentum is conserved only
 if the corresponding component of $\Vzeta_i$ is zero.
These observations are actually a consequence
 of existing work
 \cite{Gratus-OT-2012ap,Dereli-GT-2007jpa},
 where it was shown that in the presence of a static background field,
 there is a conserved Noether current
 where there is a Killing symmetry that is simultaneously
 a symmetry of the metric
 and
 of the background field as it appears in the Lagrangian.
In our case the background field appears as\footnote{Note
  that the use of $\axipot$ in the Lagrangian does not imply
  the $\axipot$ is a physical field,
  since it is merely a potential for \xpair{\Vzeta}{\axivat}
  and Lagrangians often contain potentials.
 For example,
   the Lagrangian for the electromagnetic field
  is written in terms of the electromagnetic potential.}
%[
\begin{align}
\axipot \, \VE \cdot \VB
.
\label{Stress_Lag}
\end{align}
%]
Thus time is a Killing symmetry of the Minkowski metric
 and a Lie symmetry of the axion current if
 $\axivat=\partial_t \axipot=0$, 
 in this case the corresponding Noether current is energy.
The results of \cite{Gratus-OT-2012ap}
 can also be used to predict the conservation of angular momentum
 if $\Vzeta = \Vzeta_r(r) \, \AxisR$
 in the \xtrio{r}{\theta}{\phi} coordinate system,
 and of angular momentum about the $z$-axis if
 $\Vzeta = \Vzeta_r(r,z) \AxisR + \Vzeta_z(r,z) \AxisZ$
 in the cylindrical \xtrio{r}{\theta}{z} case.
The additional axionic $\axivat \VE \cdot \VB$
 contribution to this EM {energy} flux equation acts comparably to
 the standard energy storage term
 (cf. \eqref{AxionPoynting_Value})
 or the work-done-on-charges term $\VE \cdot \VJ$.
Depending on the local fields,
 it acts as a source or sink for EM energy.
In the usual case of an interaction with an
 axion particle field,
 this would be interpreted as a transfer of EM energy to or from
 those axions,
 where $\axivat$ can be identified with $\dTime \axiphi$.
Here,
 it refers to energy transfer to or from the constitutive mechanism
 causing the {\pAxionResponse}.

%
% ======================================================================
\section{Conclusion}\label{ch_Conclusion}

We have presented a minimal extension to
 the standard constitutive relations for Maxwell's equations,
 and have achieved this by
 combining the Maxwell-Amp{\`e}re equation with
 the constitutive properties of an electromagnetic medium.
This 
 means that we eliminate the need for the excitation fields
 {\xpair{\VD}{\VH}},
 and can permit a wider range of responses
 than the standard constitutive tensor approach.
Although constraints mean that most likely only
 4 new axionic (``axion-like'')
 parameters are permitted,
 these nevertheless represent media impossible to treat
 in the traditional approach,
 even if we allow inhomogeneity or dispersion.
As such our new CMCR can be made to look like standard EMCR,
 but with the addition of extra
 {\pAxionResponse}s.

Notably,
 there are two cases when we \emph{cannot} express
 our new {\pAxionResponse}s as an extension of standard EMCR,
 these being if we require the constitutive properties to be homogeneous,
 or in the case of non-trivial topology.
In particular,
 homogeneous blocks of axionic materials represented
 using our CMCR
 appear as \emph{inhomogeneous} materials if the representation
 is converted over to the standard EMCR.
This is why the CMCR can represent axionic effects without relying --
 as is usually expected --
 on boundary effects.

This means that there are cases where
 it is advantageous --
 or even required --
 to write the {\pAxionResponse} in our way.
These results are linked to related discoveries
 \cite{Gratus-KM-2019foop-nocharge}
 enabled by removing {\xpair{\VD}{\VH}},
 such as non-conservation of charge in topologically non-trivial spaces,
 and the treatment of charges passing through wormholes.
We also remarked on the possibility of there being
 an additional 16 more exotic constitutive parameters,
 for a grand total of 55 in all.
If these exist,
 we either require additional new types of field and charge,
 or need to reconsider what we aim to represent
 by the excitation fields {\xpair{\VD}{\VH}}.

Another direction is to enumerate the plane eigenmodes
 of uniform media associated with our theory.
Solving the eigen-problem of a conventional biaxial medium
 leads to the familiar self-intersecting Fresnel phase surface
 carrying four singular points
 \cite{BornWolf-Principles}. 
Recent generalisation from biaxial media
 to media with independent dielectric,
 magnetic,
 and magneto-electric tensors
 (a total of 20 material parameters)
 has been shown to give rise to a much richer Kummer surface,
 containing up to sixteen singular directions
 \cite{Favaro-H-2014arxiv}. 
With our proposed media containing a total of 55 parameters,
 even more contorted and exotic Fresnel phase surfaces
 are to be anticipated.

In future work we also aim to go beyond the minimal extension
 and theoretical discussion presented here.
Our examples suggest
 a range of opportunities for further work
 based on these results
 in topological and periodic systems,
 in attempting a metamaterial implementation of the {\pAxionResponse},
 or in extending the treatment to second order to treat
 Ohmic effects.

As a final remark,
 we wish to emphasise that this paper has only been written
 in standard vector calculus notation
 so as to make it more widely accessible.
In fact,
 the CMCR minimal extension presented here
 can be written --
 and was originally written --
 much more succinctly in coordinate
 free notation using exterior forms.
This is briefly presented in appendix \ref{ch_Coordfree},
 where the equations do not include an explicit metric,
 and are therefore also consistent with a
 pre-metric formulation of electromagnetism \cite{HehlObukhov}.

%
% ======================================================================
\acknowledgments

{Both JG and PK are grateful for the support provided
by STFC (the Cockcroft Institute ST/G008248/1 and ST/P002056/1) and
EPSRC (the Alpha-X project EP/N028694/1). PK would like to acknowledge
the hospitality of Imperial College London.}

%
% ======================================================================
%\bibliographystyle{unsrt}
%\bibliography{bibtex}
%\bibliography{Gratus-MK-2019arxiv-area51}

\begin{thebibliography}{68}
\expandafter\ifx\csname natexlab\endcsname\relax\def\natexlab#1{#1}\fi
\expandafter\ifx\csname bibnamefont\endcsname\relax
  \def\bibnamefont#1{#1}\fi
\expandafter\ifx\csname bibfnamefont\endcsname\relax
  \def\bibfnamefont#1{#1}\fi
\expandafter\ifx\csname citenamefont\endcsname\relax
  \def\citenamefont#1{#1}\fi
\expandafter\ifx\csname url\endcsname\relax
  \def\url#1{\texttt{#1}}\fi
\expandafter\ifx\csname urlprefix\endcsname\relax\def\urlprefix{URL }\fi
\providecommand{\bibinfo}[2]{#2}
\providecommand{\eprint}[2][]{\url{#2}}

\bibitem[{\citenamefont{Jackson}(1999)}]{Jackson-ClassicalED}
\bibinfo{author}{\bibfnamefont{J.~D.} \bibnamefont{Jackson}},
  \\ \emph{\bibinfo{title}{Classical Electrodynamics}}
  \\ (\bibinfo{publisher}{Wiley}, \bibinfo{year}{1999}),
   \bibinfo{edition}{3rd} ed.,
  \\ ISBN \bibinfo{isbn}{978-0-471-30932-1}.

\bibitem[{\citenamefont{Reitz et~al.}(1980)\citenamefont{Reitz, Milford, and
  Christy}}]{RMC}
\bibinfo{author}{\bibfnamefont{J.~R.} \bibnamefont{Reitz}},
  \bibinfo{author}{\bibfnamefont{F.~J.} \bibnamefont{Milford}},
  \bibnamefont{and} \bibinfo{author}{\bibfnamefont{R.~W.}
  \bibnamefont{Christy}}, 
  \\ \emph{\bibinfo{title}{Foundations of electromagnetic
  theory}} 
  \\ (\bibinfo{publisher}{Addison-Wesley}, \bibinfo{year}{1980}),
  \bibinfo{edition}{3rd} ed.,
 \\ ISBN \bibinfo{isbn}{0-201-06332-8}.

\bibitem[{\citenamefont{Gratus et~al.}(2019{\natexlab{a}})\citenamefont{Gratus,
  Kinsler, and McCall}}]{Gratus-KM-2019ejp-dhfield}
\bibinfo{author}{\bibfnamefont{J.}~\bibnamefont{Gratus}},
  \bibinfo{author}{\bibfnamefont{P.}~\bibnamefont{Kinsler}}, \bibnamefont{and}
  \bibinfo{author}{\bibfnamefont{M.~W.} \bibnamefont{McCall}},
  \\ \bibinfo{journal}{Eur. J. Phys.} \textbf{\bibinfo{volume}{40}},
  \bibinfo{pages}{025203} (\bibinfo{year}{2019}{\natexlab{a}}),
  \\ \XARXIV{1903.01957},
  \\ \XDOI{10.1088/1361-6404/ab009c}.

\bibitem[{\citenamefont{Gratus et~al.}(2019{\natexlab{b}})\citenamefont{Gratus,
  Kinsler, and McCall}}]{Gratus-KM-2019foop-nocharge}
\bibinfo{author}{\bibfnamefont{J.}~\bibnamefont{Gratus}},
  \bibinfo{author}{\bibfnamefont{P.}~\bibnamefont{Kinsler}}, \bibnamefont{and}
  \bibinfo{author}{\bibfnamefont{M.~W.} \bibnamefont{McCall}},
  \\ \bibinfo{journal}{Found. Phys.} \textbf{\bibinfo{volume}{49}},
  \bibinfo{pages}{330} (\bibinfo{year}{2019}{\natexlab{b}}),
  \\ \XARXIV{1904.04103},
  \\ \XDOI{10.1007/s10701-019-00251-5}.

\bibitem[{\citenamefont{Tobar et~al.}(2019)\citenamefont{Tobar, McAllister, and
  Goryachev}}]{Tobar-MG-2019pdu}
\bibinfo{author}{\bibfnamefont{M.~E.} \bibnamefont{Tobar}},
  \bibinfo{author}{\bibfnamefont{B.~T.} \bibnamefont{McAllister}},
  \bibnamefont{and}
  \bibinfo{author}{\bibfnamefont{M.}~\bibnamefont{Goryachev}},
  \\ \bibinfo{journal}{Physics of the Dark Universe}
  \textbf{\bibinfo{volume}{26}}, \bibinfo{pages}{100339}
  (\bibinfo{year}{2019}), \\ \XARXIV{1809.01654},
  \\ \XDOI{10.1016/j.dark.2019.100339}.

\bibitem[{\citenamefont{Visinelli}(2013)}]{Visinelli-2013mpla}
\bibinfo{author}{\bibfnamefont{L.}~\bibnamefont{Visinelli}},
  \\ \bibinfo{journal}{Mod. Phys. Lett. A} \textbf{\bibinfo{volume}{28}},
  \bibinfo{pages}{1350162} (\bibinfo{year}{2013}), \\ \XARXIV{1401.0709},
  \\ \XDOI{10.1142/S0217732313501629}.

\bibitem[{\citenamefont{Hehl and Obukhov}(2003)}]{HehlObukhov}
\bibinfo{author}{\bibfnamefont{F.~W.} \bibnamefont{Hehl}} \bibnamefont{and}
  \bibinfo{author}{\bibfnamefont{Y.~N.} \bibnamefont{Obukhov}},
  \\ \emph{\bibinfo{title}{Foundations of Classical Electrodynamics: Charge, Flux,
  and Metric}},% vol.~\bibinfo{volume}{33} 
 \\ (\bibinfo{publisher}{Birkh{\"a}user},
  \bibinfo{address}{Boston}, \bibinfo{year}{2003}),\\ ISBN
  \bibinfo{isbn}{0-8176-4222-6}.
%, \\ \bibinfo{note}{cf
%  http://arxiv.org/abs/physics/0005084}.

\bibitem[{\citenamefont{Post}(1997)}]{Post-FSEM}
\bibinfo{author}{\bibfnamefont{E.~J.} \bibnamefont{Post}},
  \emph{\\ \bibinfo{title}{Formal Structure of Electromagnetics}}
  \\ (\bibinfo{publisher}{Dover Publications}, \bibinfo{address}{Mineola, N.Y.},
  \bibinfo{year}{1997}).

\bibitem[{\citenamefont{Lakhtakia and Weiglhofer}(1996)}]{Lakhtakia-W-1996pla}
\bibinfo{author}{\bibfnamefont{A.}~\bibnamefont{Lakhtakia}} \bibnamefont{and}
  \bibinfo{author}{\bibfnamefont{W.~S.} \bibnamefont{Weiglhofer}},
  \\ \bibinfo{journal}{Phys. Lett. A} \textbf{\bibinfo{volume}{213}},
  \bibinfo{pages}{107} (\bibinfo{year}{1996}),
  \\ \XDOI{10.1016/0375-9601(96)00121-1}.

\bibitem[{\citenamefont{Lakhtakia and Mackay}(2015)}]{Lakhtakia-M-2015spie}
\bibinfo{author}{\bibfnamefont{A.}~\bibnamefont{Lakhtakia}} \bibnamefont{and}
  \bibinfo{author}{\bibfnamefont{T.~G.} \bibnamefont{Mackay}},
  \\ \bibinfo{journal}{Proceedings of SPIE} \textbf{\bibinfo{volume}{9558}},
  \bibinfo{pages}{95580C} (\bibinfo{year}{2015}), \\ \XARXIV{1506.00123},
  \\ \XDOI{10.1117/12.2190105}.

\bibitem[{\citenamefont{Lakhtakia and Mackay}(2016)}]{Lakhtakia-M-2016jn}
\bibinfo{author}{\bibfnamefont{A.}~\bibnamefont{Lakhtakia}} \bibnamefont{and}
  \bibinfo{author}{\bibfnamefont{T.~G.} \bibnamefont{Mackay}},
  \\ \bibinfo{journal}{J. Nanophotonics} \textbf{\bibinfo{volume}{10}},
  \bibinfo{pages}{033004} (\bibinfo{year}{2016}), \\ \XARXIV{1601.02525},
  \\ \XDOI{10.1117/1.JNP.10.033004}.

\bibitem[{\citenamefont{Obukhov and Hehl}(2005)}]{Obukhov-H-2005pla}
\bibinfo{author}{\bibfnamefont{Y.~N.} \bibnamefont{Obukhov}} \bibnamefont{and}
  \bibinfo{author}{\bibfnamefont{F.~W.} \bibnamefont{Hehl}},
  \\ \bibinfo{journal}{Phys. Lett. A} \textbf{\bibinfo{volume}{341}},
  \bibinfo{pages}{357} (\bibinfo{year}{2005}),
  \\ \XDOI{10.1016/j.physleta.2005.05.006}.

\bibitem[{\citenamefont{Lindell and Sihvola}(2013)}]{Lindell-S-2013ieeetap}
\bibinfo{author}{\bibfnamefont{I.}~\bibnamefont{Lindell}} \bibnamefont{and}
  \bibinfo{author}{\bibfnamefont{A.}~\bibnamefont{Sihvola}},
  \\ \bibinfo{journal}{IEEE Trans. Antennas and Propagation}
  \textbf{\bibinfo{volume}{61}}, \bibinfo{pages}{768 } (\bibinfo{year}{2013}),
  \\ \XDOI{10.1109/TAP.2012.2223445}.

\bibitem[{\citenamefont{Hehl et~al.}(2008)\citenamefont{Hehl, Obukhov, Rivera,
  and Schmid}}]{Hehl-ORS-2008pla-relativisitic}
\bibinfo{author}{\bibfnamefont{F.~W.} \bibnamefont{Hehl}},
  \bibinfo{author}{\bibfnamefont{Y.~N.} \bibnamefont{Obukhov}},
  \bibinfo{author}{\bibfnamefont{J.-P.} \bibnamefont{Rivera}},
  \bibnamefont{and} \bibinfo{author}{\bibfnamefont{H.}~\bibnamefont{Schmid}},
  \\ \bibinfo{journal}{Physics Lett. A} \textbf{\bibinfo{volume}{372}},
  \bibinfo{pages}{1141 } (\bibinfo{year}{2008}),
  \\ \XDOI{10.1016/j.physleta.2007.08.069}.

\bibitem[{\citenamefont{Li et~al.}(2010)\citenamefont{Li, Wang, Qi, and
  Zhang}}]{Li-WQZ-2010np-dynamical}
\bibinfo{author}{\bibfnamefont{R.}~\bibnamefont{Li}},
  \bibinfo{author}{\bibfnamefont{J.}~\bibnamefont{Wang}},
  \bibinfo{author}{\bibfnamefont{X.-L.} \bibnamefont{Qi}}, \bibnamefont{and}
  \bibinfo{author}{\bibfnamefont{S.-C.} \bibnamefont{Zhang}},
  \\ \bibinfo{journal}{Nat. Phys.} \textbf{\bibinfo{volume}{6}},
  \bibinfo{pages}{284} (\bibinfo{year}{2010}),
  \\ \XDOI{10.1038/NPHYS1534}.

\bibitem[{\citenamefont{Wu et~al.}(2016)\citenamefont{Wu, Salehi, Koirala,
  Moon, and Oh}}]{Wu-SKMO-2016s-quantised}
\bibinfo{author}{\bibfnamefont{L.}~\bibnamefont{Wu}},
  \bibinfo{author}{\bibfnamefont{M.}~\bibnamefont{Salehi}},
  \bibinfo{author}{\bibfnamefont{N.}~\bibnamefont{Koirala}},
  \bibinfo{author}{\bibfnamefont{J.}~\bibnamefont{Moon}}, \bibnamefont{and}
  \bibinfo{author}{\bibfnamefont{S.}~\bibnamefont{Oh}},
  \\ \bibinfo{journal}{Science} \textbf{\bibinfo{volume}{354}},
  \bibinfo{pages}{1124 } (\bibinfo{year}{2016}),
  \\ \XDOI{10.1126/science.aaf5541}.

\bibitem[{\citenamefont{Feng}(2010)}]{Feng-2010araa-dark}
\bibinfo{author}{\bibfnamefont{J.~L.} \bibnamefont{Feng}},
  \\ \bibinfo{journal}{Annu. Rev. Astron. Astrophys.} 
  \textbf{\bibinfo{volume}{48}},
  pp.
  \bibinfo{pages}{495--545} (\bibinfo{year}{2010}),
  \\ \XDOI{10.1146/annurev-astro-082708-101659}.

\bibitem[{\citenamefont{Heras and Baez}(2008)}]{Heras-B-2009ejp}
\bibinfo{author}{\bibfnamefont{J.~A.} \bibnamefont{Heras}} \bibnamefont{and}
  \bibinfo{author}{\bibfnamefont{G.}~\bibnamefont{Baez}},
  \\ \bibinfo{journal}{Eur. J. Phys.} \textbf{\bibinfo{volume}{30}},
  \bibinfo{pages}{23} (\bibinfo{year}{2008}), \\ \XARXIV{0901.0194},
  \\ \XDOI{10.1088/0143-0807/30/1/003}.

\bibitem[{\citenamefont{Gratus et~al.}(2015)\citenamefont{Gratus, Perlick, and
  Tucker}}]{Gratus-PT-2015jpa}
\bibinfo{author}{\bibfnamefont{J.}~\bibnamefont{Gratus}},
  \bibinfo{author}{\bibfnamefont{V.}~\bibnamefont{Perlick}}, \bibnamefont{and}
  \bibinfo{author}{\bibfnamefont{R.~W.} \bibnamefont{Tucker}},
  \\ \bibinfo{journal}{J. Phys. A} \textbf{\bibinfo{volume}{48}},
  \bibinfo{pages}{435401} (\bibinfo{year}{2015}),
  \\ \XDOI{10.1088/1751-8113/48/43/435401}.

\bibitem[{\citenamefont{Nye}(1985)}]{Nye-PROPX}
\bibinfo{author}{\bibfnamefont{J.~F.} \bibnamefont{Nye}},
  \\ \emph{ \bibinfo{title}{Physical Properties of Crystals: Their Representation
  by Tensors and Matrices}} 
  \\(\bibinfo{publisher}{Oxford University Press},
  \bibinfo{address}{Oxford, England}, \bibinfo{year}{1985}),\\ ISBN
  \bibinfo{isbn}{978-0-19-851165-6}.

\bibitem[{\citenamefont{Fiebig}(2005)}]{Fiebig-2005jpd}
\bibinfo{author}{\bibfnamefont{M.}~\bibnamefont{Fiebig}},
  \\ \bibinfo{journal}{Journal of Physics D: Applied Physics}
  \textbf{\bibinfo{volume}{38}}, \bibinfo{pages}{R123} (\bibinfo{year}{2005}),
  \\ \XDOI{10.1088/0022-3727/38/8/R01}.

\bibitem[{\citenamefont{Eerenstein et~al.}(2006)\citenamefont{Eerenstein,
  Mathur, and Scott}}]{Eerenstein-MS-2006n}
\bibinfo{author}{\bibfnamefont{W.}~\bibnamefont{Eerenstein}},
  \bibinfo{author}{\bibfnamefont{N.~D.} \bibnamefont{Mathur}},
  \bibnamefont{and} \bibinfo{author}{\bibfnamefont{J.~F.} \bibnamefont{Scott}},
  \\ \bibinfo{journal}{Nature} \textbf{\bibinfo{volume}{442}},
  \bibinfo{pages}{759} (\bibinfo{year}{2006}),
  \\ \XDOI{10.1038/nature05023}.

\bibitem[{\citenamefont{Hillion}(1993)}]{Hillion-1993pre}
\bibinfo{author}{\bibfnamefont{P.}~\bibnamefont{Hillion}},
  \\ \bibinfo{journal}{Phys. Rev. E} \textbf{\bibinfo{volume}{47}},
  \bibinfo{pages}{1365} (\bibinfo{year}{1993}),
  \\ \XDOI{10.1103/PhysRevE.47.1365}.

\bibitem[{\citenamefont{Wang et~al.}(2009)\citenamefont{Wang, Zhou, Koschny,
  Kafesaki, and Soukoulis}}]{Wang-ZKKS-2009joa}
\bibinfo{author}{\bibfnamefont{B.}~\bibnamefont{Wang}},
  \bibinfo{author}{\bibfnamefont{J.}~\bibnamefont{Zhou}},
  \bibinfo{author}{\bibfnamefont{T.}~\bibnamefont{Koschny}},
  \bibinfo{author}{\bibfnamefont{M.}~\bibnamefont{Kafesaki}}, %\bibnamefont{and}
  \bibinfo{author}{\bibfnamefont{C.~M.} \bibnamefont{Soukoulis}},
  \\ \bibinfo{journal}{J. Opt. A} \textbf{\bibinfo{volume}{11}},
  \bibinfo{pages}{114003} (\bibinfo{year}{2009}), 
%\\ \bibinfo{note}{conference:
%  1st International Workshop on Theoretical and Computational
%  Nano-Photonics;Bad Honnef, Germany, 3rd -- 5th December, 2008},
  \\ \XDOI{10.1088/1464-4258/11/11/114003}.

\bibitem[{\citenamefont{McCall}(2009)}]{McCall-2009joa}
\bibinfo{author}{\bibfnamefont{M.~W.} \bibnamefont{McCall}},
  \\ \bibinfo{journal}{J. Opt. A} \textbf{\bibinfo{volume}{11}},
  \bibinfo{pages}{074006} (\bibinfo{year}{2009}),
  \\ \XDOI{10.1088/1464-4258/11/7/074006}.

\bibitem[{\citenamefont{New}(2011)}]{New-INLO}
\bibinfo{author}{\bibfnamefont{G.~H.~C.} \bibnamefont{New}},
  \\ \emph{\bibinfo{title}{Introduction to nonlinear optics}}
  \\(\bibinfo{publisher}{Cambridge University Press},
  \bibinfo{address}{Cambridge}, \bibinfo{year}{2011}),\\ ISBN
  \bibinfo{isbn}{978-0-521-87701-5}.

\bibitem[{\citenamefont{Boyd}(2008)}]{Boyd-NLO}
\bibinfo{author}{\bibfnamefont{R.~W.} \bibnamefont{Boyd}},
  \\ \emph{\bibinfo{title}{Nonlinear Optics}} 
  \\ (\bibinfo{publisher}{Academic Press
  Inc.}, \bibinfo{address}{New York}, \bibinfo{year}{2008}),
  \bibinfo{edition}{3rd} ed.,\\ ISBN \bibinfo{isbn}{978-0-12-369470-6}.
%,
%  \\ \bibinfo{note}{1st ed. 1994, 2nd ed. 2003}.

\bibitem[{\citenamefont{Agranovich and Ginzburg}(1984)}]{AgranoGinsberg}
\bibinfo{author}{\bibfnamefont{V.~M.} \bibnamefont{Agranovich}}
  \bibnamefont{and} \bibinfo{author}{\bibfnamefont{V.}~\bibnamefont{Ginzburg}},
  \\ \emph{\bibinfo{title}{Crystal Optics with Spatial Dispersion, and Excitons}},
  \\ Springer Series in Solid-State Sciences
  \\ (\bibinfo{publisher}{Springer-Verlag}, \bibinfo{address}{Berlin Heidelberg},
  \bibinfo{year}{1984}),\\ ISBN \bibinfo{isbn}{978-3-662-02408-9},
  \\ \XDOI{10.1007/978-3-662-02406-5}.

\bibitem[{\citenamefont{Belov et~al.}(2002)\citenamefont{Belov, Tretyakov, and
  Viitanen}}]{Belov-TV-2002jewa}
\bibinfo{author}{\bibfnamefont{P.~A.} \bibnamefont{Belov}},
  \bibinfo{author}{\bibfnamefont{S.~A.} \bibnamefont{Tretyakov}},
  \bibnamefont{and} \bibinfo{author}{\bibfnamefont{A.~J.}
  \bibnamefont{Viitanen}}, \\ \bibinfo{journal}{J. Electromang. Waves Appl.}
  \textbf{\bibinfo{volume}{16}}, \bibinfo{pages}{1153} (\bibinfo{year}{2002}),
  \\ \XDOI{10.1163/156939302X00688}.

\bibitem[{\citenamefont{Ciraci et~al.}(2013)\citenamefont{Ciraci, Pendry, and
  Smith}}]{Ciraci-PS-2013cphc}
\bibinfo{author}{\bibfnamefont{C.}~\bibnamefont{Ciraci}},
  \bibinfo{author}{\bibfnamefont{J.~B.} \bibnamefont{Pendry}},
  \bibnamefont{and} \bibinfo{author}{\bibfnamefont{D.~R.} \bibnamefont{Smith}},
  \\ \bibinfo{journal}{ChemPhysChem} \textbf{\bibinfo{volume}{14}},
  \bibinfo{pages}{1109} (\bibinfo{year}{2013}),
  \\ \XDOI{10.1002/cphc.201200992}.

\bibitem[{\citenamefont{Kinsler}(2019)}]{Kinsler-2019arxiv-spatype}
\bibinfo{author}{\bibfnamefont{P.}~\bibnamefont{Kinsler}},
  \\\emph{\bibinfo{title}{A new introduction to spatial dispersion: reimagining the basic concepts}}\\
  %\\ \bibinfo{journal}{Arxiv}  (\bibinfo{year}{2019}),
  \bibinfo{journal}{Photon. Nanostructures Fundam. Appl.} \textbf{\bibinfo{volume}{43}},
    \bibinfo{pages}{100897}  (\bibinfo{year}{2021}),
    \\ \XDOI{10.1016/j.photonics.2021.100897}, 
    \\ \XARXIV{1904.11957}.
  %\\ \XWEB{https://arxiv.org/abs/1904.11957}.


\bibitem[{\citenamefont{Bohren}(2010)}]{Bohren-2010ejp}
\bibinfo{author}{\bibfnamefont{C.~F.} \bibnamefont{Bohren}},
  \\ \bibinfo{journal}{Eur. J. Phys.} \textbf{\bibinfo{volume}{31}},
  \bibinfo{pages}{573} (\bibinfo{year}{2010}),
  \\ \XDOI{10.1088/0143-0807/31/3/014}.

\bibitem[{\citenamefont{Kinsler}(2011)}]{Kinsler-2011ejp}
\bibinfo{author}{\bibfnamefont{P.}~\bibnamefont{Kinsler}},
  \\ \bibinfo{journal}{Eur. J. Phys.} \textbf{\bibinfo{volume}{32}},
  \bibinfo{pages}{1687} (\bibinfo{year}{2011}), 
  %\\ \bibinfo{note}{the arXiv version has additional appendices}, 
  \\ \XARXIV{1106.1792},
  \\ \XDOI{10.1088/0143-0807/32/6/022}.

\bibitem[{\citenamefont{Heras}(2011)}]{Heras-2011ajp}
\bibinfo{author}{\bibfnamefont{J.~A.} \bibnamefont{Heras}},
  \\ \bibinfo{journal}{Am. J. Phys.} \textbf{\bibinfo{volume}{79}},
  \bibinfo{pages}{409} (\bibinfo{year}{2011}),
  \\ \XDOI{10.1119/1.3533223}.

\bibitem[{\citenamefont{Ehrenberg and Siday}(1949)}]{Ehrenberg-S-1949prsb}
\bibinfo{author}{\bibfnamefont{W.}~\bibnamefont{Ehrenberg}} \bibnamefont{and}
  \bibinfo{author}{\bibfnamefont{R.~E.} \bibnamefont{Siday}},
  \\ \bibinfo{journal}{Proc. Phys. Soc. B} \textbf{\bibinfo{volume}{62}},
  \bibinfo{pages}{8} (\bibinfo{year}{1949}), 
%\\ \bibinfo{note}{(first publication
%  of ``Aharonov-Bohm effect'')},
  \\ \XDOI{10.1088/0370-1301/62/1/303\%0A}.

\bibitem[{\citenamefont{Aharonov and Bohm}(1959)}]{Aharonov-B-1959pr}
\bibinfo{author}{\bibfnamefont{Y.}~\bibnamefont{Aharonov}} \bibnamefont{and}
  \bibinfo{author}{\bibfnamefont{D.}~\bibnamefont{Bohm}},
  \\ \bibinfo{journal}{Phys. Rev.} \textbf{\bibinfo{volume}{115}},
  \bibinfo{pages}{485} (\bibinfo{year}{1959}),
  \\ \XDOI{10.1103/PhysRev.115.485}.

\bibitem[{\citenamefont{Matteucci et~al.}(2003)\citenamefont{Matteucci,
  Iencinella, and Beeli}}]{Matteucci-IB-2003fp}
\bibinfo{author}{\bibfnamefont{G.}~\bibnamefont{Matteucci}},
  \bibinfo{author}{\bibfnamefont{D.}~\bibnamefont{Iencinella}},
  \bibnamefont{and} \bibinfo{author}{\bibfnamefont{C.}~\bibnamefont{Beeli}},
  \\ \bibinfo{journal}{Foundations of Physics} \textbf{\bibinfo{volume}{33}},
  \bibinfo{pages}{577} (\bibinfo{year}{2003}),
  \\ \XDOI{10.1023/A:1023766519291}.

\bibitem[{\citenamefont{Rindler}(1989)}]{Rindler-1989ajp}
\bibinfo{author}{\bibfnamefont{W.}~\bibnamefont{Rindler}},
  \\ \bibinfo{journal}{Am. J. Phys.} \textbf{\bibinfo{volume}{57}},
  \bibinfo{pages}{993} (\bibinfo{year}{1989}),
  \\ \XDOI{10.1119/1.15782}.

\bibitem[{\citenamefont{Kinsler}(2010)}]{Kinsler-2010pra-lfiadc}
\bibinfo{author}{\bibfnamefont{P.}~\bibnamefont{Kinsler}},
  \\ \bibinfo{journal}{Phys. Rev. A} \textbf{\bibinfo{volume}{82}},
  \bibinfo{pages}{055804} (\bibinfo{year}{2010}), \\ \XARXIV{1008.2088},
  \\ \XDOI{10.1103/PhysRevA.81.013819}.

\bibitem[{\citenamefont{Itin}(2008)}]{Itin-2008grg}
\bibinfo{author}{\bibfnamefont{Y.}~\bibnamefont{Itin}}, \\ \bibinfo{journal}{Gen.
  Relativ. Gravit.} \textbf{\bibinfo{volume}{40}}, \bibinfo{pages}{1219}
  (\bibinfo{year}{2008}),
  \\ \XDOI{10.1007/s10714-007-0599-8}.

\bibitem[{\citenamefont{Gratus et~al.}(2017)\citenamefont{Gratus, Kinsler,
  Letizia, and Boyd}}]{Gratus-KLB-2017apa-malaga}
\bibinfo{author}{\bibfnamefont{J.}~\bibnamefont{Gratus}},
  \bibinfo{author}{\bibfnamefont{P.}~\bibnamefont{Kinsler}},
  \bibinfo{author}{\bibfnamefont{R.}~\bibnamefont{Letizia}}, \bibnamefont{and}
  \bibinfo{author}{\bibfnamefont{T.}~\bibnamefont{Boyd}},
  \\ \bibinfo{journal}{Appl. Phys. A} \textbf{\bibinfo{volume}{123}},
  \bibinfo{pages}{108} (\bibinfo{year}{2017}), 
%\\ \bibinfo{note}{also was
%  gratus2016Electromagnetic, in "Proceedings of META16 Malaga - Spain. The 7th
%  International Conference on Metamaterials,
% Photonic Crystals and Plasmonics", 2016.}, 
 \\ \XDOI{10.1007/s00339-016-0649-8}.

\bibitem[{\citenamefont{Boyd et~al.}(2018{\natexlab{a}})\citenamefont{Boyd,
  Gratus, Kinsler, and Letizia}}]{Boyd-GKL-2018oe-tbwire}
\bibinfo{author}{\bibfnamefont{T.}~\bibnamefont{Boyd}},
  \bibinfo{author}{\bibfnamefont{J.}~\bibnamefont{Gratus}},
  \bibinfo{author}{\bibfnamefont{P.}~\bibnamefont{Kinsler}}, \bibnamefont{and}
  \bibinfo{author}{\bibfnamefont{R.}~\bibnamefont{Letizia}},
  \\ \bibinfo{journal}{Opt. Express} \textbf{\bibinfo{volume}{26}},
  \bibinfo{pages}{2478} (\bibinfo{year}{2018}{\natexlab{a}}),
  \\ \XARXIV{1801.04927}, \\ \XDOI{10.1364/OE.26.002478}.

\bibitem[{\citenamefont{Boyd et~al.}(2018{\natexlab{b}})\citenamefont{Boyd,
  Gratus, Kinsler, Letizia, and Seviour}}]{Boyd-GKLS-2018as-mdpi}
\bibinfo{author}{\bibfnamefont{T.}~\bibnamefont{Boyd}},
  \bibinfo{author}{\bibfnamefont{J.}~\bibnamefont{Gratus}},
  \bibinfo{author}{\bibfnamefont{P.}~\bibnamefont{Kinsler}},
  \bibinfo{author}{\bibfnamefont{R.}~\bibnamefont{Letizia}}, \bibnamefont{and}
  \bibinfo{author}{\bibfnamefont{R.}~\bibnamefont{Seviour}},
  \\ \bibinfo{journal}{Appl. Sci.} \textbf{\bibinfo{volume}{8}},
  \bibinfo{pages}{1276} (\bibinfo{year}{2018}{\natexlab{b}}),
  \\ \XARXIV{1807.09041}, \\ \XDOI{10.3390/app8081276}.

\bibitem[{\citenamefont{Surjadi et~al.}(2019)\citenamefont{Surjadi, Gao, Du,
  Li, Xiong, Fang, and Lu}}]{Surjadi-GDLXFL-2019aem}
\bibinfo{author}{\bibfnamefont{J.~U.} \bibnamefont{Surjadi}},
  \bibinfo{author}{\bibfnamefont{L.}~\bibnamefont{Gao}},
  \bibinfo{author}{\bibfnamefont{H.}~\bibnamefont{Du}},
  \bibinfo{author}{\bibfnamefont{X.}~\bibnamefont{Li}},
  \bibinfo{author}{\bibfnamefont{X.}~\bibnamefont{Xiong}},
  \bibinfo{author}{\bibfnamefont{N.~X.} \bibnamefont{Fang}}, %\bibnamefont{and}
  \bibinfo{author}{\bibfnamefont{Y.}~\bibnamefont{Lu}},
  \\ \bibinfo{journal}{Advanced Engineering Materials}
  \textbf{\bibinfo{volume}{21}}, \bibinfo{pages}{1800864}
  (\bibinfo{year}{2019}),
  \\ \XDOI{10.1002/adem.201800864}.

\bibitem[{\citenamefont{Yu et~al.}(2018)\citenamefont{Yu, Zhou, Liang, Jiang,
  Yu, Zhou, Liang, Jiang, and Wu}}]{Yu-LJW-2018pms}
\bibinfo{author}{\bibfnamefont{X.}~\bibnamefont{Yu}},
  \bibinfo{author}{\bibfnamefont{J.}~\bibnamefont{Zhou}},
  \bibinfo{author}{\bibfnamefont{H.}~\bibnamefont{Liang}},
  \bibinfo{author}{\bibfnamefont{Z.}~\bibnamefont{Jiang}},
  \bibinfo{author}{\bibfnamefont{L.~W.} \bibnamefont{Yu}},
  \bibinfo{author}{\bibfnamefont{J.}~\bibnamefont{Zhou}},
  \bibinfo{author}{\bibfnamefont{H.}~\bibnamefont{Liang}},
  \bibinfo{author}{\bibfnamefont{Z.}~\bibnamefont{Jiang}}, \bibnamefont{and}
  \bibinfo{author}{\bibfnamefont{L.}~\bibnamefont{Wu}},
  \\ \bibinfo{journal}{Progress in Materials Science}
  \textbf{\bibinfo{volume}{94}}, \bibinfo{pages}{114} (\bibinfo{year}{2018}),
  \\ \XDOI{10.1016/j.pmatsci.2017.12.003}.

\bibitem[{\citenamefont{Kinsler and
  McCall}(2014{\natexlab{a}})}]{Kinsler-M-2014adp-scast}
\bibinfo{author}{\bibfnamefont{P.}~\bibnamefont{Kinsler}} \bibnamefont{and}
  \bibinfo{author}{\bibfnamefont{M.~W.} \bibnamefont{McCall}},
  \\ \bibinfo{journal}{Ann. Phys. (Berlin)} \textbf{\bibinfo{volume}{526}},
  \bibinfo{pages}{51} (\bibinfo{year}{2014}{\natexlab{a}}), 
 \\ \XARXIV{1308.3358},
  \\ \XDOI{10.1002/andp.201300164}.

\bibitem[{\citenamefont{Baev et~al.}(2015)\citenamefont{Baev, Prasad, Agren,
  Samoc, and Wegener}}]{Baev-PASW-2015pr}
\bibinfo{author}{\bibfnamefont{A.}~\bibnamefont{Baev}},
  \bibinfo{author}{\bibfnamefont{P.~N.} \bibnamefont{Prasad}},
  \bibinfo{author}{\bibfnamefont{H.}~\bibnamefont{Agren}},
  \bibinfo{author}{\bibfnamefont{M.}~\bibnamefont{Samoc}}, \bibnamefont{and}
  \bibinfo{author}{\bibfnamefont{M.}~\bibnamefont{Wegener}},
  \\ \bibinfo{journal}{Phys. Rep.} \textbf{\bibinfo{volume}{594}},
  \bibinfo{pages}{1} (\bibinfo{year}{2015}),
  \\ \XDOI{10.1016/j.physrep.2015.07.002}.

\bibitem[{\citenamefont{Kinsler and McCall}(2015)}]{Kinsler-M-2015pnfa-tofu}
\bibinfo{author}{\bibfnamefont{P.}~\bibnamefont{Kinsler}} \bibnamefont{and}
  \bibinfo{author}{\bibfnamefont{M.~W.} \bibnamefont{McCall}},
  \\ \bibinfo{journal}{Photon. Nanostruct. Fundam. Appl.}
  \textbf{\bibinfo{volume}{15}}, \bibinfo{pages}{10} (\bibinfo{year}{2015}),
 % \\ \bibinfo{note}{tTOFU},
  \\ \XDOI{10.1016/j.photonics.2015.04.005}.

\bibitem[{\citenamefont{McCall et~al.}(2018)\citenamefont{McCall, Pendry,
  Galdi, Lai, Horsley, Li, Zhu, Mitchell-Thomas, Quevedo-Teruel, Tassin
  et~al.}}]{McCall-etal-2018jo-roadmapto}
\bibinfo{author}{\bibfnamefont{M.}~\bibnamefont{McCall}},
  \bibinfo{author}{\bibfnamefont{J.~B.} \bibnamefont{Pendry}},
  \bibinfo{author}{\bibfnamefont{V.}~\bibnamefont{Galdi}},
  \bibinfo{author}{\bibfnamefont{Y.}~\bibnamefont{Lai}},
  \bibinfo{author}{\bibfnamefont{S.~A.~R.} \bibnamefont{Horsley}},
  \bibinfo{author}{\bibfnamefont{J.}~\bibnamefont{Li}},
  \bibinfo{author}{\bibfnamefont{J.}~\bibnamefont{Zhu}},
  \bibinfo{author}{\bibfnamefont{R.~C.} \bibnamefont{Mitchell-Thomas}},
  \bibinfo{author}{\bibfnamefont{O.}~\bibnamefont{Quevedo-Teruel}},
  \bibinfo{author}{\bibfnamefont{P.}~\bibnamefont{Tassin}},
  \bibnamefont{et~al.}, \\ \bibinfo{journal}{J. Opt.}
  \textbf{\bibinfo{volume}{20}}, \bibinfo{pages}{063001}
  (\bibinfo{year}{2018}),
  \\ \XDOI{10.1088/2040-8986/aab976}.

\bibitem[{\citenamefont{Dolin}(1961)}]{Dolin-1961ivuzr}
\bibinfo{author}{\bibfnamefont{L.~S.} \bibnamefont{Dolin}},
  \\ \bibinfo{journal}{Izv. Vyssh. Uchebn. Zaved. Radiofizika}
  \textbf{\bibinfo{volume}{4}}, \bibinfo{pages}{964} (\bibinfo{year}{1961}),
  \\ \XWEB{https://www.math.utah.edu/\%7Emilton/DolinTrans2.pdf}.

\bibitem[{\citenamefont{Pendry et~al.}(2006)\citenamefont{Pendry, Schurig, and
  Smith}}]{Pendry-SS-2006sci}
\bibinfo{author}{\bibfnamefont{J.~B.} \bibnamefont{Pendry}},
  \bibinfo{author}{\bibfnamefont{D.}~\bibnamefont{Schurig}}, \bibnamefont{and}
  \bibinfo{author}{\bibfnamefont{D.~R.} \bibnamefont{Smith}},
  \\ \bibinfo{journal}{Science} \textbf{\bibinfo{volume}{312}},
  \bibinfo{pages}{1780} (\bibinfo{year}{2006}),
  \\ \XDOI{10.1126/science.1125907}.

\bibitem[{\citenamefont{McCall et~al.}(2011)\citenamefont{McCall, Favaro,
  Kinsler, and Boardman}}]{McCall-FKB-2011jo}
\bibinfo{author}{\bibfnamefont{M.~W.} \bibnamefont{McCall}},
  \bibinfo{author}{\bibfnamefont{A.}~\bibnamefont{Favaro}},
  \bibinfo{author}{\bibfnamefont{P.}~\bibnamefont{Kinsler}}, \bibnamefont{and}
  \bibinfo{author}{\bibfnamefont{A.}~\bibnamefont{Boardman}},
  \\ \bibinfo{journal}{J. Opt.} \textbf{\bibinfo{volume}{13}},
  \bibinfo{pages}{024003} (\bibinfo{year}{2011}),
  \\ \XDOI{10.1088/2040-8978/13/2/024003}.

\bibitem[{\citenamefont{Gratus et~al.}(2016)\citenamefont{Gratus, Kinsler,
  McCall, and Thompson}}]{Gratus-KMT-2016njp-stdisp}
\bibinfo{author}{\bibfnamefont{J.}~\bibnamefont{Gratus}},
  \bibinfo{author}{\bibfnamefont{P.}~\bibnamefont{Kinsler}},
  \bibinfo{author}{\bibfnamefont{M.~W.} \bibnamefont{McCall}},
  \bibnamefont{and} \bibinfo{author}{\bibfnamefont{R.~T.}
  \bibnamefont{Thompson}}, \\ \bibinfo{journal}{New J. Phys.}
  \textbf{\bibinfo{volume}{18}}, \bibinfo{pages}{123010}
  (\bibinfo{year}{2016}), \\ \XARXIV{1608.00496},
  \\ \XDOI{10.1088/1367-2630/18/12/123010}.

\bibitem[{\citenamefont{Li and Pendry}(2008)}]{Li-P-2008prl}
\bibinfo{author}{\bibfnamefont{J.}~\bibnamefont{Li}} \bibnamefont{and}
  \bibinfo{author}{\bibfnamefont{J.~B.} \bibnamefont{Pendry}},
  \\ \bibinfo{journal}{Phys. Rev. Lett.} \textbf{\bibinfo{volume}{101}},
  \bibinfo{pages}{203901} (\bibinfo{year}{2008}),
  \\ \XDOI{10.1103/PhysRevLett.101.203901}.

\bibitem[{\citenamefont{Kinsler and
  McCall}(2014{\natexlab{b}})}]{Kinsler-M-2014pra}
\bibinfo{author}{\bibfnamefont{P.}~\bibnamefont{Kinsler}} \bibnamefont{and}
  \bibinfo{author}{\bibfnamefont{M.~W.} \bibnamefont{McCall}},
  \\ \bibinfo{journal}{Phys. Rev. A} \textbf{\bibinfo{volume}{89}},
  \bibinfo{pages}{063818} (\bibinfo{year}{2014}{\natexlab{b}}),
  \\ \XARXIV{1311.2287},
  \\ \XDOI{10.1103/PhysRevA.89.063818}.

\bibitem[{\citenamefont{Mitchell-Thomas
  et~al.}(2013)\citenamefont{Mitchell-Thomas, McManus, Quevedo-Teruel, Horsley,
  and Hao}}]{MitchellThomas-MQHH-2013prl}
\bibinfo{author}{\bibfnamefont{R.~C.} \bibnamefont{Mitchell-Thomas}},
  \bibinfo{author}{\bibfnamefont{T.~M.} \bibnamefont{McManus}},
  \bibinfo{author}{\bibfnamefont{O.}~\bibnamefont{Quevedo-Teruel}},
  \bibinfo{author}{\bibfnamefont{S.~A.~R.} \bibnamefont{Horsley}},
  \bibnamefont{and} \bibinfo{author}{\bibfnamefont{Y.}~\bibnamefont{Hao}},
  \\ \bibinfo{journal}{Phys. Rev. Lett.} \textbf{\bibinfo{volume}{111}},
  \bibinfo{pages}{213901} (\bibinfo{year}{2013}),
  \\ \XDOI{10.1103/PhysRevLett.111.213901}.

\bibitem[{\citenamefont{Lai et~al.}(2009)\citenamefont{Lai, Chen, Zhang, and
  Chan}}]{Lai-CZC-2009prl}
\bibinfo{author}{\bibfnamefont{Y.}~\bibnamefont{Lai}},
  \bibinfo{author}{\bibfnamefont{H.}~\bibnamefont{Chen}},
  \bibinfo{author}{\bibfnamefont{Z.-Q.} \bibnamefont{Zhang}}, \bibnamefont{and}
  \bibinfo{author}{\bibfnamefont{C.~T.} \bibnamefont{Chan}},
  \\ \bibinfo{journal}{Phys. Rev. Lett.} \textbf{\bibinfo{volume}{102}},
  \bibinfo{pages}{093901} (\bibinfo{year}{2009}), \\ \XARXIV{0811.0458},
  \\ \XDOI{10.1103/PhysRevLett.102.093901}.

\bibitem[{\citenamefont{Kinsler and McCall}(2018)}]{Kinsler-M-2015raytail}
\bibinfo{author}{\bibfnamefont{P.}~\bibnamefont{Kinsler}} \bibnamefont{and}
  \bibinfo{author}{\bibfnamefont{M.~W.} \bibnamefont{McCall}},
  \\ \bibinfo{journal}{Wave Motion} \textbf{\bibinfo{volume}{77}},
  \bibinfo{pages}{91} (\bibinfo{year}{2018}), \\ \XARXIV{1510.06890},
  \\ \XDOI{10.1016/j.wavemoti.2017.11.002}.

\bibitem[{\citenamefont{Fathi and Thompson}(2016)}]{Fathi-T-2016prd}
\bibinfo{author}{\bibfnamefont{M.}~\bibnamefont{Fathi}} \bibnamefont{and}
  \bibinfo{author}{\bibfnamefont{R.~T.} \bibnamefont{Thompson}},
  \\ \bibinfo{journal}{Phys. Rev. D} \textbf{\bibinfo{volume}{93}},
  \bibinfo{pages}{124026} (\bibinfo{year}{2016}),
  \\ \XDOI{10.1103/PhysRevD.93.124026}.

\bibitem[{\citenamefont{Thompson and Fathi}(2015)}]{Thompson-F-2015pra}
\bibinfo{author}{\bibfnamefont{R.~T.} \bibnamefont{Thompson}} \bibnamefont{and}
  \bibinfo{author}{\bibfnamefont{M.}~\bibnamefont{Fathi}},
  \\ \bibinfo{journal}{Phys. Rev. A} \textbf{\bibinfo{volume}{92}},
  \bibinfo{pages}{013834} (\bibinfo{year}{2015}), \\ \XARXIV{1506.08507},
  \\ \XDOI{10.1103/PhysRevA.92.013834}.

\bibitem[{\citenamefont{Kinsler et~al.}(2009)\citenamefont{Kinsler, Favaro, and
  McCall}}]{Kinsler-FM-2009ejp}
\bibinfo{author}{\bibfnamefont{P.}~\bibnamefont{Kinsler}},
  \bibinfo{author}{\bibfnamefont{A.}~\bibnamefont{Favaro}}, \bibnamefont{and}
  \bibinfo{author}{\bibfnamefont{M.~W.} \bibnamefont{McCall}},
  \\ \bibinfo{journal}{Eur. J. Phys.} \textbf{\bibinfo{volume}{30}},
  \bibinfo{pages}{983} (\bibinfo{year}{2009}), \\ \XARXIV{0908.1721},
  \\ \XDOI{10.1088/0143-0807/30/5/007}.

\bibitem[{\citenamefont{Gratus et~al.}(2012)\citenamefont{Gratus, Obukhov, and
  Tucker}}]{Gratus-OT-2012ap}
\bibinfo{author}{\bibfnamefont{J.}~\bibnamefont{Gratus}},
  \bibinfo{author}{\bibfnamefont{Y.~N.} \bibnamefont{Obukhov}},
  \bibnamefont{and} \bibinfo{author}{\bibfnamefont{R.~W.}
  \bibnamefont{Tucker}}, \\ \bibinfo{journal}{Annals Physics}
  \textbf{\bibinfo{volume}{327}}, \bibinfo{pages}{2560} (\bibinfo{year}{2012}),
  \\ \XDOI{10.1016/j.aop.2012.07.006}.

\bibitem[{\citenamefont{Dereli et~al.}(2007)\citenamefont{Dereli, Gratus, and
  Tucker}}]{Dereli-GT-2007jpa}
\bibinfo{author}{\bibfnamefont{T.}~\bibnamefont{Dereli}},
  \bibinfo{author}{\bibfnamefont{J.}~\bibnamefont{Gratus}}, \bibnamefont{and}
  \bibinfo{author}{\bibfnamefont{R.~W.} \bibnamefont{Tucker}},
  \\ \bibinfo{journal}{J. Phys. A} \textbf{\bibinfo{volume}{40}},
  \bibinfo{pages}{5695} (\bibinfo{year}{2007}), \\ \XARXIV{math-ph/0610082},
  \\ \XDOI{10.1088/1751-8113/40/21/016}.

\bibitem[{\citenamefont{Born and Wolf}(2013)}]{BornWolf-Principles}
\bibinfo{author}{\bibfnamefont{M.}~\bibnamefont{Born}} \bibnamefont{and}
  \bibinfo{author}{\bibfnamefont{E.}~\bibnamefont{Wolf}},
  \\ \emph{\bibinfo{title}{Principles of Optics: Electromagnetic Theory of
  Propagation, Interference and Diffraction of Light}}
  \\ (\bibinfo{publisher}{Elsevier}, \bibinfo{year}{2013}),
 \\ ISBN \bibinfo{isbn}{978-1-48310-320-4}.

\bibitem[{\citenamefont{Favaro and Hehl}(2014)}]{Favaro-H-2014arxiv}
\bibinfo{author}{\bibfnamefont{A.}~\bibnamefont{Favaro}} \bibnamefont{and}
  \bibinfo{author}{\bibfnamefont{F.~W.} \bibnamefont{Hehl}},
  \\ \emph{\bibinfo{title}{Fresnel versus Kummer surfaces: geometrical optics in dispersionless linear (meta)materials and vacuum}}
%  \\ \bibinfo{journal}{ArXiv}  (\bibinfo{year}{2014}), 
%\\ \bibinfo{note}{invited
%  lecture at "Electromagnetic Spacetimes", Wolfgang Pauli Institute, Vienna,
%  19--23 Nov 2012}, 
\\ \XARXIV{1401.4077}.
%  \\ \XWEB{https://arxiv.org/abs/1401.4077}.

\bibitem[{\citenamefont{Gratus and Banaszek}(2018)}]{Gratus-B-2018rspa}
\bibinfo{author}{\bibfnamefont{J.}~\bibnamefont{Gratus}} \bibnamefont{and}
  \bibinfo{author}{\bibfnamefont{T.}~\bibnamefont{Banaszek}},
  \\ \bibinfo{journal}{Proc. Royal Soc. A} \textbf{\bibinfo{volume}{474}},
  \bibinfo{pages}{20170652} (\bibinfo{year}{2018}),
  \\ \XDOI{10.1098/rspa.2017.0652}.

\bibitem[{\citenamefont{Flanders}(1963)}]{Flanders1963}
\bibinfo{author}{\bibfnamefont{H.}~\bibnamefont{Flanders}},
  \\ \emph{\bibinfo{title}{Differential Forms with Applications to the Physical
  Sciences}} 
  \\(\bibinfo{publisher}{Academic Press}, \bibinfo{address}{New York},
  \bibinfo{year}{1963}), 
  \\ \bibinfo{note}{(Dover, 2003)}.

\bibitem[{\citenamefont{Deschamps}(1981)}]{Deschamps-1981ieee}
\bibinfo{author}{\bibfnamefont{G.~A.} \bibnamefont{Deschamps}},
  \\ \bibinfo{journal}{Proc. IEEE} \textbf{\bibinfo{volume}{69}},
  \bibinfo{pages}{676676} (\bibinfo{year}{1981}),
  \\ \XDOI{10.1109/PROC.1981.12048}.

\end{thebibliography}

%
% ======================================================================
\appendix

\def\LeviCiv{\epsilon}

%
% ----------------------------------------------------------------------
\section{In relativistic tensor notation}
\label{ch_GR}

In both special and general relativity
 it is usual to
 combine the electromagnetic fields {\xpair{\VE}{\VB}}
 into a single antisymmetric tensor $F_{ab}$
  \cite{Post-FSEM}.
Here indices $a,b,\ldots$
 span the range $\{0,1,2,3\}$,
 where for some coordinates $x^a$
 we have that $x^0=t$; 
 also we use indices $i,j,\ldots$  
 to represent only the spatial $\{1,2,3\}$.
For some $F_{ab}$,
 we can extract the electric and magnetic fields
 with
%[
\begin{align}
  {\pE}_i = F_{0i}
\qquadand
  {\pB}_\ell = g_{\ell i}\LeviCiv^{ijk} F_{jk}.
.
\label{GR_E_B}
\end{align}
%]
 where the extraction of the magnetic field ${\pB}_\ell$
 requires both the Levi-Civita symbol $\LeviCiv^{ijk}$
 and the metric $g_{\ell i}$.

In this notation
 Maxwell's equations \eqref{GH_Max_NoMono}, \eqref{GH_Max_Faraday} become
%[
\begin{align}
  \partial_a F_{bc} + \partial_b F_{ca} + \partial_c F_{ab}
=
  0
,
\label{GR_dF}
\end{align}
%]
which is automatically satisfied using
 the electromagnetic potential,
 which is a 4-vector ${\pA}_a$
 such that
%[
\begin{align}
  F_{ab}
&=
  \partial_a {\pA}_b
 -
  \partial_b {\pA}_a
.
\label{GR_F_dA}
\end{align}
%]

All the proofs for results in this appendix
 are given in detail in appendix \ref{ch_Proofs}.

The advantage of using this relativistic notation
 is that the four separate CMCR operators
 {\xquad{\PsiErho}{\PsiBrho}{\PsiEJ}{\PsiBJ}}
 can now be combined
 into a single CMCR operator $\Psi$.
This operator
 takes the tensor $F_{ab}$ to the 4-vector density
 ${\pJ}^a$ and so
 the two equations \eqref{M_New_Max_CR_rho}, \eqref{M_New_Max_CR_J}
 now combine into the single CMCR equation
%[
\begin{align}
  {\PsiAny} \VAct{F} = {\pJ}
\label{GR_Max-CR}
\end{align}
%]
The operator ${\PsiAny}$ obeys the same three properties.
It is linear,
 as in \eqref{GH_Psi_+-linear} and \eqref{GH_Psi_R-linear},
 so that
%[
\begin{align}
  {\PsiAny} \VAct{\alpha + \beta}
&=
  {\PsiAny} \VAct{\alpha} + {\PsiAny} \VAct{\beta}
\quadand
  {\PsiAny} \VAct{\lambda \, \alpha}
=
  \lambda \, {\PsiAny}\VAct{\alpha}
\label{FOO_def_+lin}
\end{align}
%]
where $\lambda \in \Real$.
It is also first order,
 as in \eqref{M_MHOOF_Phi} %\eqref{M_MHOOF_PsiErho})
%[
\begin{align}
  {\PsiAny} \VAct{{f}^2 \, F}
&=
  2{f} \, {\PsiAny} \VAct{{f} \, F}
 -
  {f}^2 \, {\PsiAny} \VAct{F}
,
\label{FOO_def_scalar}
\end{align}
%]
for all scalar fields ${f}$ and all antisymmetric tensors $F$.
Finally,
 the conservation of charge equation,
 corresponding to \eqref{M_New_Max_Consev},
 is now written
%[
\begin{align}
  \partial_a( {\PsiAny} \VAct{F} )^a = 0
\quadtext{where}
  F_{ab} = \partial_a {\pA}_b - \partial_b {\pA}_a
,
\label{GR_1st_order}
\end{align}
%]
 for all ${\pA}$.

\noindent\textbullet~~
All operators ${\PsiAny}$ which satisfy
\eqref{FOO_def_+lin}, \eqref{FOO_def_scalar} can be written
%[
\begin{align}
  \left( {\PsiAny} \VAct{F} \right)^a
=
  \tfrac{1}{2} {\PsiAny}^{abc} \, F_{bc}
 +
  \tfrac{1}{2} {\PsiAny}^{abcd} \left( \partial_b F_{cd} \right)
\label{FOO_CR_coord_alt}
\end{align}
%]
where clearly we demand\footnote{Here
  the brackets refer to the symmetric component,
  for example
  ${\PsiAny}^{a(bc)}
   =\tfrac{1}{2} \left({\PsiAny}^{abc} + {\PsiAny}^{acb}\right)$.}
%[
\begin{align}
 {\PsiAny}^{a(bc)} = 0
\qquadand
 {\PsiAny}^{ab(cd)} = 0
.
\label{GR_sym_last_ind}
\end{align}
%]
We can extract the components ${\PsiAny}^{abc}$ and ${\PsiAny}^{abcd}$
via
%[
\begin{align}
  {\PsiAny}^{abc}
&=
  \left( {\PsiAny} \VAct{dx^{bc}} \right)^a
\\
  {\PsiAny}^{abcd}
&=
  \left(
    {\PsiAny} \VAct{x^b \, dx^{cd}}
   -
    x^b \, {\PsiAny} \VAct{dx^{cd}}
  \right)^a
\label{GR_Coordfree_Comp}
\end{align}
%]
where $dx^{ab}$ is the antisymmetric tensor with components
%[
\begin{align}
  (dx^{ab})_{cd} 
&=
  \delta^a_c\,\delta^b_d - \delta^a_d\,\delta^b_c
\label{GR_def_dx^ab}
\end{align}
%]
It is trivial to see that ${\PsiAny}\VAct{F}$
 given by \eqref{FOO_CR_coord_alt} obeys
 \eqref{FOO_def_+lin}, \eqref{FOO_def_scalar}.
The converse is demonstrated in the appendix \ref{ch_Proofs},
 which also contains the proof of all the statements in this section.
Thus without the charge conservation equation
 there are 120 components for ${\PsiAny}$.

\noindent\textbullet~~
The  ${\PsiAny}^{abc}$ and ${\PsiAny}^{abcd}$
 used above are simply a rewriting of
 the components of {\xquad{\PsiErho}{\PsiBrho}{\PsiEJ}{\PsiBJ}}.
For example,
 we have
%[
\begin{align}
 (\PsiErho)^i = {\PsiAny}^{00i}
\,,
\quad
 (\PsiErho)^{0i} = {\PsiAny}^{000i}
\quadand
 (\PsiErho)^{ji} = {\PsiAny}^{0j0i}
.
\label{GR_comp_extract}
\end{align}
%]

\noindent\textbullet~~
Imposing charge conservation using \eqref{GR_1st_order}
 means that
 the components ${\PsiAny}^{abc}$ and ${\PsiAny}^{abcd}$ also obey
%[
\begin{align}
\partial_a {\PsiAny}^{abc} &= 0
\,,
\label{GR_comp_scalar_cond1}
\\
{\PsiAny}^{(ab)c}
+
\partial_d{\PsiAny}^{d(ab)c} &= 0
\label{GR_comp_scalar_cond2}
\\
\quadand
{\PsiAny}^{(abc)d} &= 0.
\label{GR_comp_scalar_cond3}
\end{align}
%]

\noindent\textbullet~~
In appendix \ref{ch_Proofs} we show that \eqref{GR_sym_last_ind}
 and \eqref{GR_comp_scalar_cond1}--\eqref{GR_comp_scalar_cond3} 
 imply there are 4 independent components of
 ${\PsiAny}^{abc}$,
 corresponding to \xpair{\axivav}{\axivat} 
 and 51 independent components of ${\PsiAny}^{abcd}$.

\noindent\textbullet~~
Clearly in the homogeneous case
 \eqref{GR_comp_scalar_cond1}--\eqref{GR_comp_scalar_cond3}  reduce  to
%[
\begin{align}
{\PsiAny}^{(ab)c} = 0
\qquadand
{\PsiAny}^{(abc)d}=0
\label{GR_comp_scalar_homo}
\end{align}
%]

\noindent\textbullet~~
The null equivalent condition \eqref{ccrestrictions_Null_invar}
 relevant for the non-axionic extension terms
 becomes
%[
\begin{align}
 \PsiAny \to \PsiAny + \NullPsi
\label{GR_Null_action}
\end{align}
%]
where $\NullPsi$ satisfies
 \eqref{FOO_def_+lin},  \eqref{FOO_def_scalar}; 
 and in addition
%[
\begin{align}
  \NullPsi \VAct{F} = 0
\quadtext{where}
  F_{ab} = \partial_a {\pA}_b-\partial_b {\pA}_a
\quadtext{for all}
{\pA}
\label{GR_Null_constraint}
\end{align}
%]
In terms of components \eqref{GR_Null_constraint} becomes
%[
\begin{align}
  \NullPsiIndices{abcd}
 +\NullPsiIndices{acbd}
 -\NullPsiIndices{abdc}
 -\NullPsiIndices{adbc}
=
  0
.\label{GR_Coords_Null_constraints}
\end{align}
%]
\noindent\textbullet~~
In appendix \ref{ch_Proofs} we show that
 \eqref{GR_Coords_Null_constraints} implies
 that $\NullPsiIndices{abcd}$ has 16 components.

%---------------------

\noindent\textbullet~~
Observe that although the components ${\PsiAny}^{abcd}$
 transform as a tensor density under change of coordinates,
 the components ${\PsiAny}^{abc}$ do not.
If $(x^0,x^1,x^2,x^3)$ and $(\xhat^0,\xhat^1,\xhat^2,\xhat^3)$
 are two coordinate systems then
%[
\begin{equation}
\begin{aligned}
 \hat{\PsiAny}^{\hat{a}\hat{b}\hat{c}}
&=
  \Big(
    {\PsiAny}^{abc}
    ~
    \pfrac{\xhat^\bhat}{x^b}
    ~
    \pfrac{\xhat^\chat}{x^c}
   +
    {\PsiAny}^{abcd}
    ~
    \pqfrac{\xhat^\bhat}{x^b}{x^c}
    ~
    \pfrac{\xhat^\chat}{x^d}
\\
&\qquad\qquad\qquad
 +
    {\PsiAny}^{abcd}
    ~
    \pfrac{\xhat^\bhat}{x^c}
    ~
    \pqfrac{\xhat^\chat}{x^b}{x^d}
  \Big)
  \pfrac{\xhat^\ahat}{x^a}
  ~
  \dfrac{x}{\xhat}
\end{aligned}
\label{GR_Change_coords_abc}
\end{equation}
%]
and
%[
\begin{align}
  \hat{\PsiAny}^{\hat{a}\hat{b}\hat{c}\hat{d}}
=
  {\PsiAny}^{abcd}
    ~
  \pfrac{\xhat^\bhat}{x^b}
    ~
  \pfrac{\xhat^\chat}{x^c}
    ~
  \pfrac{\xhat^\dhat}{x^d}
    ~
  \pfrac{\xhat^\ahat}{x^a}
    ~
  \dfrac{x}{\xhat}
,
\label{GR_Change_coords_abcd}
\end{align}
%]
where $\hat{\PsiAny}^{abc},\hat{\PsiAny}^{abcd}$
 are the component with respect to $(\xhat^0,\xhat^1,\xhat^2,\xhat^3)$
 and $dx/d\xhat$ is the Jacobian.
\ResubOne{If higher order operators were to be considered, 
 in an extension of our CMCR, 
 they would lead to more unusual
 non tensorial changes of coordinates \cite{Gratus-B-2018rspa}.}

%
% ----------------------------------------------------------------------
\section{Proofs}
\label{ch_Proofs}

\newcommand{\HerePsi}[1]{\PsiAny^{#1}}

\begin{proof}[Proof that
    \textup{\eqref{FOO_def_+lin}, \eqref{FOO_def_scalar}} implies
    \eqref{FOO_CR_coord_alt},
  i.e. linearity]~\\
Let $\lambda$ and $\mu$ be scalar fields.
By considering ${\PsiAny}\VAct{(\lambda+\mu)^2 F}$
 and ${\PsiAny}\VAct{(\lambda-\mu)^2 F}$
 then we can show \eqref{FOO_def_scalar}
 implies
%[
\begin{align}
  {\PsiAny}\VAct{\lambda\mu\,F}
 -
  \lambda{\PsiAny}\VAct{\mu\,F}
 -
  \mu\,{\PsiAny}\VAct{\lambda\,F}
 +
  \lambda\mu\,{\PsiAny}\VAct{F}
=
  0
\label{Proofs_1st_order}
\end{align}
%]
Let $p$ be an event in spacetime with coordinates $(p^0,p^1,p^2,p^3)$
then expanding $F_{cd}$ about $p$ to second order gives
%[
\begin{align*}
  F_{cd} &= F_{cd}|_p + (x^b-p^b)\,(\partial_b F_{cd})|_p +
\\&\qquad
  (x^a-p^a)(x^b-p^b)\,\alpha_{abcd}
\end{align*}
%]
where $\alpha_{abcd}$ a set of (indexed) scalar fields.
%[
\begin{align*}
  2\left({\PsiAny}\VAct{F}\right)^a\big|_p
&=
  {\PsiAny}\VAct{(F_{cd}|_p)\,dx^{ab}}\big|_p
\\&\qquad
 +
  {\PsiAny}\VAct{(x^c-p^c)\,(\partial_cF_{cd})|_p\,dx^{ab}}\big|_p
\\&\qquad
 +
  {\PsiAny}\VAct{(x^c-p^c)(x^d-p^d)\,\alpha_{abcd}}\big|_p
\end{align*}
%]
Now using \eqref{GR_Coordfree_Comp} and \eqref{FOO_def_+lin} we have
%[
\begin{align*}
  \left({\PsiAny}\VAct{(F_{cd}|_p)\,dx^{cd}}\right)^a\big|_p
&=
  (F_{cd}|_p)\,\left({\PsiAny}\VAct{dx^{cd}}\right)^a\big|_p
\\&=
  (F_{cd}|_p)\,\left({\PsiAny}^{acd}\right)^a\big|_p
\\&=
  \left(F_{cd}\,{\PsiAny}^{acd}\right)\big|_p
\end{align*}
%]
and
%[
\begin{align*}
\lefteqn{
  \left({\PsiAny}\VAct{(x^b-p^b)\,(\partial_bF_{cd})|_p\,dx^{cd}}\right)^a\big|_p
}
\qquad&
\\&=
  (\partial_bF_{cd})|_p\Big({\PsiAny}\VAct{x^b\,dx^{cd}}\big|_p
 -
  p^b\,{\PsiAny}\VAct{dx^{cd}}\Big)^a\Big|_p
\\
&=
  (\partial_bF_{cd})|_p\Big({\PsiAny}\VAct{x^b\,dx^{cd}}\big|_p
 -
  x^b\,{\PsiAny}\VAct{dx^{cd}}\Big)^a\Big|_p
\\
&=
  (\partial_bF_{cd})|_p\,({\PsiAny}^{abcd})\big|_p
=
  \left({\PsiAny}^{abcd}\,\partial_bF_{cd}\right)\big|_p
\end{align*}
%]
Lastly,
  using \eqref{Proofs_1st_order} we have
%[
\begin{align*}
\lefteqn{
  {\PsiAny}\VAct{(x^a-p^a)(x^b-p^b)\,\alpha_{abcd}}\big|_p
}
\qquad&
\\
&=
  \left(
    ( x^a-p^a )
    {\PsiAny} \VAct{(x^b-p^b) \, \alpha_{abcd}}
  \right)
  \big|_p
\\&\qquad
 +\left(
    ( x^b-p^a )
    {\PsiAny} \VAct{(x^a-p^a) \, \alpha_{abcd}}
  \right)
  \big|_p
\\&\qquad
 +\left(
    ( x^a-p^a ) ( x^b-p^b ) \,
    {\PsiAny} \VAct{\alpha_{abcd}}
  \right)
  \big|_p
=
  0
\end{align*}
%]
Hence
%[
\begin{align*}
  2\left({\PsiAny}\VAct{F}\right)^a\big|_p
&=
  \Big(F_{cd}\,{\PsiAny}^{acd}+{\PsiAny}^{abcd}\,\partial_bF_{cd}\Big)\Big|_p
\end{align*}
%]
Hence 
 \eqref{FOO_CR_coord_alt}.
\end{proof}

\begin{proof}[Proof of 
 \eqref{GR_comp_scalar_cond1}--\eqref{GR_comp_scalar_cond3},
 i.e. the effect of charge conservation on $\PsiAny$]~\\
Let $F_{ab} = \partial_a {\pA}_b - \partial_b {\pA}_a$
 then from \eqref{GR_1st_order}
 we have $\partial_a({\PsiAny}\VAct{F})^a=0$.
Thus using \eqref{FOO_CR_coord_alt} we have,
 using \eqref{GR_sym_last_ind},
%[
\begin{align*}
  0
&=
  \partial_a({\PsiAny}\VAct{F})^a
=
  \tfrac12
  \partial_a
  \left[
    {\PsiAny}^{abc}\, F_{bc} + {\PsiAny}^{abcd} (\partial_b F_{cd})
  \right]
\\&=
  \tfrac12 \partial_a
  \left[
    {\PsiAny}^{abc}\, (\partial_b {\pA}_c - \partial_c {\pA}_b)
   +
    {\PsiAny}^{abcd} (\partial_{bc} {\pA}_d - \partial_{bd} {\pA}_c)
  \right]
\\&=
  \partial_a
  \left[
    {\PsiAny}^{abc}\, \partial_b {\pA}_c
   +
    {\PsiAny}^{abcd} \partial_{bc} {\pA}_d
  \right]
\\&=
  \left( \partial_a{\PsiAny}^{abc} \right)
  \,
  \partial_b {\pA}_c
 +
  {\PsiAny}^{abc}
  \,
  \partial_{ab} {\pA}_c
\\
&\qquad
 +\left( \partial_a{\PsiAny}^{abcd} \right)
  \partial_{bc} {\pA}_d
 +{\PsiAny}^{abcd} \partial_{abc} {\pA}_d
\\&=
(\partial_a{\PsiAny}^{abc})\, \partial_b {\pA}_c
+ \left({\PsiAny}^{abc} + \partial_d{\PsiAny}^{dabc}\right) \partial_{ab} {\pA}_c
\\&\qquad
 + {\PsiAny}^{abcd} \partial_{abc} {\pA}_d
\\&=
  \left( \partial_a{\PsiAny}^{abc} \right)
  \,
  \partial_b {\pA}_c
 +
  \left(
    {\PsiAny}^{(ab)c}
   + \partial_d{\PsiAny}^{d(ab)c}
  \right)
  \partial_{ab} {\pA}_c
\\
&\qquad
 +
  {\PsiAny}^{(abc)d} \partial_{abc} {\pA}_d
\end{align*}
%]
Since at each point the first,
 second and third derivatives of $A$ are independent
 this gives \eqref{GR_comp_scalar_cond1}--\eqref{GR_comp_scalar_cond3}.
\end{proof}

\begin{proof}[Proof of \eqref{GR_Coords_Null_constraints}),
 i.e. constraints on the non-axionic terms]%~\\
%[
\begin{align*}
  0
&
%=\big(\Psi(dA)\big)^a
=
  \Psi^{abcd}
  \left[
    \partial_b(\partial_c {\pA}_d - \partial_d {\pA}_c)
  \right]
=
  \Psi^{abcd}
  \left[
    \partial_{bc} {\pA}_d - \partial_{bd} {\pA}_c
  \right]
\\&=
  \tfrac12
  (\partial_{bc} {\pA}_d)
  \left(
    \Psi^{abcd} + \Psi^{acbd} - \Psi^{abdc} - \Psi^{acdb}
  \right)
.
\end{align*}
%]
\end{proof}

\begin{proof}[Proof: Counting the 51 non-axionic terms.]~\\
These terms include all of the standard EMCR terms,
 as well as the extension terms denoted $\NullPsi$;
 these all contain differentials.
Here we need to calculate all $\HerePsi{abcd}$
  such that from \eqref{GR_sym_last_ind}, \eqref{GR_comp_scalar_cond3}, 
 $\HerePsi{(abc)d} = 0$ and $\HerePsi{ab(cd)} =0$.
For the following we do not use any summation convention,
 and assume all $\Set{abcd}$ are different.

\begin{description}

\item
[Case $\Set{aaaa}$, 0 terms:]
 $\HerePsi{aaaa} = 0$.

\item
[Case $\Set{aaab}$, 0 terms:]
  $\HerePsi{baaa} =   \HerePsi{abaa}=0$.\\
  $\HerePsi{aaba} = - \HerePsi{aaab}$,
 but $\HerePsi{aaba} + \HerePsi{abaa} + \HerePsi{baaa} = 0$
hence $\HerePsi{aaba} = 0$.

\item
[Case $\Set{aabb}$, 6 terms:]
 $\HerePsi{aabb}=0$.\\
 $\HerePsi{abab}+\HerePsi{baab}=0$,
 hence
 $\HerePsi{abab}=-\HerePsi{baab}=\HerePsi{baba}$.

\item
[Case $\Set{aabc}$, 36 terms:]
 $\HerePsi{aabc}+\HerePsi{abac}+\HerePsi{baac}=0$.

 $\HerePsi{abca} + \HerePsi{acba} + \HerePsi{baca}
    + \HerePsi{bcaa} + \HerePsi{caba} + \HerePsi{cbaa}=0$;
i.e.
 $\HerePsi{abca} + \HerePsi{acba} + \HerePsi{baca} + \HerePsi{caba} = 0$,
 so\\
 $\HerePsi{abac} + \HerePsi{acab} + \HerePsi{baac} + \HerePsi{caab} = 0$.\\
Hence
 $\HerePsi{aabc} + \HerePsi{aacb}=0$ (Obviously).

Number $\Set{aabc}$ is $4\times 3=12$ however not all are independent.

 $\HerePsi{1002}$, $\HerePsi{0102}$, $\HerePsi{0201}$,
 $\HerePsi{1003}$, $\HerePsi{0103}$, $\HerePsi{0301}$,
 $\HerePsi{2003}$, $\HerePsi{0203}$, $\HerePsi{0302}$.
This times four equals 36.

\item
[Case $\Set{abcd}$, 9 terms:]~\\
%[
%\begin{align*}
 $\HerePsi{abcd} + \HerePsi{bcad} + \HerePsi{cabd} +
 \HerePsi{bacd} + \HerePsi{cbad} + \HerePsi{acbd} = 0$.\\
%\end{align*}
%]
If $d=3$ this gives 5 terms:
 $\HerePsi{0123}$, $\HerePsi{1203}$, $\HerePsi{2013}$, $\HerePsi{1023}$,
 $\HerePsi{2103}$;
 here $\HerePsi{0213}$ is given by the others.

With $d=2$ this gives 3 terms:
 $\HerePsi{0312}$, $\HerePsi{3012}$, $\HerePsi{3102}$;
 here $\HerePsi{1302}$ is given by the others.

With $d=1$ this gives 1 term:
 $\HerePsi{2301}$

\end{description}

\end{proof}

\begin{proof}[Proof: Counting the 4 axionic terms.]~\\
The {\pAxionResponse} terms are those with only three indices,
 i.e. ${\PsiAny}^{abc}$, using
 \eqref{GR_sym_last_ind} and \eqref{GR_comp_scalar_cond2} 
With no summation, 
 and assuming all $\Set{abc}$ are different:
\begin{description}

\item
[Case $\Set{aaa}$, 0 terms:]
 $\HerePsi{aaa} = 0$.

\item
[Case $\Set{aab}$, 0 terms:]
Since
${\PsiAny}^{aba}=-{\PsiAny}^{aab}=\sum_c\partial_c{\PsiAny}^{caab}$
and
${\PsiAny}^{baa}=0$.
So there are no additional terms arising from repeated terms.

\item
[Case $\Set{abc}$, 4 terms:]
%[
\begin{align*}
\PsiAny^{abc}
&=
-\PsiAny^{acb} 
=
\PsiAny^{cab} 
-
2\sum_d\partial_d{\PsiAny}^{d(ac)b}
\\&=
-\PsiAny^{cba} 
-
2\sum_d\partial_d{\PsiAny}^{d(ac)b}
\\&=
\PsiAny^{bca} 
-
2\sum_d\partial_d{\PsiAny}^{d(ac)b}
-
2\sum_d\partial_d{\PsiAny}^{d(bc)a}
\\&=
-\PsiAny^{bac} 
-
2\sum_d\partial_d{\PsiAny}^{d(ac)b}
-
2\sum_d\partial_d{\PsiAny}^{d(bc)a}
\\&=
\PsiAny^{abc} 
-
2\sum_d\partial_d{\PsiAny}^{d(bc)a}
-
\\&\qquad\qquad
2\sum_d\partial_d{\PsiAny}^{d(ac)b}
-
2\sum_d\partial_d{\PsiAny}^{d(bc)a}
\\&=
\PsiAny^{abc} 
-
2\sum_d\partial_d\big({\PsiAny}^{d(bc)a}
+
{\PsiAny}^{d(ac)b}
+
{\PsiAny}^{d(bc)a}\big)
\\&=
\PsiAny^{abc} 
-
\sum_d\partial_d\big(
{\PsiAny}^{dbca}
+
{\PsiAny}^{dcba}
+
{\PsiAny}^{dacb}
\\&\qquad\qquad\qquad
+
{\PsiAny}^{dcab}
+
{\PsiAny}^{dbca}
+
{\PsiAny}^{dcba}
\big)
\end{align*}
%]
This is consistant since $\PsiAny^{dbca}+\PsiAny^{dbac}=0$.
\end{description}
Therefore there are 4 independent terms.
%[
\begin{align*}
{\PsiAny}^{012}=\axivav_3,\
{\PsiAny}^{013}=-\axivav_2,\
{\PsiAny}^{023}=\axivav_1\text{ and }
{\PsiAny}^{123}=-\axivat
\end{align*}
%]
 although they are still constrained by the first equation in
 \eqref{GR_comp_scalar_cond1}. 

\end{proof}

\begin{proof}[Proof: Counting the 16 null terms.]~\\
First note that \eqref{GR_Coords_Null_constraints}
 implies  the second equation of \eqref{GR_comp_scalar_cond3}.
This follows since
  \eqref{GR_Null_constraint} implies \eqref{GR_1st_order}.
After manipulating
 \eqref{GR_Coords_Null_constraints} and $\NullPsiIndices{ab(cd)}=0$
 we see that $\NullPsiIndices{abcd}$
 is antisymmetric in the last three indices.
Thus there are only 4 possible values of $bcd$
 and 4 possible values of $a$,
 giving 16 components.
\end{proof}

\begin{proof}[Proof of change of coordinates]~\\
Observe that ${\pJ}^a$ is a vector density since it transforms as
%[
\begin{align}
  \Jhat^a
=
  {\pJ}^b
  \pfrac{\xhat^a}{x^b}
  \,
  \dfrac{x}{\xhat}
.
\label{Proofs_J^a_density}
\end{align}
%]
%[
\begin{align*}
\textrm{Let}\qquad\qquad\qquad\quad
  {\pJ}^a
&=
  \tfrac{1}{2}
  {\PsiAny}^{abc} \,
  F_{bc}
 +
  \tfrac{1}{2}
  {\PsiAny}^{abcd} ~
  \pfrac{F_{cd}}{x^b}
 \qquad
\\\textup{and}\qquad
  \Jhat^a
&=
  \tfrac{1}{2}\Psihat^{abc}\, \Fhat_{bc}
 +
  \tfrac{1}{2}\Psihat^{abcd}\ \pfrac{\Fhat_{cd}}{\xhat^b}
.
\end{align*}
%]
Then
%[
\begin{align*}
&
  \tfrac{1}{2} \Psihat^{\ahat\bhat\chat} \,
  \Fhat_{\bhat\chat}
 +
  \tfrac{1}{2} \Psihat^{\ahat\bhat\chat\dhat} ~
  \pfrac{\Fhat_{\chat\dhat}}{\xhat^\bhat}
=
  \Jhat^\ahat
=
  {\pJ}^a ~
  \pfrac{\xhat^\ahat}{x^a} ~
  \dfrac{x}{\xhat}
\\
&=
  \left(
    \tfrac{1}{2}{\PsiAny}^{abc}\, F_{bc}
 +
  \tfrac{1}{2}{\PsiAny}^{abcd} \pfrac{F_{cd}}{x^b}
  \right)
  \pfrac{\xhat^\ahat}{x^a} ~
  \dfrac{x}{\xhat}
\\&=
  \tfrac{1}{2} {\PsiAny}^{abc} \, \Fhat_{\bhat\chat}
  \pfrac{\xhat^\bhat}{x^b}  \pfrac{\xhat^\chat}{x^c}
  \pfrac{\xhat^\ahat}{x^a} ~ \dfrac{x}{\xhat}
\\
&\qquad
 +
  \tfrac{1}{2}{\PsiAny}^{abcd} \pfrac{}{x^b}
  \left(
    \Fhat_{\chat\dhat}
    \pfrac{\xhat^\chat}{x^c} \pfrac{\xhat^\dhat}{x^d}
  \right)
  \pfrac{\xhat^\ahat}{x^a} ~
  \dfrac{x}{\xhat}
\\
&=
  \tfrac{1}{2}{\PsiAny}^{abc} \,
  \Fhat_{\bhat\chat}
  \pfrac{\xhat^\bhat}{x^b}
  \pfrac{\xhat^\chat}{x^c}
  \pfrac{\xhat^\ahat}{x^a} ~
  \dfrac{x}{\xhat}
\\
&\qquad
 +
  \tfrac{1}{2}
  {\PsiAny}^{abcd}
  \Big(
    \pfrac{\Fhat_{\chat\dhat}}{x^b}
    \pfrac{\xhat^\chat}{x^c}
    \pfrac{\xhat^\dhat}{x^d}
   +
    \Fhat_{\chat\dhat}
    \pqfrac{\xhat^\chat}{x^b}{x^c}
    \pfrac{\xhat^\dhat}{x^d}
\\
&\hspace{12em}
 +
    \Fhat_{\chat\dhat}
    \pfrac{\xhat^\chat}{x^c}
    \pqfrac{\xhat^\dhat}{x^b}{x^d}
  \Big)
  \pfrac{\xhat^\ahat}{x^a} ~
  \dfrac{x}{\xhat}
\\
&=
  \tfrac{1}{2} \Fhat_{\bhat\chat}
  \Big(
    {\PsiAny}^{abc} \,
    \pfrac{\xhat^\bhat}{x^b}
    \pfrac{\xhat^\chat}{x^c}
   +
    \pqfrac{\xhat^\bhat}{x^b}{x^c}
    \pfrac{\xhat^\chat}{x^d}
\\
&\hspace{10em}
 +
    \pfrac{\xhat^\bhat}{x^c}
    \pqfrac{\xhat^\chat}{x^b}{x^d}
  \Big)
  \pfrac{\xhat^\ahat}{x^a} ~
  \dfrac{x}{\xhat}
\\
&\hspace{2em}
 +
  \tfrac{1}{2} {\PsiAny}^{abcd}
  \pfrac{\Fhat_{\chat\dhat}}{\xhat^\bhat}
  \pfrac{\xhat^\bhat}{x^b}
  \pfrac{\xhat^\chat}{x^c}
  \pfrac{\xhat^\dhat}{x^d}
  \pfrac{\xhat^\ahat}{x^a} ~
  \dfrac{x}{\xhat}
.
\end{align*}
%]

\end{proof}

%
% ----------------------------------------------------------------------
\section{Coordinate-free notation and pre-metric electromagnetism}
\label{ch_Coordfree}

The construction of our CMCR in the main text
 can be greatly simplified using exterior differential forms
 \cite{Flanders1963,Deschamps-1981ieee,HehlObukhov}.
In this notation \eqref{GR_dF} and \eqref{GR_Max-CR} become
%[
\begin{align}
 d \, F = 0
\qquadand
 {\PsiAny} \VAct{F} = \mathcal{J}
.
\label{GR_Coordfree_Max}
\end{align}
%]
 where the first order operator ${\PsiAny}$
 maps the 2--form $F$ to the 3--form $\mathcal{J}$,
 and obeys the first order operator axioms
 \eqref{FOO_def_+lin} and \eqref{FOO_def_scalar}.
The 3--form source density $\mathcal{J}$ is
%[
\begin{align*}
 \mathcal{J}
=
  \tfrac{1}{6}
  {\pJ}^a \, \LeviCiv_{abcd} \, dx^{bcd}
=
  J^a i_{a} dx^{0123}
.
\end{align*}
%]
Here the notation is $dx^{bcd}=dx^b\wedge dx^c\wedge dx^d$.
This is consistent with
 the basis of electromagnetic field $dx^{ab}=dx^a\wedge dx^b$.
This explains why ${\pJ}^a$
 is a vector density \eqref{Proofs_J^a_density}
 which follows from
\begin{align*}
  \pfrac{}{x^b}
=
  \pfrac{\xhat^a}{x^b}\, \pfrac{}{\xhat^a}
\textrm{{\quad}and{\quad}}
  dx^{0123}
=
  \dfrac{x}{\xhat}\ d\xhat^{0123}
.
\end{align*}

Charge conservation
\eqref{GR_1st_order} simply becomes
%[
\begin{align}
d({\PsiAny}\VAct{d {\pA}})=0
\qquad\textrm{for all 1--forms~}{\pA}
,
\label{GR_Coordfree_Conti}
\end{align}
%]
and the equation for the null terms \eqref{GR_Null_constraint} becomes
%[
\begin{align}
\NullPsi\VAct{d {\pA}}=0
\qquad\textrm{for all 1--forms~}{\pA}
.
\label{GR_Coordfree_Null_constraint}
\end{align}
%]

Looking at \eqref{GR_Coordfree_Max}, \eqref{GR_Coordfree_Conti}
 it is clear that these equations do not explicitly include a metric,
 as was stated in section \ref{S-other-TO}.
The metric dependence of Maxwell's equations
 would instead be included in the CMCR operator ${\PsiAny}$,
 which is fully consistent
 with the pre-metric formulation of Maxwell's equations \cite{HehlObukhov}.

\clearpage 

\widetext

\section{Popular summary}\label{S-popular}

\large

\begin{center}
{\huge{Charge and the Light Brigade: \\
a challenge to Maxwell's Electromagnetism?}}\\
~\\
\emph{\small{``Electromagnetism, Axions, and Topology: \\
 a first-order operator approach
          to constitutive responses provides greater freedom''}}\\
~\\
{Paul Kinsler, JG, MM}
\end{center}

\setlength{\parskip}{1ex}

~\\
~

Maxwell's theory of light, electricity, and magnetism, remains one of the
greatest breakthroughs in physics. Now 150 years old, it combines electricity
and magnetism into a unified ``electromagnetism'' that anticipated both
Einstein's relativity and the field theories of 20th century physics. However,
even respectable physicists have to sometimes question the basics, and in a
pair of recent papers, culminating with the most recent in Physical Review A,
Jonathan Gratus and Paul Kinsler of Lancaster University, along with Martin
McCall of Imperial College London (GKM), have advanced a program to take a
fresh -- and perhaps controversial -- look at Maxwell's magnum opus. This
theory includes not only the familiar everyday electric field ``E'' and
magnetic field ``B'' that we learn about in school, but also a pair of
so-called ``excitation'' fields labelled D and H. These excitation fields are
used by scientists and engineers for helping with the practical task of
representing how light interacts with material objects. ``But since only E and
B can be measured'', says Prof. Martin McCall, ``we wondered what unexpected
loopholes we might be able to expose''.

GKM got
the ball rolling in the journal Foundations of Physics\footnote{See
     {{\lq\lq}Evaporating black-holes, wormholes,
     vacuum polarisation: must they always conserve charge?{\rq\rq}},\\
      J. Gratus, P. Kinsler, M.W. McCall, \qquad
        {Found. Phys. \textbf{49}, 330 (2019)},\\
      {http://doi.org/10.1007/s10701-019-00251-5}, \qquad
        {https://arxiv.org/abs/1904.04103}.},
 by describing some circumstances where Maxwell's equations
need not guarantee global charge conservation -- the sort of conclusion that
usually makes physicists nervous.  Their first step was to turn D and H --
figuratively -- into shadows, so that they need no longer be unique. The next
was more exotic -- to consider how this re-interpreted Maxwell theory behaves
around evaporating black holes, or near wormholes in spacetime. With both the
shadow fields and exotic spacetime in place, GKM found that \emph{global} charge
conservation is no longer mathematically guaranteed by Maxwell's equations,
even though it still remained in place locally.  Fortunately the scenarious
involved either evaporating black holes or wormholes, taking any observable
effects well outside of everyday experience.

Now, in recent work in the journal Physical Review A\footnote{See
     {{\lq\lq}Electromagnetism, Axions, and Topology: \\
 a first-order operator approach
          to constitutive responses provides greater freedom{\rq\rq}},\\
      J. Gratus, M.W. McCall, P. Kinsler, \qquad
        {Phys. Rev. A \textbf{101}, 043804 (2020)},\\
      {http://doi.org/10.1103/PhysRevA.101.043804}, \qquad
        {https://arxiv.org/abs/1911.12631}.},
 GKM have shown that if
these ``shadow'' fields are not unique, then they might even be dispensed
with entirely. ``We can combine the excitation fields with our understanding of
how the electric and magnetic fields interact with matter'', says Dr Jonathan
Gratus, ``so we devised a way to transform Maxwell's equations into something
more general''. This upgraded theory now allows the electromagnetic fields to
interact with new types of material, and in particular with materials that
mimic axions. Axions are ghostly particles with curious symmetry properties,
that have so far have never been detected. In the standard view of
electromagnetism, axions only see the surfaces and boundaries between
different types of material, but in GKM's new theory, axions are also allowed
to interact \emph{inside} blocks of material.

The best thing about this new work is that axionic interactions are much
easier to investigate in tabletop experiments than are evaporating black holes
(or wormholes). Nevertheless, axions are still fairly esoteric -- we do not
even know if axions exist, despite the existence of various proposals for
trying to detect them. Consequently, GKM turned to the hot topic area of
material design known as ``metamaterials'', where existing substances, such as
dielectrics, metals, and semiconductors, are interwoven in a latticework
combination to create new -- and specific -- types of material response. ``We
have even designed some axion-like metamaterial elements'', says Dr Paul
Kinsler, ``such tiny electro-mechanical machines could be arranged in an array,
respond to applied fields, and so create the micro-currents needed for the
axion-like response''.  It is even possible to imagine an experimental set up
around an electrically charged cylinder, where the topological arrangement
forbids any description based on the standard macroscopic Maxwell equations".

Maxwell's equations for classical electromagnetism have stood unchanged and
unchallenged for over 150 years. But these latest results from Gratus,
Kinsler, and McCall present opportunities for new -- and unexpected --
advances in our understanding of light, electromagnetism, and material
properties.

\end{document}